\documentclass[twocolumn,tighten,twocolappendix]{aastex631}
\usepackage[encapsulated]{CJK}
\usepackage{ucs}
\usepackage[utf8x]{inputenc}
\newcommand{\jpntext}[1]{\begin{CJK}{UTF8}{min}#1\end{CJK}}
\usepackage{amsmath,mathtools}

\newcommand{\te}{$T_{\rm e}$([N\,{\sc ii}])}
\newcommand{\Ne}{$n_{\rm e}$([S\,{\sc ii}])}
\newcommand{\tee}{$T_{\rm e}$}
\newcommand{\Nee}{$n_{\rm e}$}
\newcommand{\chb}{$c({\rm H}\beta)$}
\newcommand{\ha}{H$\alpha$}
\newcommand{\hb}{H$\beta$}
\newcommand{\hg}{H$\gamma$}
\newcommand{\hii}{H\,{\sc ii}}
\newcommand{\heii}{He\,{\sc ii}}
\newcommand{\sii}{[S\,{\sc ii}]}
\newcommand{\siii}{[S\,{\sc iii}]}
\newcommand{\nii}{[N\,{\sc ii}]}
\newcommand{\niii}{[N\,{\sc iii}]}

\newcommand{\oiii}{[O\,{\sc iii}]}

\received{June 27, 2021}
\revised{August 2, 2021}
\accepted{August 24, 2021}

\submitjournal{PASP Tutorial}

\shorttitle{Proper Plasma Analysis Practice (PPAP)}
\shortauthors{Ueta and Otsuka}
\graphicspath{{./}{figures/}}
\begin{document}

\title{Proper Plasma Analysis Practice (PPAP),\\
an Integrated Procedure of the Extinction Correction and Plasma Diagnostics:\\
a Demo with an HST/WFC3 Image Set of NGC\,6720}

\author[0000-0003-0735-578X]{Toshiya Ueta (\jpntext{植田稔也})}
\affiliation{Department of Physics and Astronomy,
University of Denver,
2112 E.\ Wesley Ave., 
Denver, CO 80208, USA}
\affiliation{Okayama Observatory, Kyoto University,
Honjo, Kamogata, Asakuchi, Okayama, 719-0232, Japan}
\affiliation{JSPS Invitation Fellow for Research in Japan (FY2020, long-term)}

\author[0000-0001-7076-0310]{Masaaki Otsuka (\jpntext{大塚雅昭})}
\affiliation{Okayama Observatory, Kyoto University,
Honjo, Kamogata, Asakuchi, Okayama, 719-0232, Japan}

\begin{abstract}
In this work, we propose a proper plasma analysis practice (PPAP),
an updated procedure of plasma diagnostics in the era of 
spatially-resolved spectroscopy.
In particular, we emphasize the importance of performing 
both of the extinction correction and the direct method of 
plasma diagnostics simultaneously as an integrated process.
This approach is motivated by the reciprocal dependence 
between critical parameters in these analyses,
which can be resolved by iteratively seeking a converged solution.
The use of PPAP allows us to eliminate unnecessary assumptions 
that prevent us from obtaining an exact solution
at each element of the spectral imaging data.
Using a suite of {\sl HST\/}/WFC3 narrowband images of the 
planetary nebula, NGC\,6720, 
we validate PPAP by
(1) simultaneously and self-consistently deriving the extinction, 
{\chb}, and electron density/temperature distribution, ({\Ne}, {\te}),
maps that are consistent with each other, and 
(2) obtaining identical metal abundance distribution maps, 
($n({\rm N}^+)/n({\rm H}^+)$, $n({\rm S}^+)/n({\rm H}^+)$),
from multiple emission line maps at different wavelengths/transition energies.
We also determine that the derived {\chb} consists both of the ISM and 
circumsource components and that the ionized gas-to-dust mass ratio 
in the main ring is at least 437 and as high as about 1600.
We find that, unless we deliberately seek self-consistency,
uncertainties at tens of \% can easily arise in outcomes,
making it impossible to discern actual spatial variations that 
occurs at the same level, defeating the purpose of conducting
spatially resolved spectroscopic observations.
\end{abstract}


\section{Introduction} 
\label{sect:intro}

Plasma diagnostics are fundamental to 
understanding the physical conditions of various gaseous systems
(e.g.\ \citealt{osterbrock2006,Kewley2019}).
The relative strengths of various diagnostic emission lines 
determine the excitation states of specific gaseous species, 
yielding their electron densities and temperatures,
and subsequently, metal abundances
(e.g.\ \citealt{peimbert2017,Nicholls2020}).
However, these emission lines required in plasma diagnostics must first 
be corrected for both the interstellar and circumsource extinction
by adopting a suitable extinction law,
especially when the amount of extinction is not really negligible
(e.g.\ \citealt{draine2003,salim2020}).
This is the quintessence of observational astronomy,
in which all measurements made from a distance are affected
by extinction.

The determination of extinction is far from a trivial task.
Practically, the amount of extinction is usually determined,
for example, by comparing the observed diagnostic H\,{\sc i} 
recombination line ratio with its theoretical expectation
(i.e.\ the intrinsic line ratio without extinction).
The theoretical line ratios can be computed  
for specific electron density ({\Nee}) and temperature ({\tee})
of the target emitting gas (e.g.\ \citealt{Hummer1987,storey1995}).
Obviously, {\Nee} and {\tee} are the very quantities to be figured out 
by plasma diagnostics using {\em extinction-corrected} line ratios.
Hence, this is a classic catch-22 situation.
Therefore, to overcome this conundrum, 
plasma diagnostics ought to be performed 
{\em together with} the extinction determination/correction 
as {\em a single integrated process}.

However, what is traditionally exercised in the literature  
has been to introduce a number of simplifications.
For example, {\em ad hoc} {\Nee} and {\tee} may be chosen
to force a value of extinction to get the subsequent plasma diagnostics 
going, or an {\em ad hoc} extinction value may even be adopted.
In such cases, 
one should at least guarantee consistency between the {\em assumed} 
{\Nee} and {\tee} values as the bases for the {\em assumed} extinction
and the {\Nee} and {\tee} values as the actual outcomes
of the subsequent plasma diagnostics.
In other words, the initially assumed {\Nee} and {\tee} values
cannot be very different from the final {\Nee} and {\tee} values
to assure that these {\Nee} and {\tee} values represent the ionized
gas in the target object.
In practice, the subtlety of such consistency tends to be lost in translation,
because the extinction determination and plasma diagnostics
are often dealt with as two separate issues.
Hence, consistency between these two sets of {\Nee} and {\tee} 
is rarely scrutinized, let alone guaranteed.
Consequently, such inconsistencies would usually invite 
uncertainties, albeit inadvertently.

Meanwhile, spatially resolved 2-D plasma diagnostics have been becoming 
very relevant lately in many branches of astronomy and astrophysics, 
especially with the increasing availability of integral field 
spectrographs (e.g.\ \citealt{walsh2020}).
When measurements of extended target sources are made 
in a spatially resolved manner,
both the extinction determination and plasma diagnostics 
ought to be performed at each detector element. 
It is to ensure that the spatially extended nature 
of the target sources is fully appreciated at the end
of the analyses.

However, the spatial variation of relevant parameters 
in extended objects has rarely been considered carefully enough
in both the extinction determination and plasma diagnostics.
Often a single-valued extinction is adopted by assuming uniform 
{\Nee} and {\tee} across the whole extent of an extended object
(e.g.\ \citealt{lame1994,Guerrero1997}).
Such simplifications may be permissible as long as the aim is to 
sample representative quantities of an extended object in an integrated sense.
Regrettably, such a deed defeats the purpose of spatially resolved observations,
by forcing an extended object with artificial uniformity, i.e.,
imposing an absolutely unnecessary source of uncertainties.

In the present work, therefore, 
we propose a procedure that streamlines
both the extinction correction and plasma diagnostics 
as a single integrated process of iterative data reduction.
By seeking convergence at each spatial element
through this iterative process,
we can simultaneously and self-consistently determine 
the extinction map
as well as the electron density/temperature ({\Nee} and {\tee}) maps,
which allow us to carry out a plethora of spatially-resolved analyses.
Below, we demonstrate and validate the process in detail
using a full suite of the archived 
Hubble Space Telescope ({\sl HST\/})/WFC3 narrowband 
images of the Galactic planetary nebula (PN), NGC\,6720.

First, we briefly describe the adopted data
and the line calibration applied,
before we detail the iterative procedure (\S\,\ref{sect:analyses}).
We then derive
the extinction map (\S\,\ref{sect:ext}),
{\Ne} and {\te} maps (\S\,\ref{sect:denstemp}),
and extinction-corrected line emission maps (\S\,\ref{sect:linemaps}),
as well as
metal abundance distribution maps (\S\,\ref{sect:abund}),
while giving detailed discussion of the outcomes 
and contrasting between the present results and 
those obtained with typical simplifications (\S\,\ref{sect:others}).
In the end, we summarize and promote one of the most 
self-consistent and fully 2-D plasma diagnostics ever performed 
(\S\,\ref{sect:summary}).

\section{Analyses}
\label{sect:analyses}

\subsection{WFC3 Narrowband Image Set of NGC\,6720}
\label{sect:data}

For the present study, we adopt images of the planetary nebula, NGC\,6720,
stored in the Hubble Legacy Archive\footnote{https://hla.stsci.edu/}.
These images were taken with the WFC3 camera 
on 2011 September 19 and 25 
as part of the program 12309 (PI: C.\ R.\ O’Dell; 
\citealt{odell2013a,odell2013b,odell2013c}).
We adopt this data set 
because this program is one of the few that used the exquisite suite 
of the WFC3 narrowband filters most extensively,
providing an excellent opportunity 
to perform self-consistent plasma diagnostics in full 2-D
based on narrowband images
and compare the results with those of the previous investigations.

Table\,\ref{tab:filters} lists all 
the filters used in this program, 
with their average wavelength and rectangular width
(according to {\sc pysynphot}; \citealt{pysynphot})
and ``official'' description \citep{dressel2019}.
However, the F953N image is not used because of 
the known severe fringe pattern
(\citealt{wong2010}; also \S\,5.4.4 of \citealt{dressel2019}).
The ``Q'' in a filter name stands for a ``QUAD'' filter,
which refers to one of the $2 \times 2$ mosaic of four filters
providing four different bandpasses simultaneously
with each band covering a quarter of the nominal 
field of view (FoV: \citealt{dressel2019}).
Because of the smaller field coverage of the QUAD filters,
the subsequent analyses are all restricted within the NW quadrant 
of the main ring structure of NGC\,6720
(see below; 
also see the Shared Field of View of All Filters in 
Fig.\,2 of \citealt{odell2013a} and 
Figs.\,4d,e of \citealt{ueta2019}).

\begin{table}[t]
    \centering
    \begin{tabular}{lccl}
    \hline\hline
    Filter & $\lambda$ & $\Delta\lambda$ & Official Description \\
           & (\AA) & (\AA) & \\
    \hline
    FQ436N & 4367.21 & \phantom{1}43.35 & {\hg}\,4340\,{\AA}$+${\oiii}\,4363\,{\AA} \\
    FQ437N & 4371.09 & \phantom{1}29.99 & {\oiii}\,4363\,{\AA} \\
    F469N  & 4688.14 & \phantom{1}49.68 & {\heii}\,4686\,{\AA} \\
    F487N  & 4871.42 & \phantom{1}60.40 & {\hb}\,4861\,{\AA} \\
    F502N  & 5009.70 & \phantom{1}65.29 & {\oiii}\,5007\,{\AA} \\
    F547M  & 5451.08 & 649.14           & Str\"{o}mgren\,y (continuum) \\
    FQ575N & 5757.87 & \phantom{1}18.37 & {\nii}\,5755\,{\AA} \\
    F645N  & 6453.71 & \phantom{1}84.22 & continuum \\
    F656N  & 6561.45 & \phantom{1}17.65 & {\ha}\,6563\,{\AA} \\
    F658N  & 6584.95 & \phantom{1}27.56 & {\nii}\,6583\,{\AA} \\
    FQ672N & 6716.62 & \phantom{1}19.37 & {\sii}\,6717\,{\AA} \\
    F673N  & 6766.05 & 117.77           & {\sii}\,6717/31\,{\AA} \\
    FQ674N & 6730.77 & \phantom{1}17.63 & {\sii}\,6731\,{\AA} \\
    FQ750N & 7502.55 & \phantom{1}70.43 & continuum \\
    F953N  & 9530.92 & \phantom{1}97.04 & {\siii}\,9532\,{\AA} \\
    \hline
    \end{tabular}
    \caption{{\sl HST\/}/WFC3 filters used in the Program 12309.
    The quoted values are the bandpass average wavelength ($\lambda$)
    and bandpass rectangular width ($\Delta\lambda$) defined in
    {\sc pysynphot} \citep{pysynphot}.}
    \label{tab:filters}
\end{table}

\subsection{The QP Method}
\label{sect:qp}

Many WFC3 filters can isolate the target emission line reasonably well.
However, some filters suffer from unavoidable blending of neighboring lines
at comparable strengths (Appendix A.2 of \citealt{dressel2019}).
The most critical is the F656N and F658N filter pair.
Even though the F656N and F658N filters are officially described as
the {\ha} and {\nii} filters, respectively (Table\,\ref{tab:filters}),
their transmission profiles cover both the {\ha} and {\nii} lines
\citep{pysynphot}.
As a result, 
both of the F656N and F658N images are blends of {\ha} and {\nii} lines
at different proportions.

To address this line-blending issue of narrowband filter images, 
we developed a new calibration method based on Quadratic Programming (QP), 
dubbed the QP line extraction method \citep{ueta2019}.
Using the {\sl HST} images of NGC\,6720 in the F656N and F658N filters
in comparison with the ground-based slit-scan spectral imaging
data cube around {\ha}, 
we demonstrated that the QP method properly extracted 
the individual {\ha}-only and {\nii}-only line maps
from the line-blended raw F656N and F658N images \citep{ueta2019}.

The QP method recognizes narrowband images that are affected by line blending 
as linear combinations of emission maps of blended lines,
each of which is modulated by the system throughput at the wavelength
of the blended lines.
Provided that the underlying continuum is properly subtracted,
and assuming that all the lines involved are sufficiently narrow
(i.e.\ a line can be specified by a single wavelength),
the solution for such a set of linear equations can then be sought
as a QP problem in a least-squares sense 
under the presence of appropriate constraints.
For the present work, 
we isolate the {\ha} and {\nii} 6548/83\,{\AA} line maps from the F656N and F658N images,
the {\sii} 6717/31\,{\AA} line maps from the FQ672N, F673N, and FQ674N images,
and 
the {\oiii} 4363\.{\AA} and {\hg} maps from the FQ436N and FQ437N images.
As the QP processing itself is already described fully elsewhere \citep{ueta2019},
we outline how these images are processed in Appendix\,\ref{sect:fullqp}.

Here, we emphasize that the QP line calibration method is more advantageous 
than other methods that attempt to remove the fractional contribution 
of the unwanted blended lines by just scaling the raw images
(e.g.\ Appendix of \citealt{odell2013a}).
Such scaling methods are strictly dependent 
on the line emission distribution of the raw line-blended images.
That is, 
the line emission distribution of an isolated target line
is always the same as 
that of the raw line-blended image that mimics the the distribution of 
the target line.

In reality, however, the line emission distribution of the blended lines
represents neither that of the target line nor that of the blended line 
(as it is a mix of the two):
the resulting de-blended line emission maps do not have to appear 
the same as the raw line-blended images.
Instead, the QP method can recover
the emission distribution of individual lines
as the optimized solution of a QP problem on a pixel-to-pixel basis.
This is what makes the QP method unique 
in comparison with other line calibration methods
for narrowband images.

\subsection{Iterative Determination of {\chb} and {\rm (}{\Nee}, {\tee}{\rm )}}
\label{sect:iteration}

The apparent interdependence among 
the extinction and ({\Nee}, {\tee}) 
may not have been taken into consideration
with a sufficient amount of attention 
that it deserves for some unknown reason(s).
As outlined in \S\,\ref{sect:intro},
to derive the extinction and ({\Nee}, {\tee}) self-consistently 
by breaking the circular logic behind the theories of extinction
and astrophysical plasma,
both the extinction determination
and plasma diagnostics must be performed 
simultaneously as one integrated process.
In the following subsections,
we establish an iterative process 
through which 
this interdependence among the extinction and ({\Nee}, {\tee}) 
is carefully addressed.

\subsubsection{Extinction Correction}
\label{sect:extcorr}

We start by defining $c(\lambda)$, the extinction at some wavelength, 
$\lambda$, as the base-10 power-law index to describe the reduction of  
the intrinsic flux, $I_0(\lambda)$, to the observed flux, $I(\lambda)$, by
\begin{eqnarray}
I(\lambda) = I_0(\lambda) \times 10^{-c(\lambda)}.
\label{eq:extdef}
\end{eqnarray}
Then, we can determine $c({\rm H}\beta)$,
the extinction at {\hb},
using, for example, 
the observed-to-intrinsic {\ha}-to-{\hb} flux ratio
via
\begin{eqnarray}
c({\rm H}\beta)
=
\frac{-\log_{10}\frac{I({\rm H}\alpha)/I({\rm H}\beta)}{I_0({\rm H}\alpha)/I_0({\rm H}\beta)}}{\frac{A_{{\rm H}\alpha}-A_{{\rm H}\beta}}{A_{{\rm H}\beta}}}.
\label{eq:chb}
\end{eqnarray}
Here, 
$c($\ha$)/c($\hb$)$ has been replaced by $A($\ha$)/A($\hb$)$,
where $A(\lambda)$ is the total extinction at $\lambda$
(the extinction on the magnitude basis)
and $A(\lambda) = 2.5 \times c(\lambda)$ because
\begin{eqnarray}
I(\lambda) = I_0(\lambda) \times 10^{-\frac{A(\lambda)}{2.5}}.
\end{eqnarray}

We can thus define
the observed flux at $\lambda$ relative to the {\hb} flux as
\begin{eqnarray}
\frac{I(\lambda)}{I({\rm H}\beta)}
&=&
\frac{I_{0}(\lambda) \times 10^{-c(\lambda)}}{I_{0}({\rm H}\beta) \times 10^{-c({\rm H}\beta)}} \nonumber\\
&=&
\frac{I_{0}(\lambda)}{I_{0}({\rm H}\beta)}
10^{-\left(c(\lambda)-c({\rm H}\beta)\right)},
\end{eqnarray}
from which $c(\lambda)$ can be determined via
\begin{eqnarray}
c(\lambda) 
&=&
-\log_{10}
\left(
\frac{I(\lambda)/I({\rm H}\beta)}{I_0(\lambda)/I_0({\rm H}\beta)}
\right)
+
c({\rm H}\beta)\nonumber\\
&=& c({\rm H}\beta)
\left(
\frac{A_{\lambda}-A_{{\rm H}\beta}}{A_{{\rm H}\beta}} + 1
\right)\nonumber\\
&=&
c({\rm H}\beta)
\left(
\frac{A_{\lambda}}{A_{{\rm H}\beta}}
\right)\nonumber\\
&\approx&
c({\rm H}\beta) \cdot
\left<
\frac{A_{\lambda}}{A_{V}}
\right>/
\left<
\frac{A_{{\rm H}\beta}}{A_{V}}\right>,
\label{eq:extlaw}
\end{eqnarray}
given that we know {\chb} and the extinction curve,
$\left<A_{\lambda}/A_{V}\right>$
(the average extinction at $\lambda$ relative to $V$),
which is provided by some extinction law
(e.g.\ \citealt{draine2003,salim2020}).
Then, we can recover the intrinsic line flux map at any $\lambda$
from the observed line flux map at $\lambda$ via Eq.\,(\ref{eq:extdef}).

An obvious caveat here is 
the choice of the extinction law,
and the total-to-selective extinction, $R_V$,
which scales the selected extinction curve.
Thus, the value of $R_V$ along the line of sight to 
the target object must be evaluated properly
before we go any further.
For the present study, we opt to adopt the Galactic extinction law 
of \citet[][CCM, hereafter]{ismccm},
as the present analyses are concerned with data in the optical,
in which there is little difference among the existing
extinction laws.
As for $R_V$, 
we use the value of $3.13 \pm 0.03$ to the direction of NGC\,6720, 
interpolated from $R_V$ values determined for nearby stars 
within a 0.5$^{\circ}$ radius (\citealt{gon2012,gon2017}).
We assume that $R_V$ is uniform over NGC\,6720,
as the adopted data by \citet{gon2012} and \citet{gon2017} 
do not have sufficient spatial resolution.

Another even more subtle but no less insignificant caveat is 
the choice of the theoretical {\ha}-to-{\hb} line flux ratio,
which is required to derive $c(\lambda)$ 
via Eqs.\,(\ref{eq:chb}) and (\ref{eq:extlaw}).
Here, we need to remind ourselves that 
the theoretical {\ha}-to-{\hb} line flux ratio
is really a function of {\Nee} and {\tee} of the emitting gas 
of the target (e.g.\ \citealt{storey1995}). 
Hence, the determination of the interstellar extinction is actually
dependent on {\Nee} and {\tee} of the target object,
which are the very quantities that we seek.
Therefore, 
the present series of processes to determine 
{\chb} and ({\Nee}, {\tee})
must truly be performed as an integrated iterative procedure
seeking the convergence of all of these values.

However, such an iteration is hardly exercised in practice:
usually {\Nee} and {\tee} are assumed to be some ``typical'' values.
This exercise {\em may be} permissible 
as long as target sources are unresolved
and the adopted {\Nee} and {\tee} values 
are consistent with values derived as a result of the 
subsequent plasma diagnostics. 
However, when we expect these quantities to vary spatially 
in extended objects,
we ought to perform this iterative process rigorously
in each spatial element of the input images.

Therefore, for the present study, 
we first adopt $n_{\rm e} = 10^{3}$\,cm$^{-3}$ and $T_{\rm e} = 10^{4}$\,K
as {\em the initial values}.
The {\em initial} {\Nee} and {\tee} values then yield 
the {\em initial} theoretical $I_0({\rm H}\alpha)/I_0({\rm H}\beta)$ ratio
of 2.858 under the Case B/optically thick condition \citep{storey1995}
with the CCM extinction law and $R_V = 3.13$.
Here, we stress that the theoretical H\,{\sc i} line 
ratio needed in determining the extinction {\em does} 
depend on {\Nee} and {\tee}.
In the literature, 
this fundamental dependence is often neglected and 
the ratio is set to 2.85 or thereabout 
without any reference to the assumed {\Nee} and {\tee} values,
which may be different from 
$n_{\rm e} = 10^{3}$\,cm$^{-3}$ and $T_{\rm e} = 10^{4}$\,K.
In the present work, however,
these values are iteratively updated for convergence
(so is the theoretical H\,{\sc i} line ratio)
{\em to ascertain the optimum consistency}
between plasma diagnostics and extinction correction.

In addition, we note that extinction toward a target is not 
necessarily just of interstellar origin.
The target object itself may be surrounded by its own obscuring 
agents (i.e.\ dust grains) contributing to the circumsource extinction,
on top of what is caused by the interstellar medium (ISM).
For evolved stars such as NGC\,6720, 
the circumstellar matter (CSM) most likely causes
a significant amount of the CSM extinction.
Hence, if the ISM extinction is hastily adopted as the intrinsic 
extinction toward a target source with a substantial 
circumsource extinction,
the corresponding {\chb} value toward the source is 
bound to be underestimated.\footnote{For example, 
using the Galactic Dust Reddening and Extinction tool made 
available via \url{https://irsa.ipac.caltech.edu/applications/DUST/}}.

The mean $E(B-V)$ reddening toward NGC\,6720 is 
reported to be $0.0870 \pm 0.0005$ \citep{sfd1998}.
Under the CCM extinction law with $R_V=3.13$,
the reported $E(B-V)$ translates to {\chb} as  
$0.1266 \pm 0.0007$ via
\begin{eqnarray}
 R_V = \frac{A_V}{E(B-V)} 
 = \frac{2.5 c(V)}{E(B-V)} 
 = \frac{2.5 c({\rm H}\beta)}{E(B-V)}\left<\frac{A_V}{A_{{\rm H}\beta}}\right>.
\nonumber 
\end{eqnarray}
This means that {\chb} to be obtained in the present analysis 
for NGC\,6720 has to be greater than 0.1266.
Any excess extinction to be detected toward NGC\,6720
in the present work,
therefore, is the CSM extinction component contributed 
by the circumstellar dust grains, 
which is not accounted for by any of the ISM extinction studies
(e.g.\ \citealt{sfd1998}).
This is true, of course, only if the obscuring property of the 
circumstellar matter follows that of the adopted ISM dust grains 
(CCM for the present case).

\subsubsection{Plasma Diagnostics}
\label{sect:plasma}

In performing plasma diagnostics, atoms are represented 
as $n$-level energy states. 
Then, {\Nee} and {\tee} are determined by solving a set of 
equilibrium equations for the adopted $n$ levels.
In these equilibrium equations, the collisional excitation coefficient has
$n_{\rm e} T_{\rm e}^{1/2} \exp(-\Delta E/kT_{\rm e})$ dependence 
(where $\Delta E$ is the energy difference between any two levels)
and 
the collisional de-excitation coefficient has
$n_{\rm e} T_{\rm e}^{1/2}$ dependence, while
the radiation de-excitation coefficient has dependence on
neither {\Nee} nor {\tee}
to the first order \citep{osterbrock2006,atomicastro}.

Thus, 
if we take the ratio between lines whose transition energies are
close to each other (i.e.\ $\Delta E \sim 0$;
e.g.\ the {\sii} 6731-to-6717\,{\AA} ratio),
such ratios can depend {\em mostly} on {\Nee}
and only weakly on {\tee}.
On the contrary, 
if we take the ratio between lines whose transition energies are
very different
(e.g.\ the {\nii} 5755-to-6583\,{\AA} ratio),
such ratios can depend {\em mainly} on {\tee} and only weakly on {\Nee}.
Then, we can find {\Nee} and {\tee} as the point in the {\Nee}-{\tee} space 
at which these two diagnostic line ratio curves intersect
(see, also, \S\,\ref{sect:O3}).
As this point can not be computed analytically,
{\Nee} and {\tee} have to be evaluated numerically by iteration.

For this work, 
we use the {\sii}\,6731-to-6717\,{\AA} line ratio map 
and the {\nii}\,5755-to-6583\,{\AA} line ratio map
as the input diagnostic line ratio maps.
In addition, the present set of data offers
the {\oiii}\,4363-to-5007\,{\AA} line ratio
as another diagnostic for higher excitation regions 
than those probed via the {\nii}\,5755-to-6583\,{\AA} line ratio.
However, the {\sl HST\/}/WFC3 narrowband filter set does not offer 
any diagnostic ratio appropriate to prove higher excitation regions
with the {\oiii}\,4363-to-5007\,{\AA} line ratio.
Hence, we will not use the {\oiii} \,4363-to-5007\,{\AA} line ratio
(but will come back to this point later in \S\,\ref{sect:O3}).
This forced choice of diagnostic line ratios, however, 
is not necessarily bad, 
because the spatial extent of {\sii} and {\nii} is similar to
that of {\ha} and {\hb}, covering the main ring structure of
NGC\,6720.
For the subsequent discussion, we always refer to {\Nee} and {\tee}
as {\Ne} and {\te} with the diagnostic line used 
to explicitly indicate relevant energy regimes.

\subsubsection{Iterative Procedure}
\label{sect:procedure}

\begin{figure}
    \centering
	\includegraphics[width=\columnwidth]{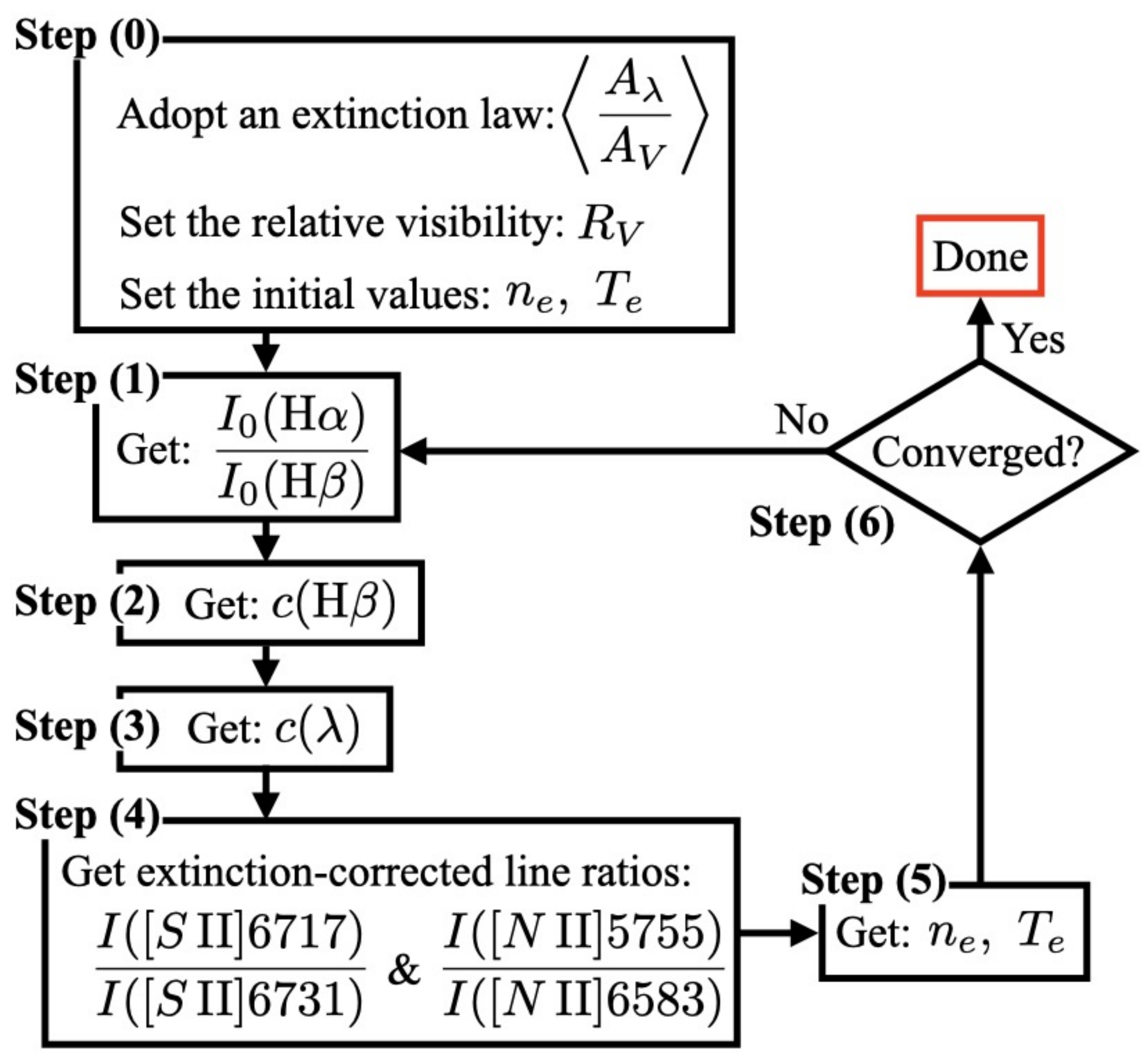}
	\caption{Schematic representation of PPAP, 
	through which {\chb} and ({\Ne}, {\te}) are determined
	by iteration.}
	\label{fig:schema}
\end{figure}

It is not too difficult to imagine that
deriving {\chb} and ({\Nee}, {\tee}) that simultaneously satisfy  
extinction correction and plasma diagnostics at each spatial element 
were too cumbersome in the past, 
especially when computational resources were not readily available.
Thus, constant {\Nee} and {\tee} might be assumed for the extinction 
correction, and plasma diagnostics were performed subsequently only 
once to yield constant {\Nee} and {\tee} even for extended objects, 
never to be retroactively checked for consistency.
This practice appears to have been adopted for years in the literature.
In many of the previous investigations of NGC\,6720,
a preset {\tee} was adopted to derive {\Nee}, and then,
a preset {\Nee}, instead of the derived {\Nee},
was used to derive {\tee}, and the computation was left at that 
(e.g.\ \citealt{lame1994,Garnett2001,odell2013b}),
while the subtle intertwined nature of these analyses seems to 
have been lost in translation. 

However, in the present time when decent computational resources
are regularly available, there is no reason not to perform iterative
searches for {\chb} and ({\Nee}, {\tee}) at each spatial element
through both the extinction determination and plasma diagnostics 
in a fully self-consistent manner.
Therefore, we propose proper plasma analysis practice (PPAP),
a fully self-consistent and spatially-resolved procedure 
of the extinction correction and plasma diagnostics 
as a streamlined iterative process
(schematically represented in Fig.\,\ref{fig:schema}):
\begin{enumerate}
    \item[(0)] Select the extinction law to use and 
    set the $R_V$ value toward the target source
    as well as the initial {\Nee} and {\tee} values;
    \item[(1)] Compute the theoretical $I_0({\rm H}\alpha)/I_0({\rm H}\beta)$
    ratio based on the initial {\Nee} and {\tee} values
    following, e.g., \citet{storey1995};
    \item[(2)] Compute {\chb} by comparing the observed 
    $I({\rm H}\alpha)/I({\rm H}\beta)$ and theoretical
    $I_0({\rm H}\alpha)/I_0({\rm H}\beta)$ maps
    using
    the PyNeb {\sc setCorr} and {\sc cHbeta} functions 
    (Eq.\,(\ref{eq:chb})),
    and then, $c(\lambda)$ using 
    the PyNeb {\sc getCorr} function (Eq.\,(\ref{eq:extlaw}))
    with the adopted extinction law;
    \item[(3)] Correct the observed line flux maps 
    for extinction using the $c(\lambda)$ map obtained in Step (2);
    \item[(4)] Determine {\Ne} and {\te} 
    using the extinction-corrected {\sii} 6717-to-6731\,{\AA} and 
    {\nii} 5755-to-6583\,{\AA} line ratio maps
    in the PyNeb {\sc getCrossTemDen} function;
    \item[(5)] Compare the old and new {\Ne} and {\te} values, 
    and terminate the process if the convergence is achieved 
     (i.e.\ the difference is negligible);
    \item[(6)] If the convergence condition is not met,
    repeat Steps (1) through (4) with the new {\Ne} and {\te} values.
\end{enumerate}

When the convergence is achieved at Step (6),
the best-fit {\chb}, {\Ne}, and {\te} distribution maps are left 
to us for further analyses.
The converged {\chb} allows us to produce optimally extinction-corrected 
line emission maps.
With the help of the converged {\Ne}, and {\te} maps,
extinction-corrected line emission maps in turn 
yield other products of plasma diagnostics
such as ionic and elemental abundance distribution maps.
In implementing PPAP for the present study, 
we employ PyNeb (\citealt{pyneb}),
a python implementation of a set of tools for analyzing emission lines
based on the method used by the IRAF {\sc nebular} package 
(\citealt{shaw1995,shaw1998}).

\section{Results}
\label{sect:results}

\subsection{The Converged {\chb} Map}
\label{sect:ext}

\begin{figure}
    \centering
	\includegraphics[width=\columnwidth]{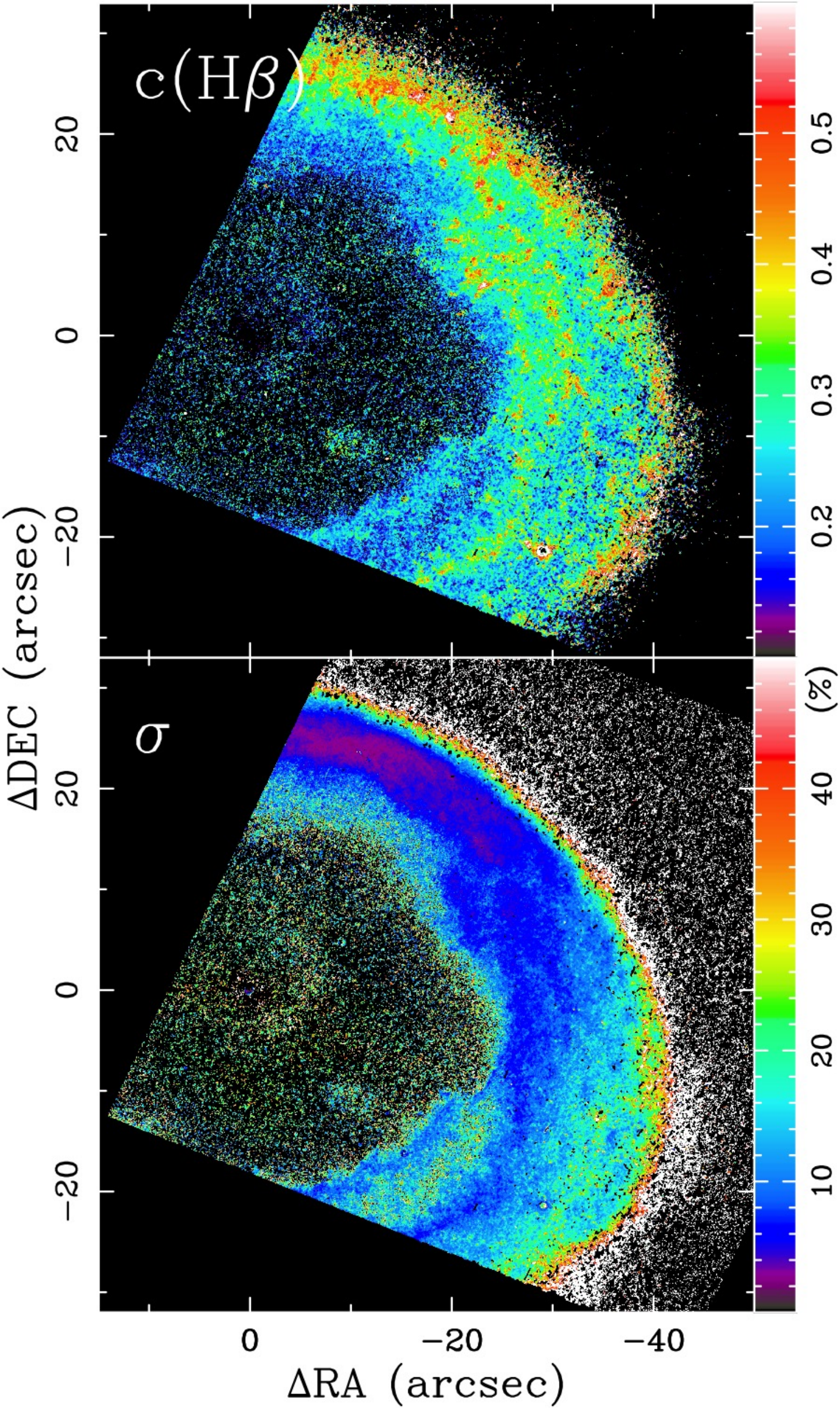}
	\caption{The {\chb} distribution map (top) and 
	the corresponding one-$\sigma$ percentage uncertainty map (bottom)
	of the NW quadrant of NGC\,6720,
	derived self-consistently through PPAP.
	Maps are shown
	in the original pixel scale of the input {\sl HST\/} images
	at $0\farcs0396\,$pix$^{-1}$
	with the relative RA and Dec offsets 
    from the position of the central star,
    (18:53:35.0970, $+$33:01:44.8831), indicated by the tickmarks.
    The color wedges show the range of the values presented.}
	\label{fig:cHb}
\end{figure}

\begin{figure}
    \centering
	\includegraphics[width=\columnwidth]{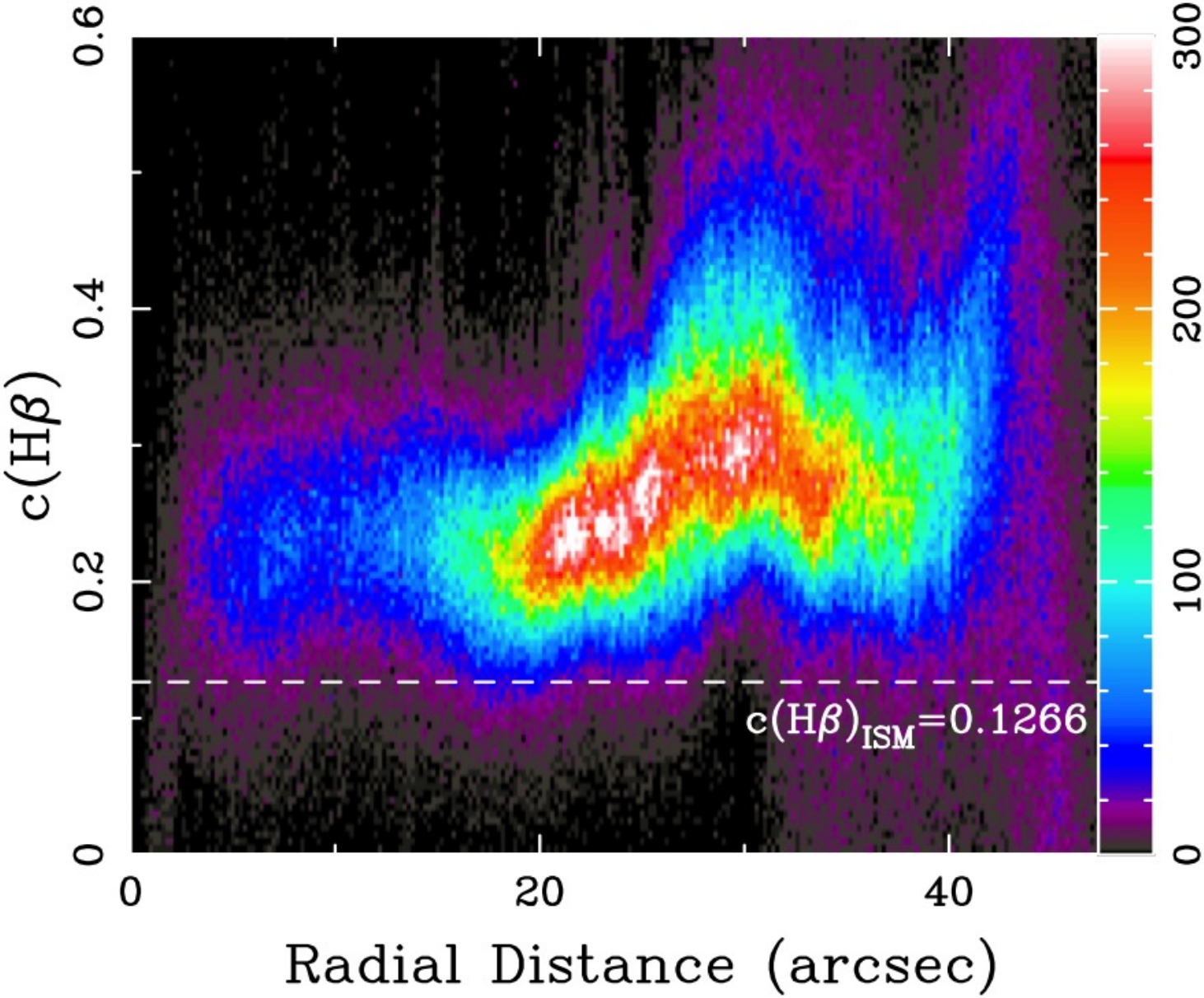}
	\caption{The radial density distribution map of {\chb}, 
    with the frequency indicated by the color,
    showing the {\chb} radial profile without 
    suppressing the azimuthal variation.
	The ISM contribution to {\chb} ($0.1266 \pm 0.0007$) 
	is marked by the dashed line.}
    \label{fig:cHbradial}
\end{figure}

\subsubsection{Observed Spatial Variations}

First and foremost, we examine the converged {\chb} map
as the basis for all the subsequent analyses.
In Fig.\,\ref{fig:cHb}, we present the {\chb} map (top)
and the corresponding one-$\sigma$ percent uncertainty map (bottom)
of the NW quadrant of NGC\,6720 converged via PPAP
based on the observed 
{\ha}-to-{\hb},
{\sii}\,6717-to-6731\,{\AA}, and
{\nii}\,5755-to-6583\,{\AA} ratio maps.
The one-$\sigma$ uncertainty is calculated at each pixel 
by varying each of the three input line ratios
by its standard deviation ($\pm\sigma_{\rm line~ratio}$)
and propagating the resulting uncertainties accordingly.
In Fig.\,\ref{fig:cHbradial}, we also present the {\chb} radial density
distribution.
We adopt this format instead of a typical azimuthally-averaged 
radial profile to better reveal the radial variation of {\chb} itself
as well as its spread.

These figures indicate that the converged {\chb} values largely populate
the characteristic ``main ring'' structure of NGC\,6720
seen between $\sim$20$^{\prime\prime}$ and $\sim$40$^{\prime\prime}$ 
from the center
slightly elongated along the position angle (PA; E from N) of $-120^{\circ}$,
but not so much in the central cavity (within $\sim$20$^{\prime\prime}$) and 
not at all in the region beyond the main ring (beyond $\sim$40$^{\prime\prime}$).
Because the effectiveness of PPAP is dictated 
by the quality of input line emission maps, 
the whole process can be bottlenecked at Step (4)
by less sensitive line maps 
(Fig.\,\ref{fig:schema}; \S\,\ref{sect:iteration}).
For the present case, we do not obtain converged results in all 
pixels within the inner cavity because S/N of {\sii} line emission
is unfortunately marginal (especially Fig.\,\ref{fig:s2}c).

Nonetheless, we see that 
{\chb} is more or less constant at slightly greater than 0.2
over $\sim$20--25$^{\prime\prime}$ and then increases radially 
to $\sim$0.25--0.35 at the outer edge of the main ring
($30-40^{\prime\prime}$),  
with the mean of $0.29 \pm 0.09$
(corresponding to $\sim$49\,\% attenuation at 4861\,{\AA}).
At many positions near the outer edge of the main ring, 
especially along the short axis (PA of $-30^{\circ}$),
{\chb} is found to be as great as 0.4 or even greater
(more than 60\,\% attenuation).
On the contrary, the maximum {\chb} at the elongated tip
of the main ring along the long axis (PA of $-120^{\circ}$)
is only marginally high ($\sim$0.25--0.35,
corresponding to 45--50\,\% attenuation).

This azimuthal trending of {\chb} is generally consistent with 
the dust distribution in NGC\,6720 revealed 
by far-IR dust continuum emission maps 
taken with the Herschel Space Observatory \citep{vanhoof2010}.
While the spatial resolution differs quite a bit in the optical
and in the far-IR, the dust distribution shows a greater degree
of dust concentration around the short axis than around the long axis.
As suggested, for example, 
by \citet{Guerrero1997} and \citet{odell2013a},
the presence of a denser molecular gas of the photo-dissociation region (PDR)
beyond the main ring 
may be directing outflows emanating from the central region
preferentially into the low density polar directions that is
slightly inclined with respect to the line of sight,
effectively generating the elongated appearance of the main ring
(cf.\ Fig.\,14 by \citealt{Guerrero1997}; Fig.\,11 by \citealt{odell2013a}).
Past kinematic studies indeed revealed
an expanding ellipsoidal shell whose long-axis is almost aligned 
with the line of sight (e.g.\ \citealt{Guerrero1997,odell2007,Martin2016}).

In the central cavity (within 20$^{\prime\prime}$), 
on the other hand, we find that 
{\chb} is roughly constant at $0.22 \pm 0.06$. 
The absence of {\chb} near the central star 
(within $\sim$5$^{\prime\prime}$)
does not mean the absence of attenuating dust grains there:
{\chb} is simply not reliably computed around the central star 
because the input line maps are affected 
by the presence of the central star
(e.g.\ imperfect subtraction of the continuum).
Again, far-IR dust continuum maps show more or less uniform
dust emission in the inner cavity \citep{vanhoof2010}, 
while it is expected that dust grains exist along the line of sight 
on the near and far sides of the rarefied high-temperature plasma 
region around the central star.

The observed general radial behavior of {\chb} 
is consistent with the expected stratification of the nebula.
That is, {\chb} values tend to be higher in the main ring 
where the bulk of the nebula material is located
than in the inner cavity 
where the higher-temperature plasma is rarefied.
Then, the radially increasing {\chb} is naturally explained 
by the radial decrease of the gas temperature, i.e., 
the degree of ionization.

The present {\chb} map at the exquisite 0\farcs0396\,pix$^{-1}$ scale 
(Fig.\,\ref{fig:cHb}) successfully reveals details of its spatial distribution.
A comparison between the {\chb} map and the observed {\ha}-to-{\hb} map
(Fig.\,\ref{fig:Ha2Hb}, top) clearly indicates that the presence of dust 
grains attenuates the bluer {\hb} line emission more than the redder
{\ha} line emission, leaving a higher {\ha}-to-{\hb} ratio at that location.
Such high {\chb} (high {\ha}-to-{\hb} ratio) regions appear to form
micro-structures that resemble radial cometary structures
and/or Rayleigh-Taylor (RT) instability fingers
as observed in the Helix Nebula \citep{helix}
and the Crab Nebula \citep{crab}, respectively.
There is a greater number of such clumps closer to the outer 
edge of the main ring.
This is probably caused by the inhomogeneous gas distribution 
in the main ring being eroded by photoevaporating radiation 
from the central star.
Ionizing radiation does not just simply travel radially 
in the nebula, but permeates through the inhomogeneous gas 
by going preferentially into low density regions, leaving
high-density clumps behind.

As discussed at the end of \S\,\ref{sect:extcorr}, 
we expect {\chb} of $0.1266 \pm 0.007$ from the ISM alone.
How the converged {\chb} compares with this expectation
is demonstrated well in Fig.\,\ref{fig:cHbradial}:
the ISM component is represented by the horizontal dashed line.
We immediately see that the derived {\chb} is greater than the ISM value
for the entire extent of the nebula where we have measurements.
This result indeed proves that NGC\,6720 itself provides 
the source of self-attenuation in its circumstellar nebula. 
The circumstellar {\chb} is about 0.1 in the inner cavity
($\sim$20\,\% reduction),
as high as 0.3 at the outer edge in the short axis direction
($\sim$50\,\% reduction), and
0.12--0.17 at the outer edge in the long axis direction
($\sim$30\,\% reduction).
Hence,  
the circumstellar contribution to {\chb} in NGC\,6720 amounts to 
about the same or even greater than the ISM contribution.
Given that NGC\,6720 is an object at a moderate Galactic latitude
reasonably away from the Bulge ($(l, b) = (+63.1701, +13.9781)$),
it may not be very appropriate, in general, to adopt the ISM {\chb} 
value indiscriminately as the total {\chb} for any target source.

The circumstellar {\chb} component of $\sim$0.1 measured 
in the inner cavity is most likely attributed to dust grains 
floating {\em in front of} the central high-temperature region 
around the central star along the line of sight.
Thus, we expect that there is another {\chb} $\sim$0.1 worth
of dust grains {\em behind} the central high-temperature region.
If so, {\chb} in the inner cavity as a whole would be
about 0.33 including the ISM component ($= 0.13+0.1+0.1$),
which is very much consistent with the amount of {\chb} 
we observe at the outer edge of the main ring.
Hence, it appears that the main ring of NGC\,6720 contain
on average {\chb} $\sim$0.2 worth of dust grains 
in any radial direction.

\subsubsection{Comparison with the Previous Results}

\begin{figure}
    \centering
    \includegraphics[width=\columnwidth]{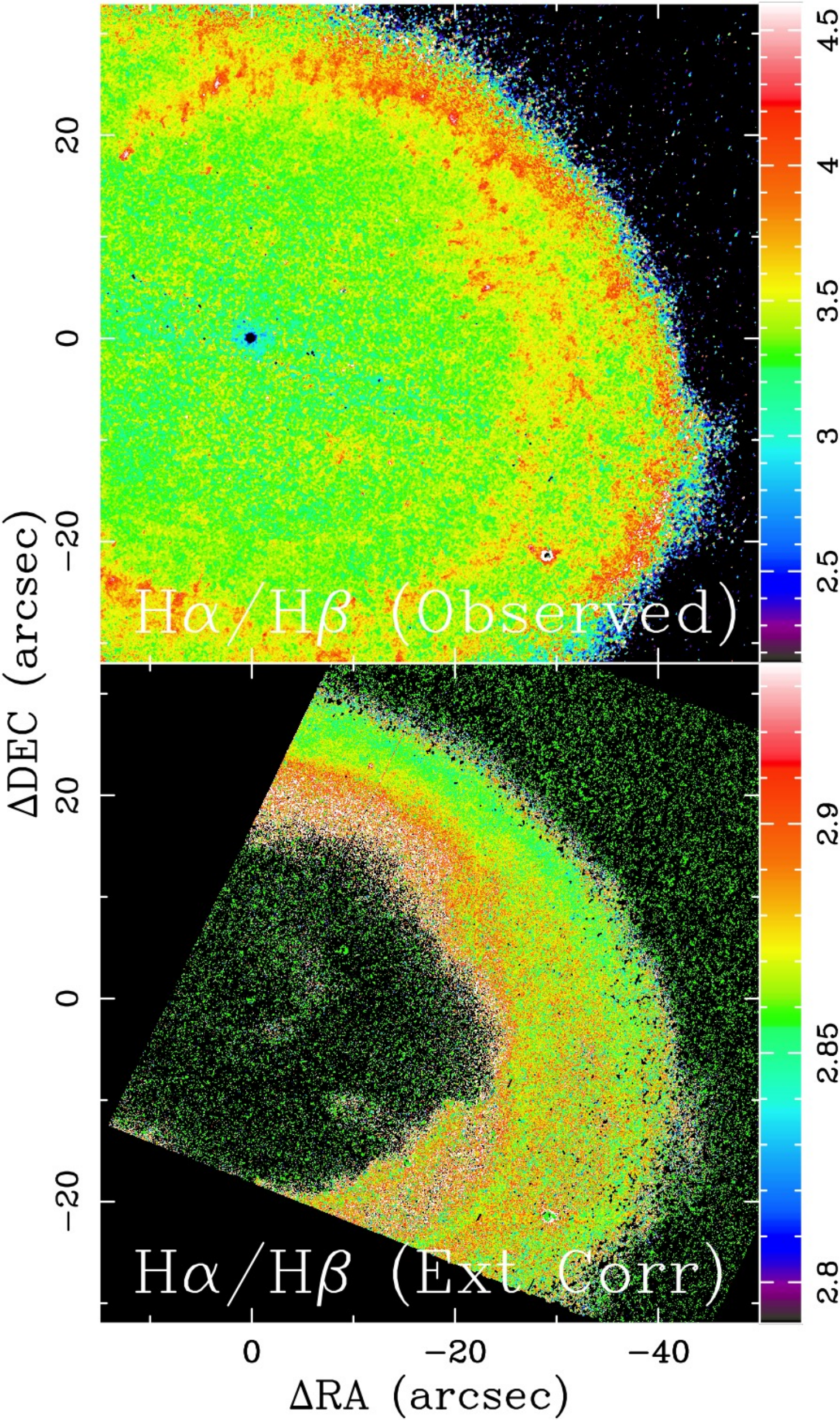}
    \caption{The observed (top) and extinction-corrected (bottom) 
    {\ha}-to-{\hb} line ratio maps of the NW quadrant of NGC\,6720.
    Image conventions follow those of Fig.\,\ref{fig:cHb}.
    The color wedge shows values in the $\pm3 \sigma$ range
    around the median.}
    \label{fig:Ha2Hb}
\end{figure}

One of the first spectral imaging investigation of NGC\,6720 
was performed by \citet{lame1994} with a Fabry-Perot imaging spectrograph.
They presented the observed {\ha}-to-{\hb} map (their Fig.\,3)
and noted a smooth radial increase of the value 
from 3.68 in the central cavity to 3.74 in the main ring,
with the maximum of $\sim$3.93 happening at around the 3/4 of the ring
(with the median of $3.74 \pm 0.22$).
These values compare reasonably well (within $\sim$10\,\%) 
with what we see in our QP-processed (observed, but line contamination corrected)
{\ha}-to-{\hb} map (Fig.\,\ref{fig:Ha2Hb}, top),
in which the ratio rises 
from $3.36\pm0.11$ in the central cavity
to $3.61\pm0.12$ in the main ring
with the median of $3.38\pm0.40$.

\citet{lame1994} went on to correct for extinction by assuming
the constant theoretical {\ha}-to-{\hb} ratio of 2.86 in the main ring, 
for which {\Nee} of $10^2$\,cm$^{-3}$ and {\tee} of $10^4$\,K
are assumed via \citealt{Hummer1987} under the same CCM extinction law.
This prompted the corrected {\ha}-to-{\hb} ratio of 2.81 in the central cavity,
which corresponds to {\tee} of 13,500\,K.
They attributed the {\tee} discrepancy to the local inhomogeneity
and deviation from the Case B condition,
and moved on without re-evaluating the theoretical {\ha}-to-{\hb} ratio
based on the updated {\tee}, never seeking consistency
between the extinction correction and plasma diagnostics 
(very typical in the literature).
So, their {\chb} values would have been 0.31--0.32, 
which are $\sim$20\,\% greater than ours.

In our case, on the other hand, {\chb}, {\Nee}, and {\tee} are 
iteratively updated for convergence, 
and out extinction-corrected {\ha}-to-{\hb} map (Fig.\,\ref{fig:Ha2Hb}, bottom) 
yields the mean ratio of $2.863 \pm 0.024$ in the main ring
out of the initial value of 2.858
(and the corresponding derived {\Ne} and {\te} maps
from the initially assumed $10^3$\,cm$^{-3}$ and $10^4$\,K;
see \S\,\ref{sect:denstemp}),
maintaining self-consistency in every single pixel
(Fig.\,\ref{fig:linemaps}a).
This means that the 20\,\% discrepancy in the analyses by 
\citet{lame1994} arose because 
(1) a {\em uniform} theoretical {\ha}-to-{\hb} ratio 
was imposed across the entire nebula
and
(2) convergence was not sought among {\chb}, {\Nee}, and {\tee}
by iteration, i.e., {\chb}, via the theoretical {\ha}-to-{\hb} ratio,
was never updated even when {\Nee} and {\tee} varied.

The same {\sl HST\/}/WFC3 data set was previously analyzed 
by \citet{odell2013b}.
They originally reported radially decreasing {\chb} along both 
the long- and short-axis of the nebula. 
We communicated with them about this discrepancy
during the early phase of writing of this manuscript.
It turned out that one of the coefficients in their 
flux calibration formula for F656N was in error,
and that their updated results now show radially increasing {\chb}
around 0.3 in the main ring \citep{odell2021}.
We note that
their derivation is based on
the constant {\ha}-to-{\hb} ratio of 2.87 across the nebula
assuming a $10^4$\,K gas at low density (the value unspecified) 
under the extinction law by \citet{Whitford1958} with $R_V = 3.1$.
Also, their WFC3 image flux calibration does not follow the
standard STScI method, but is based on their own method established 
with their own {\em ground-based} long-slit spectroscopy data
(i.e.\ not generally reproducible).

Other previous analyses were typically based on
multi-position-aperture or long-slit spectroscopy
(e.g.\
\citealt{Hawley1977,Barker1980,Guerrero1997,Garnett2001}). 
Each of these studies adopted a certain extinction law 
(plus $R_V$)
and theoretical {\ha}-to-{\hb} ratio
based on their own assumptions of {\Nee} and {\tee}.
The adopted parameters are sometimes explicitly mentioned,
and sometimes not.
The reported {\chb} (plus {\Nee} and {\tee}) values are
generally consistent with our values, some more so and others less.
The direct comparisons to assess details beyond general agreement
do not seem instructive, as none of the previous derivations 
were done self-consistently.
There is more recent spectral imaging study of NGC\,6720
\citep{Martin2016}
based on observations with 
an imaging Fourier transform spectrometer
\citep{sitelle}.
However,
their study is focused on the velocity structure of the nebula,
and no plasma diagnostics are discussed.

In light of the previous studies smothered with inconsistencies, 
the lesson to be learned here seems rather obvious.
We ought to eliminate inconsistencies where we can, 
because any inconsistencies introduced will exacerbate 
uncertainties in the outcomes.
All things considered, 
we can reiterate one major advantage of PPAP 
in comparison with other methods in the literature. 
PPAP yields the self-consistent {\chb} (and hence, {\Nee}, and {\tee}) 
distribution based on only the input line flux maps
plus the choice of the extinction law to adopt,
with no need to adopt any other parameters and {\em ad hoc} assumptions.

\subsection{The Converged {\Ne} and {\te} Maps}
\label{sect:denstemp}

\begin{figure*}
	\includegraphics[width=\columnwidth]{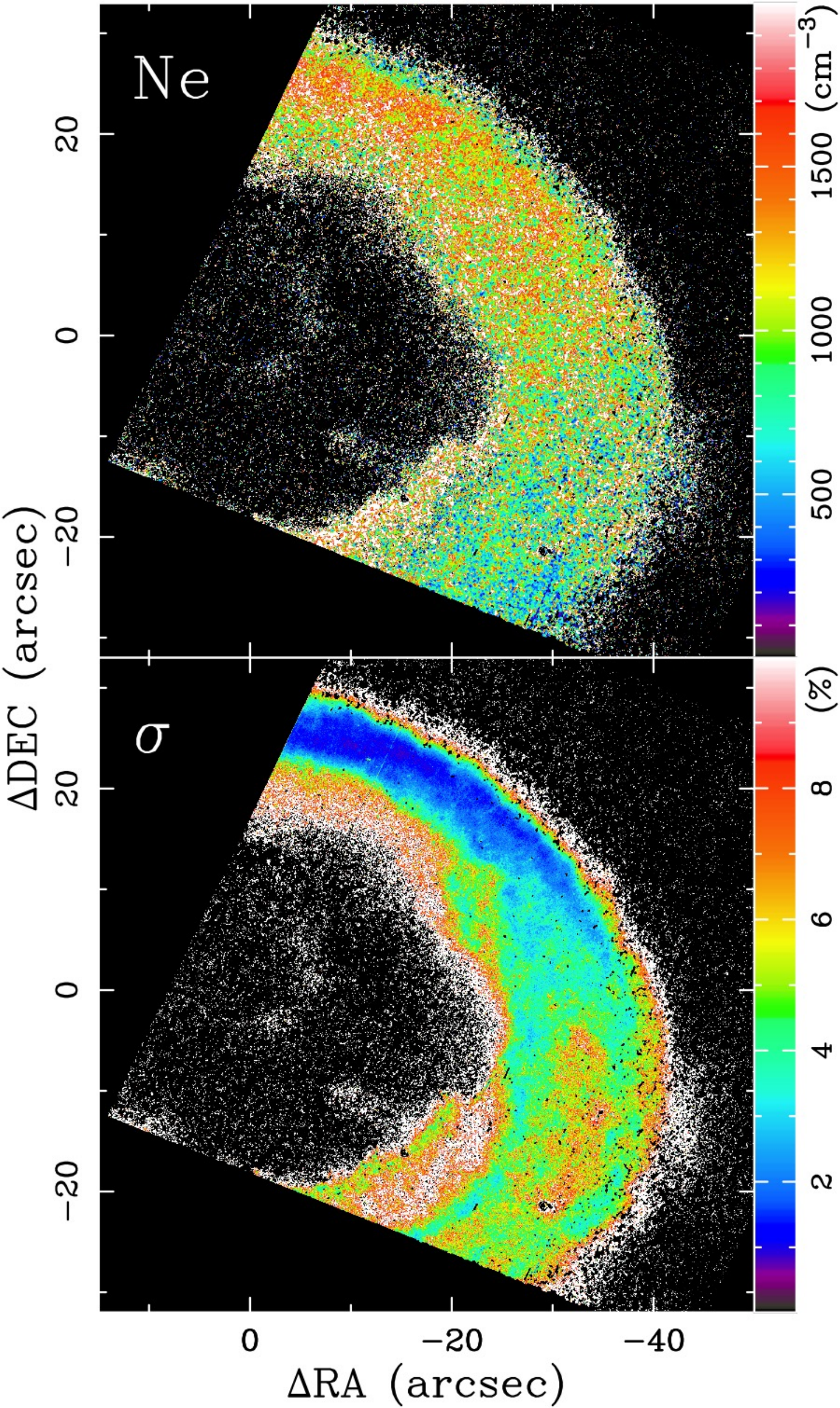}
	\hfill
	\includegraphics[width=\columnwidth]{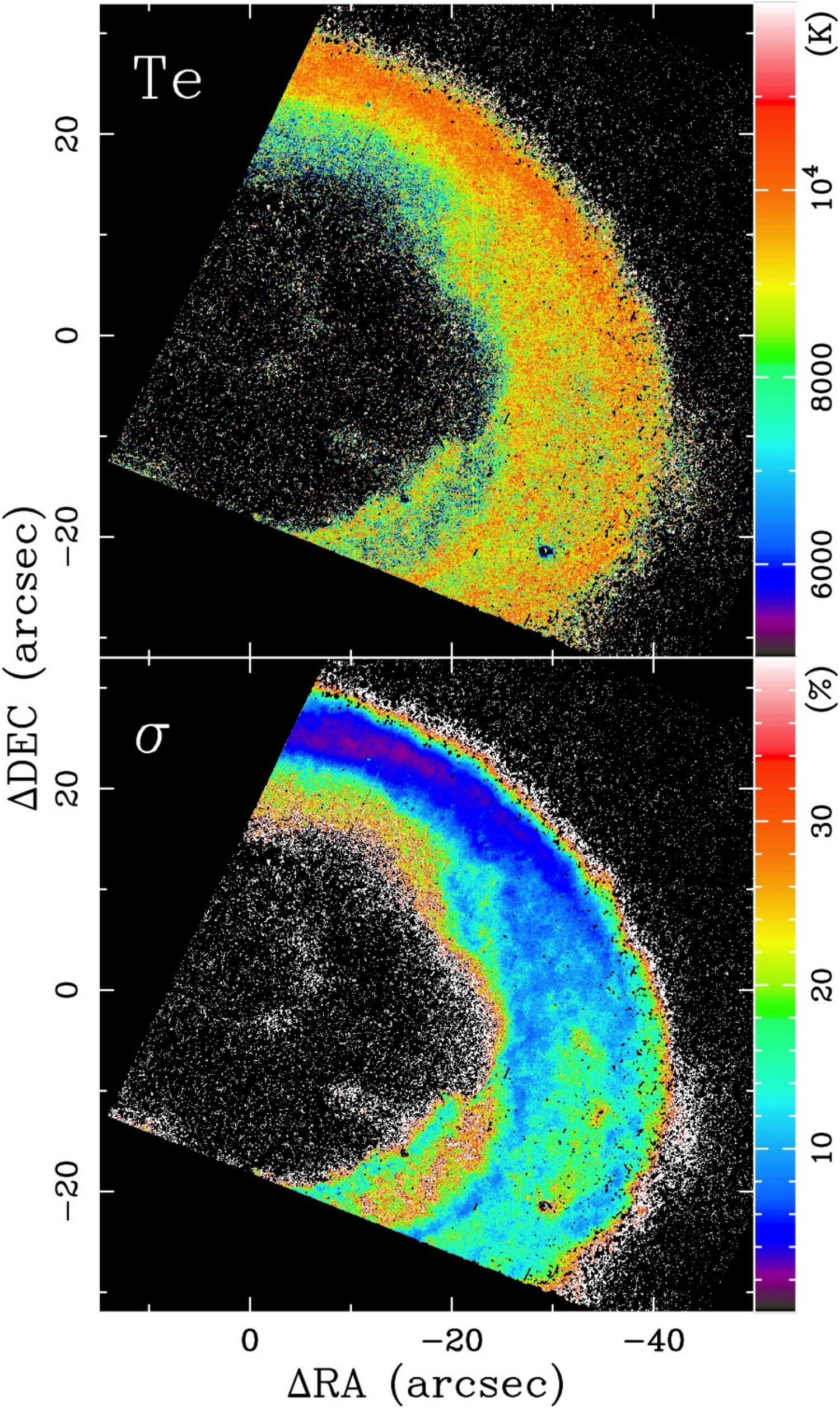}
	\caption{The {\Ne} and {\te} distribution maps with 
	the corresponding one-$\sigma$ percent uncertainty map
	derived self-consistently through PPAP
	(left and right, respectively).
	Image conventions follow those of Fig.\,\ref{fig:cHb}.}
    \label{fig:NeTe}
\end{figure*}

\begin{figure}
	\includegraphics[width=\columnwidth]{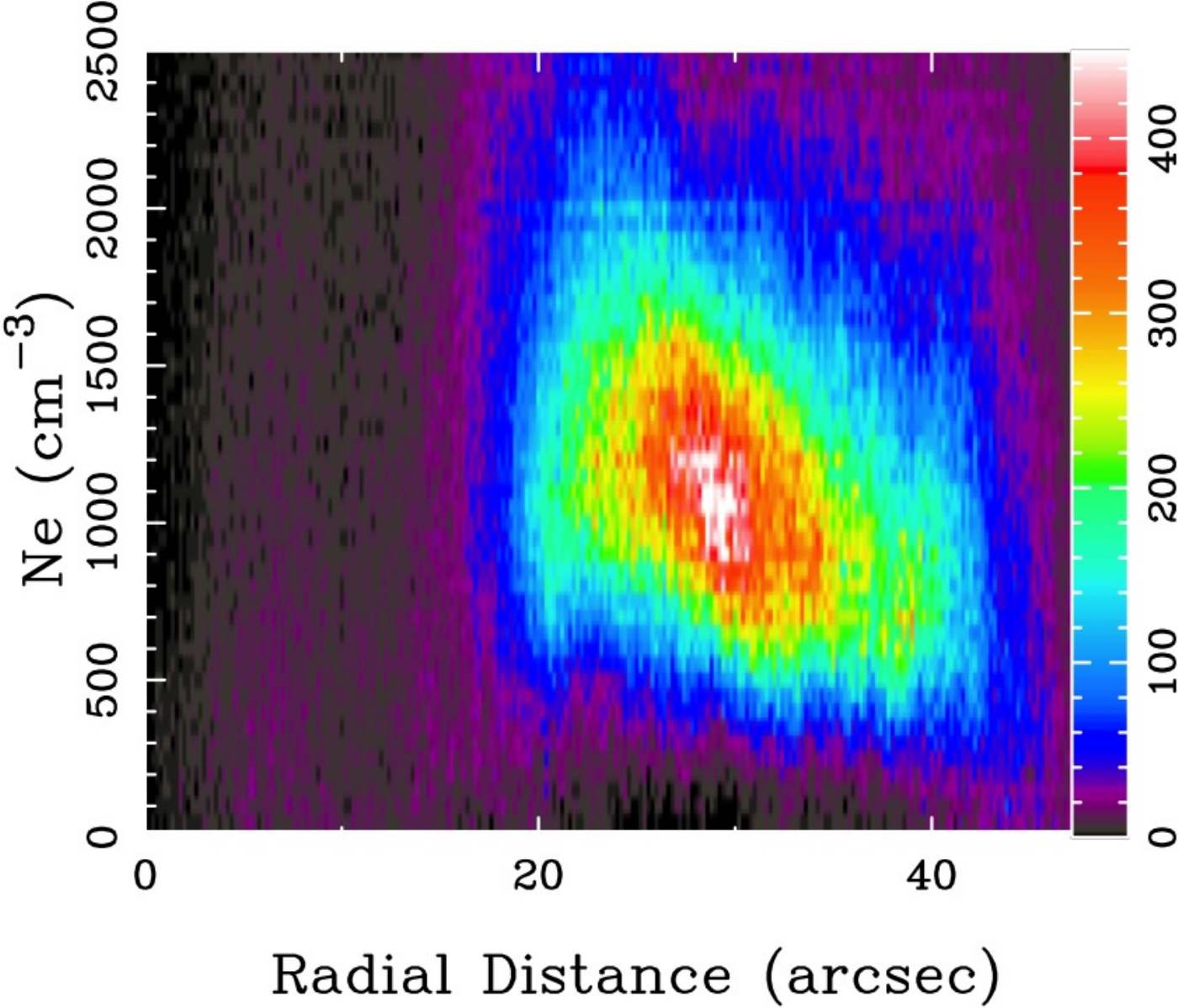}
	\includegraphics[width=\columnwidth]{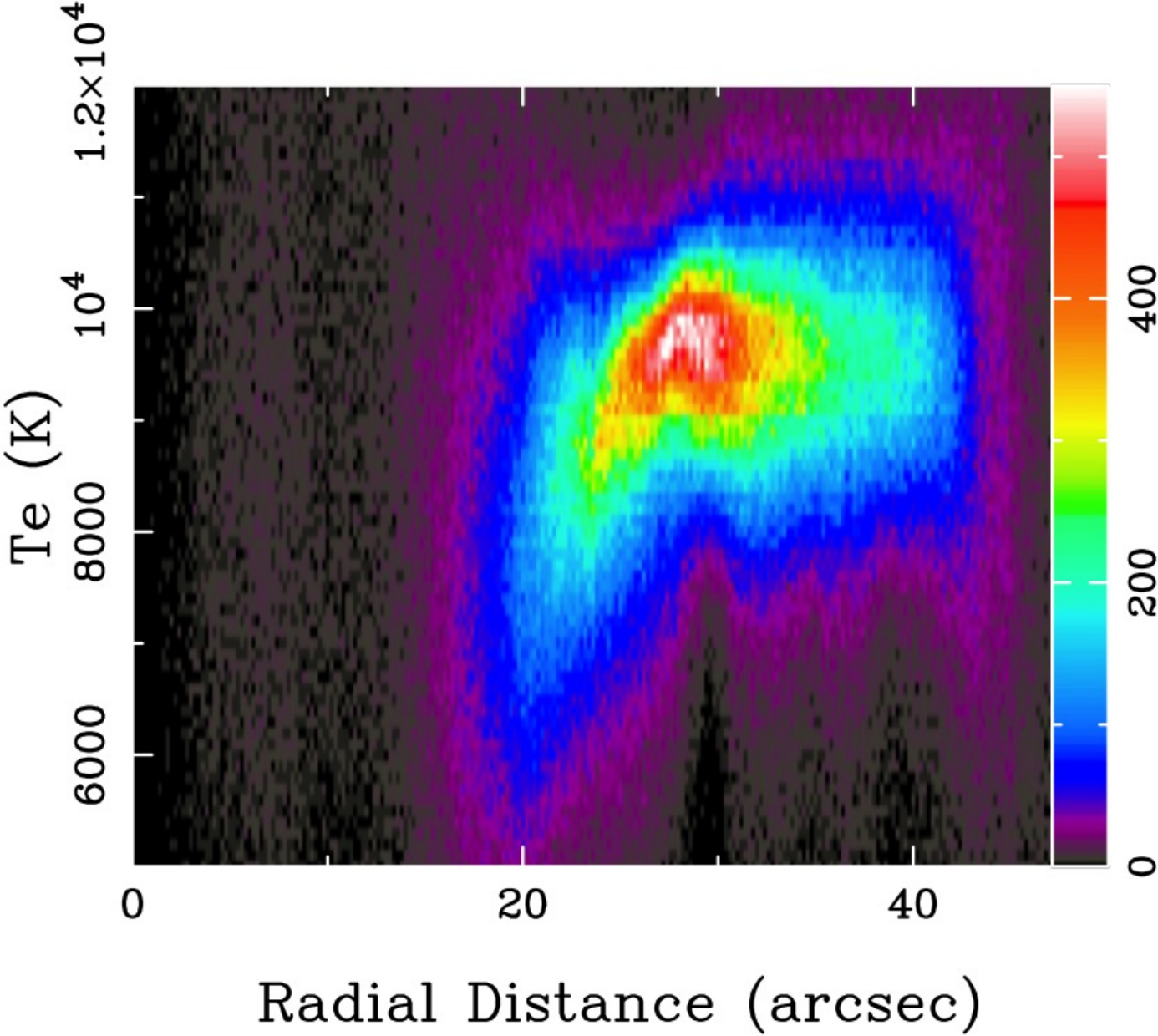}
    \caption{The radial density distribution map of {\Ne} and {\te} 
    with the frequency indicated by the color
    (top and bottom, respectively).
    Image conventions follow those of Fig.\,\ref{fig:cHbradial}.}
    \label{fig:NeTeradial}
\end{figure}

\subsubsection{Observed Spatial Variations}
\label{sect:observed}

The {\Ne} and {\te} distribution maps converged via PPAP
are summarized in Fig.\,\ref{fig:NeTe} with their corresponding
percentage uncertainty maps.
The radial density distribution maps of {\Ne} and {\te} are also 
presented in Fig.\,\ref{fig:NeTeradial}.
In computing {\Ne} and {\te} with PyNeb \citep{pyneb}, 
we use the transition probabilities (Einstein As) 
by \citet{rynkun2019} and \citet{FISCHER2004},
and collision strengths by \citet{tz2010} and \citet{Tayal2011},
respectively.
The {\Ne} and {\te} uncertainties are computed
in the same way as the {\chb} uncertainties.
As is the case for {\chb}, 
{\Ne} and {\te} are reliably determined mostly in the main ring.
This is partly because of S/N in the diagnostic {\sii} line ratio map
(Fig.\,\ref{fig:s2}), but also partly because the {\sii} and {\nii}
diagnostics are better suited to probe the moderately ionized region
(i.e.\ the main ring) than the highly ionized wind bubble region
(i.e.\ the inner cavity).

The converged {\Ne} and {\te} maps show a lesser degree of 
spatial variation in the main ring than the converged {\chb} map:
we do not recognize micro-structures in the {\Ne} and {\te} maps
as in the {\chb} map.
Closer examination suggests that there is some anti-correlation 
between {\Ne} and {\chb}:
{\Ne} tends to be smaller where {\chb} is larger.
This is indeed expected because dust grains are
more likely found where temperature is low, i.e., 
the degree of ionization is low, and vice versa.

The main ring appears to be divided into the following two parts
in terms of {\Ne} and {\te}:
(1) the inner part ($\sim$20--25$^{\prime\prime}$),
in which both {\Ne} and {\te} show a radially increasing trend
with relatively higher uncertainties ($\sim$25\,\%), and
(2) the outer part (beyond 25$^{\prime\prime}$ up to 30--40$^{\prime\prime}$),
where {\Ne} decreases and {\te} is more or less uniform.

The {\Ne} and {\te} values increase radially
from $\sim$1,000\,cm$^{-3}$ to $\sim$1500\,cm$^{-3}$ in {\Ne} 
(with the mean of $1,400 \pm 580$\,cm$^{-3}$)
and 
from $\sim$7,000\,K to $\sim$9,300\,K in {\te} 
(with the mean of $8,390 \pm 1,270$\,K)
in the inner part of the main ring.
Their spatial distribution of {\Ne} and {\te} indicates that 
this inner part is rather circular in projection
(extends up to $\sim$25$^{\prime\prime}$ in all azimuthal directions).
Thus, this region is possibly spherical in 3-D, suggesting that 
this ``spherical'' region represents the highly-ionized central wind bubble 
caused by the radiation from the central star,
eating its way into the higher density ellipsoidal wind shell
(the outer part beyond $\sim$25$^{\prime\prime}$).

Thus, the observed radial rise of {\Ne} in the inner part of the main ring
can be understood as caused by the snowplowing effect 
by the PN fast wind emanating from the central star (e.g.\,\citealt{kwok2000}).
However, the observed radial rise of {\te} may be counter-intuitive.
Such radially increasing {\te} trending is recognized numerically,
and attributed to the self-shielding effect of the ionizing radiation field:
soft ionizing photons are already absorbed by metals in the inner part
and remaining hard ionizing photons in the outer part would 
yield more heat per ionization, making {\te} higher at greater radial 
distance away from the central star \citep{Kewley2019}.

In the outer part of the ring,
{\Ne} radially decreased to about 1,000\,cm$^{-3}$
(with the mean of $1,170 \pm 540$\,cm$^{-3}$),
while {\te} remains about the same at $\sim$9,300\,K
(with the mean of $9,290 \pm 1,060$\,K).
Given that the degree of ionization is more or less the same in
this part of the shell as implied from rather uniform {\te},
the radial fall-off of {\Ne} probably reflects the density 
distribution in this part.

Azimuthally,
{\Ne} and {\te} change from $\sim$1,500\,cm$^{-3}$ and $\sim$10,500\,K
along the short axis to $\sim$500\,cm$^{-3}$ and {\te} $\sim$8,100\,K
along the long axis, respectively.
This azimuthal trending is consistent with what we see
in the {\chb} distribution (Fig.\,\ref{fig:cHb}).
Generally speaking,
this azimuthal variation is what gives the observed spread of the 
{\Ne} and {\te} values --
about 1,000\,cm$^{-3}$ for {\Ne} and
about 2,000\,K for {\te} --
in the density distribution (Fig.\,\ref{fig:NeTeradial}).

While the radial density distribution maps do not show any representative
{\Ne} and {\te} values in the inner cavity (Fig.\,\ref{fig:NeTeradial})
as in {\chb} (Fig.\,\ref{fig:cHb}),
the mean {\Ne} and {\te} values nevertheless turn out to be
$1,340 \pm 750$\,cm$^{-3}$ and 
$8,140 \pm 1,730$\,K.
Here, readers are reminded that these values are based on the
{\sii} and {\nii} diagnostics, which are suited to probe a low-excitation gas.
In other words, these measurements are obtained from the low-excitation
plasma that lies along the line of sight to the inner cavity,
i.e., {\sii} and {\nii} are most likely not co-spatial 
with the central wind bubble region.
Rather, the valid {\Ne} and {\te} values in this region 
most likely sample the near-side of the thinned-out bipolar cavities
of the ellipsoidal shell of the nebula
(cf.\ Fig.\,14 by \citealt{Guerrero1997}; Fig.\,11 by \citealt{odell2013a}).

\subsubsection{Comparison with the Previous Results}

The previous {\Nee} estimates yielded
500--700\,cm$^{-3}$ \citep{lame1994},
470--615\,cm$^{-3}$ \citep{Guerrero1997},
660--740\,cm$^{-3}$ \citep{Garnett2001}, and
350--550\,cm$^{-3}$ \citep{odell2013b},
based on the assumed {\tee} value of $10^{4}$\,K
and assumed theoretical {\ha}-to-{\hb} ratio 
(for which certain {\Nee} was assumed).
They used the derived {\Nee} values to update {\tee}
that ranged roughly from 10,000 to 13,000\,K,
but without updating the theoretical {\ha}-to-{\hb} ratio 
(hence, extinction correction) according to 
the updated {\Nee} and {\tee}.

More specifically, for example, \citet{lame1994} adopted 
the theoretical {\ha}-to-{\hb} ratio of 2.86 uniformly 
across the nebula, assuming a uniform {\Nee} of $10^3$\,cm$^{-3}$ 
and {\tee} of $10^{4}$\,K.
A comparison between the assumed theoretical {\ha}-to-{\hb} ratio of 2.86
and the observed {\ha}-to-{\hb} ratio of $\sim$3.74 yielded a certain 
{\chb} distribution.
Then, extinction corrected {\sii} and {\nii} diagnostic line ratios 
yielded the {\Ne} distribution of 500--700\,cm$^{-3}$ and 
{\te} distribution of 10,000--13,500\,K.
\citet{lame1994} concluded their analyses then, 
without re-evaluating {\chb} via the updated theoretical {\ha}-to-{\hb}
ratio based on the derived {\Ne} and {\te} distributions,
which differ from the assumed uniform {\Nee} and {\tee},
and following through the rest of the analyses for consistency.

Because both of the {\Ne} and {\te} distributions were different 
from the initial uniform assumption, the theoretical {\ha}-to-{\hb} 
ratio, and hence, the {\chb} distribution should have had been recomputed
for consistency.
Subsequently, the resulting {\Ne} and {\te} distributions
should have had been updated again to reflect changes in the 
diagnostic line ratios because of the {\chb} offsets.
Ideally, this iteration should have had been repeated 
until all values converged to the final values for the maximum consistency.
Unfortunately, consistency among these parameters was 
not sought iteratively by \citet{lame1994}.

This is very representative of how analyses were performed 
in the literature.
Hence, there does not seem much point in directly comparing the 
{\Nee} and {\tee} values found in the literature with our results.
The reciprocal dependence among {\chb} and ({\Nee}, {\tee})
were not considered carefully enough for consistency in the past.
The strangest thing in the literature is that inconsistencies 
in the final {\chb} and ({\Nee}, {\tee}) values and their derivatives
were usually attributed to local inhomogeneities, 
even though a uniform distribution of the initial parameters 
were usually assumed to begin with.
It is really this initial assumption of uniformity that imposes
inconsistencies in the first place.
There is really no need to assume {\chb} and ({\Nee}, {\tee}): 
all can be determined simultaneously by seeking convergence iteratively.
In the present analyses, only a few iterations (at most 5) were enough.

\subsubsection{Tolerable Uncertainties}
\label{sect:uncertainties}

To make PPAP work effectively, it is important to keep self-consistency 
in the whole of the analyses.
However, at some point, the propagated uncertainties from the line 
fluxes themselves would become greater than the uncertainties caused 
by assuming constant ({\Nee}, {\tee}).
Then, what uncertainties are tolerable in following PPAP?\.
While this question may sound simple, 
it is actually not simple at all.
This is because the tolerable levels of uncertainties are actually 
dependent on the actual line ratios used, the ({\Nee}, {\tee}) values
themselves, and how varying {\Nee} and {\tee} are in the actual 
spatial distribution in the target source, among other things.

For the case of NGC 6720, the {\Nee} and {\tee} values are 
spread around 1,340\,cm$^{-3}$ and 8,140\,K with a width of 750\,cm$^{-3}$
and 1,730\,K, respectively (\S\,\ref{sect:observed}).
Thus, if uncertainties in {\Nee} and {\tee} exceed these spreads, 
assuming constant ({\Nee}, {\tee}) would be as good as PPAP.
A set of experiments with PyNeb reveal that uncertainties of 10--15\,\% 
in the {\sii} ratio would cause large enough uncertainties in the resulting
{\Nee} so that considering non-uniform {\Nee} would not be so meaningful.
The same is said for {\tee} if uncertainties in the {\nii} ratio are 
25--30\,\%.
Hence, uncertainties of roughly 10\,\% and 25--30\,\% can be tolerated
in the {\sii} and {\nii} line flux measurements, respectively.
While it is difficult to generalize this result for a particular source, 
uncertainties of 10\,\% or better appear to be necessary for PPAP to 
work as a rule of thumb.

\subsection{Gas-to-Dust Ratio Map}
\label{sect:G2D}

With the {\chb} and ({\Ne}, {\te}) maps determined, 
it is possible to evaluate the gas-to-dust mass ratio map at {\hb}.
This is because dust grains are represented by {\chb}, 
while {\hb}-emitting ionized gas is represented 
by the {\hb} surface brightness map, {\Ne} map,
and the {\hb} emissivity map derived from the ({\Ne}, {\te}) maps.
The definition of the extinction allows us to relate
{\chb} and the gas-to-dust mass ratio as follows.
\begin{eqnarray}
c({\rm H}\beta) &=& \frac{A_{{\rm H}\beta}}{2.5} 
= (\log e) \tau_{{\rm H}\beta} 
\nonumber\\
&=& (\log e) (Q^{\rm ext}_{{\rm H}\beta} \pi a^2) N_{\rm dust}
\nonumber\\
&=& (\log e) (Q^{\rm ext}_{{\rm H}\beta} \pi a^2)
\frac{\rho_{\rm dust}/m_{\rm dust}}{\rho_{{\rm H}^+}/m_{{\rm H}^+}}
N_{{\rm H}^+}
\end{eqnarray}
where $N_{[{\rm dust,~N}^+]}$, $\rho_{[{\rm dust,~N}^+]}$, and
$m_{[{\rm dust,~N}^+]}$ are the column density, mass density, and 
particle mass of dust grains and {\hb}-emitting gas, respectively,
and 
$Q^{\rm ext}_{{\rm H}\beta}$ is the extinction efficiency coefficient
under the assumption of grains being spheres of radius $a$.
Hence, the gas-to-dust mass ratio is;
\begin{eqnarray}
\frac{\rho_{{\rm H}^+}}{\rho_{\rm dust}} 
&=&
(\log e) (Q^{\rm ext}_{{\rm H}\beta} \pi a^2)
\frac{m_{{\rm H}^+}}{m_{\rm dust}} \frac{N_{{\rm H}^+}}{c({\rm H}\beta)}
\nonumber\\
&=&
(\log e) (Q^{\rm ext}_{{\rm H}\beta} \pi a^2)
\frac{m_{{\rm H}^+}}{\frac{4\pi}{3}a^3\rho^{\rm bulk}_{\rm dust}}
\frac{N_{{\rm H}^+}}{c({\rm H}\beta)}
\nonumber\\
&=&
\frac{3(\log e) Q^{\rm ext}_{{\rm H}\beta} m_{{\rm H}^+}}{4 a \rho^{\rm bulk}_{\rm dust}} 
\frac{N_{{\rm H}^+}}{c({\rm H}\beta)}
\end{eqnarray}
where $\rho^{\rm bulk}_{\rm dust}$ is the bulk density of dust grain material.

\begin{figure}
    \centering
    \includegraphics[width=\columnwidth]{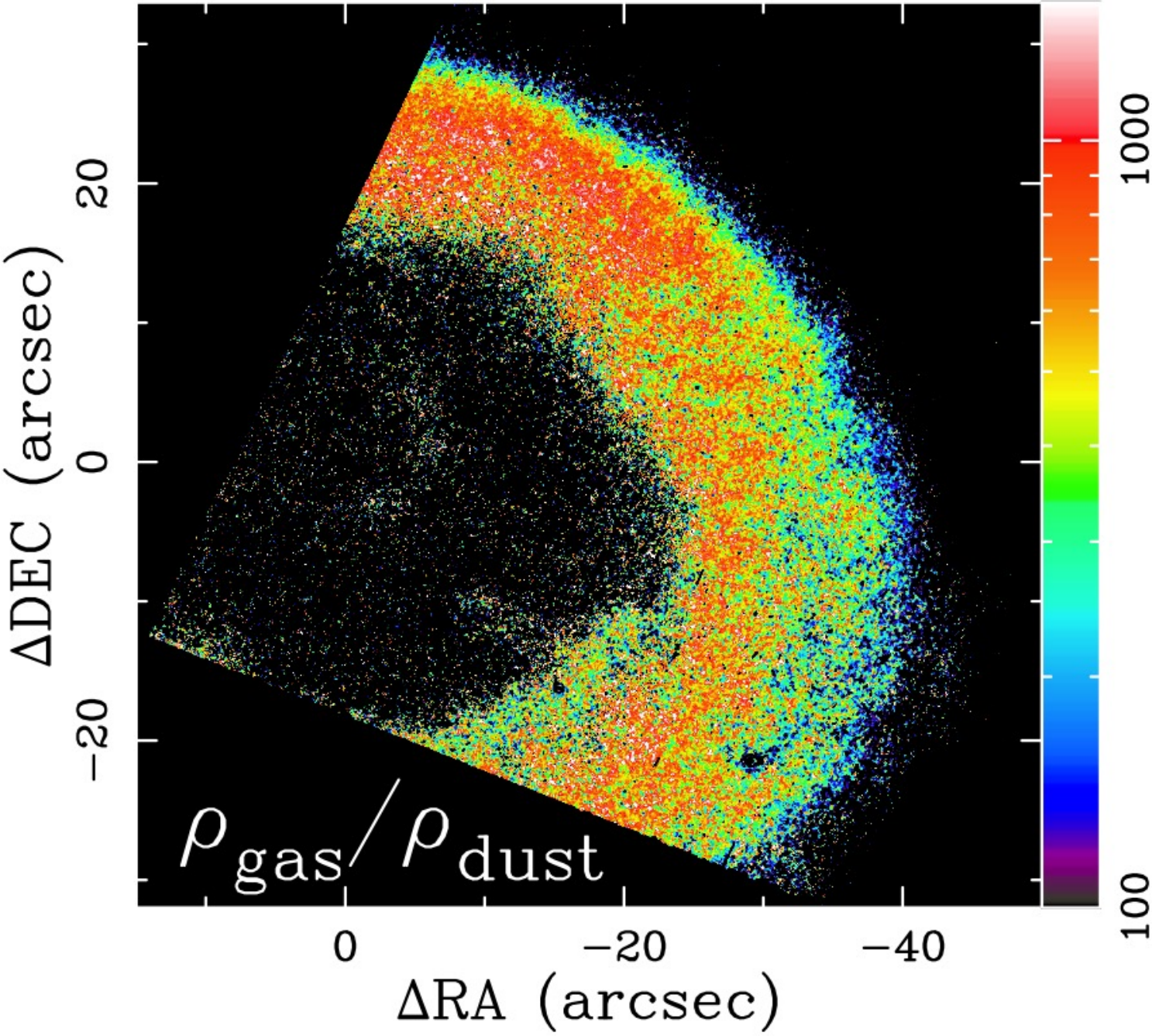}
    \caption{The gas-to-dust mass ratio map at {\hb} obtained 
    from the extinction-corrected {\hb}, ({\Ne}, {\te}), and
    {\chb} maps by assuming certain dust properties.
    Image conventions follow those of Fig.\,\ref{fig:cHb}.}
    \label{fig:G2D}
\end{figure}

Meanwhile, the {\hb} surface brightness 
is expressed in terms of the
column density of {\hb}-emitting gas as follows.
\begin{eqnarray}
I_{{\rm H}\beta} &=&
\int \frac{\epsilon_{{\rm H}\beta} n_{\rm e} n_{{\rm H}^+}}{4\pi} ds
\nonumber\\
&\approx&
\frac{\epsilon_{{\rm H}\beta} n_{\rm e}}{4\pi} \int n_{{\rm H}^+} ds
=
\frac{\epsilon_{{\rm H}\beta} n_{\rm e}}{4\pi} N_{{\rm H}^+}
\end{eqnarray}
where the integral is done through the target along the line of sight,
$\epsilon_{{\rm H}\beta}$ is the {\hb} emissivity 
as defined
by \citet{storey1995} as a function of {\Ne} and {\te}.
By combining the above two equations, we have
\begin{eqnarray}
\frac{\rho_{{\rm H}^+}}{\rho_{\rm dust}} 
&=&
\frac{3\pi(\log e) Q^{\rm ext}_{{\rm H}\beta} m_{{\rm H}^+}}{a \rho^{\rm bulk}_{\rm dust}} 
\frac{ I_{{\rm H}\beta}}{\epsilon_{{\rm H}\beta} n_{\rm e} c({\rm H}\beta)}
\end{eqnarray}
Then, we can derive the gas-to-dust mass ratio map at {\hb}
using the extinction-corrected {\hb} map (which gives $I_{{\rm H}\beta}$),
the ({\Ne}, {\te}) maps (which give $\epsilon_{{\rm H}\beta}$ and $n_{\rm e}$),
and the {\chb} maps,
by assuming spherical dust grains of ``smoothed astronomical silicate''
\citep{wd2001} of radius 0.1\,$\mu$m and bulk density 3\,g\,cm$^{-3}$
(as the dust chemistry of NGC\,6720 is still unknown).

The derived gas-to-dust mass ratio map at {\hb} is presented in Fig.\,\ref{fig:G2D}.
The distribution appears similar to the {\Ne} map (Fig.\,\ref{fig:NeTe}),
with the median of $437 \pm 357$,
which is a few times greater than the ``typical'' value of 100.
The derived values need to be considered as the lower limit
because we take into account only the {\hb}-emitting
ionized gas in the main ring, in which some non-negligible amount 
of atomic H gas component is expected.
According to a photoionization model of NGC\,6720 by \citet{vanhoof2010},
the amount of ionized and neutral (both atomic and molecular) gasses
is about the same (they quoted $\log(n_{e})=2.62$ and $\log(n_{\rm H}=2.60)$).
Thus, we account for only about 50\,\% of the gas.
Also, the derived {\chb} includes the ISM contribution, 
which amounts to at most roughly 50\,\% (Fig.\,\ref{fig:cHb}).

Therefore, the total gas-to-dust mass ratio seems to be about 1600
in the main ring.
Clearly, a ``typical'' ratio of 100 cannot be use indiscriminately
for the circumstellar dust component, especially when 
the amount of dust component may be reduced because of 
local environments.
In the literature, it is often practiced to estimate $N_{\rm H}$
from $A(V)$ by adopting an old empirical ISM relation,
$N_{\rm H} \approx (1.87 \times 10^{21}) \times A(V)$
\citep{sm1979}.
However, this relation implicitly assumes the gas-to-dust mass ratio
of 100 for the ``average'' ISM.
Thus, as we demonstrate here, it is simply wrong to 
adopt such an $A(V)$-to-$N_{\rm H}$ relation indiscriminately 
to any object for which some circumsource dust component is expected.

\subsection{Extinction-Corrected Line Emission Maps}
\label{sect:linemaps}
\begin{figure}
	\includegraphics[width=\columnwidth]{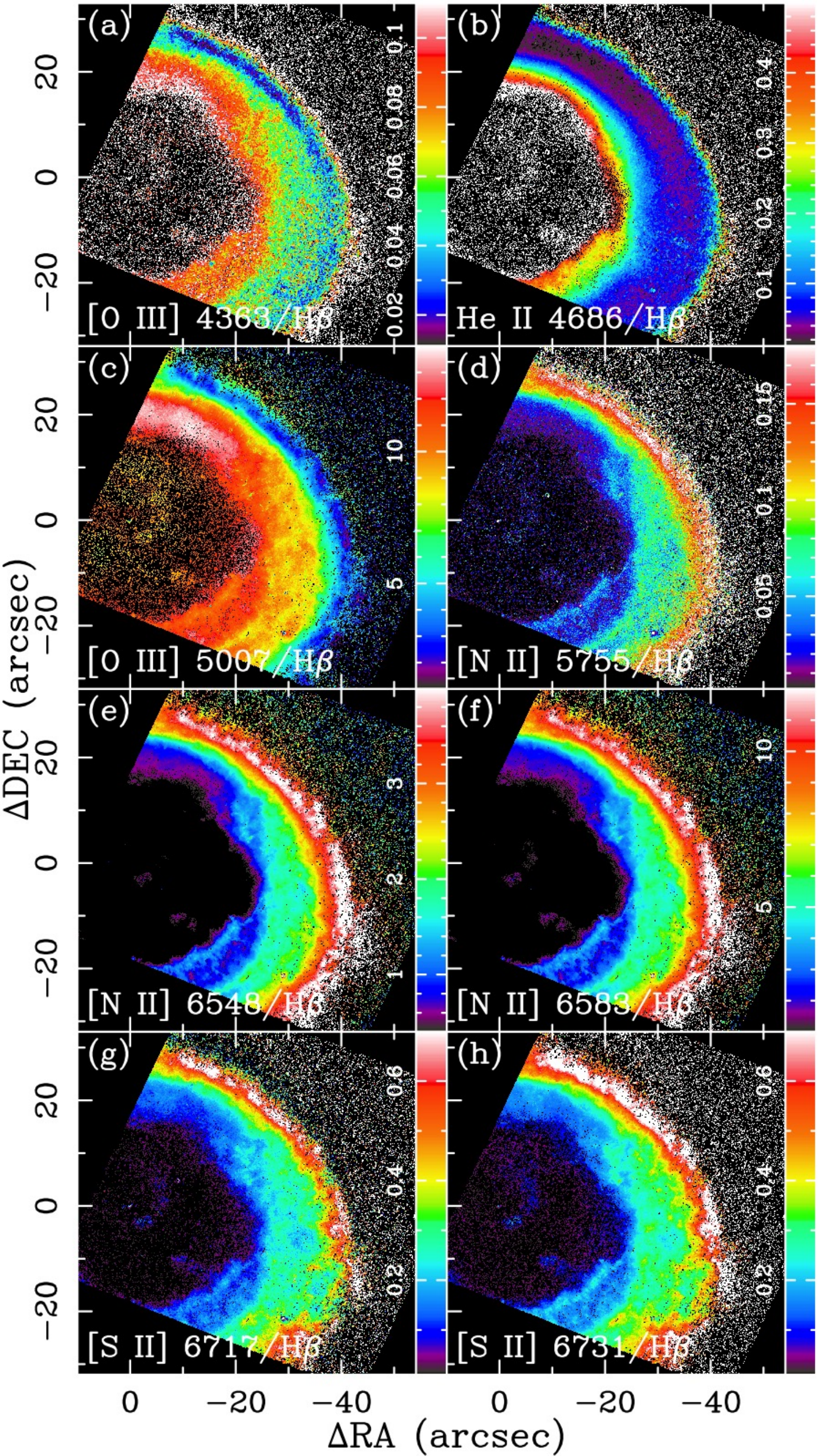}
	\caption{Various extinction-corrected line maps
	(relative to {\hb}) of the NW quadrant of NGC\,6720:
	(a) {\oiii} 4363\,{\AA}/{\hb},
	(b) {\heii} 4686\,{\AA}/{\hb},
	(c) {\oiii} 5007\,{\AA}/{\hb},
	(d) {\nii} 5755\,{\AA}/{\hb},
	(e) {\nii} 6548\,{\AA}/{\hb},
	(f) {\nii} 6583\,{\AA}/{\hb},
	(g) {\sii} 6717\,{\AA}/{\hb}, and
	(h) {\sii} 6731\,{\AA}/{\hb}.
	The wedge on the right in each panel indicates the adopted 
	linear color scale. 
	Image conventions follow from those of Fig.\,\ref{fig:cHb}.}
    \label{fig:linemaps}
\end{figure}

Fig.\,\ref{fig:linemaps} shows the extinction-corrected line maps 
relative to {\hb} extracted from the present {\sl HST\/}/WFC3 data set.
These individual line maps reveal the spatial variation of 
the relative abundance of specific ionic/elemental species 
within extended nebulae.
Quick inspection of these maps suggests that 
there are three general patterns of the emission morphology 
owing to the different excitation levels represented by these lines.

First, a strong circular emission region is seen only in the inner cavity 
($<20^{\prime\prime}$) as in the {\heii} 4686\,{\AA} map 
(Fig.\,\ref{fig:linemaps}b),
as previously reported by \citet{lame1994}.
This region can be recognized only by the fast radial decrease 
of emission around $\sim$20$^{\prime\prime}$.
Most likely this region represents the spherical PN wind bubble 
at the core of the nebula, in which the ambient temperature is 
the highest.

Second, there is another circular emission region that appears more extended
than the {\heii} 4686\,{\AA} region, 
encompassing the inner part of the main ring (up to 25--30$^{\prime\prime}$),
as seen in the {\oiii} 4363 and 5007\,{\AA} maps (Fig.\,\ref{fig:linemaps}a,c). 
This region is where the {\ha}-to-{\hb} ratio
is high (Fig.\,\ref{fig:Ha2Hb}, bottom),
and where {\chb}, {\Nee}, and {\tee} radially 
increase to the maximum (Figs.\,\ref{fig:cHbradial} and \ref{fig:NeTeradial}),
delineating the extent of the ionized region.

Third, emission is strong only in the periphery of the main ring 
(beyond $\sim$30$^{\prime\prime}$) as in the {\nii} and {\sii} maps
(Fig.\,\ref{fig:linemaps}d--h).
This region corresponds to where {\chb} is high ($\sim$0.3 and above;
Figs.\,\ref{fig:cHb}, \ref{fig:cHbradial}).
{\Ne} and {\te} are also high in this region 
(1,200\,cm$^{-3}$, Figs.\,\ref{fig:NeTe}; 
$>10^4$\,K, \ref{fig:NeTeradial}), 
most likely representing the PDR surrounding the ionized region. 

The apparent transition of morphology from the inner part 
(radially decreasing) to outer part (radially increasing) 
of the main ring is quite remarkable.
The radially increasing trending of {\te} toward the periphery of the main ring
(Fig.\,\ref{fig:NeTe}, right) is promoted by the radially increasing trending of 
{\sii} and {\nii} line emission (Figs.\,\ref{fig:linemaps}d--h).
As discussed earlier, this is interpreted as the self-shielding effect,
in which the majority of the ionizing photons is already consumed in the 
inner cavity and in the inner part of the main ring and the left-over
far-reaching high-energy photons heat the outer part of the main ring. 
Hence, this transition zone at $\sim$30$^{\prime\prime}$ from the central star
very likely corresponds to the location of the ionization front (IF) 
that separates the high-excitation emission region 
(Figs.\,\ref{fig:linemaps}a--c) and the low-excitation region (Figs.\,\ref{fig:linemaps}d--h).


\begin{figure}
	\includegraphics[width=\columnwidth]{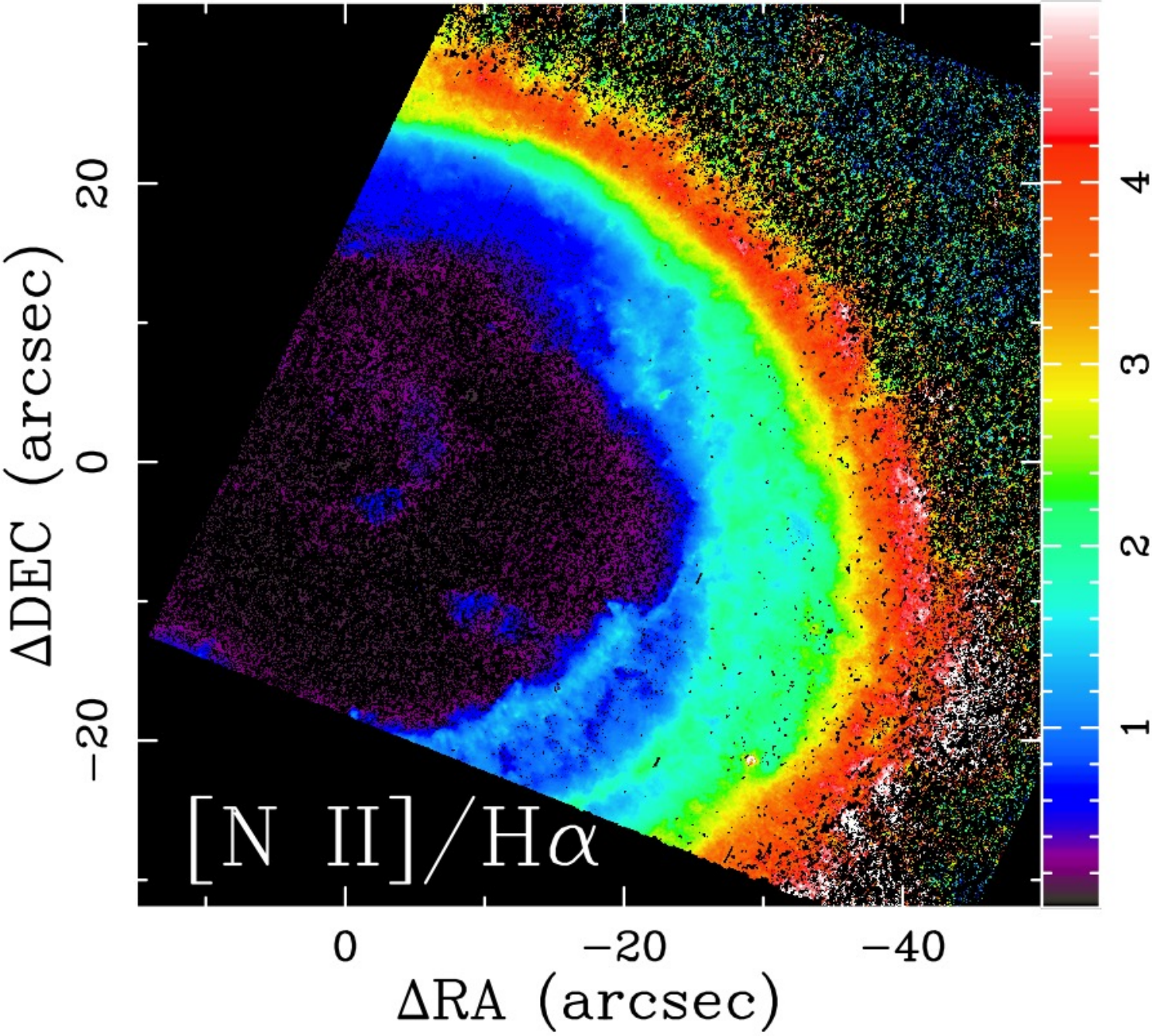}
	\includegraphics[width=\columnwidth]{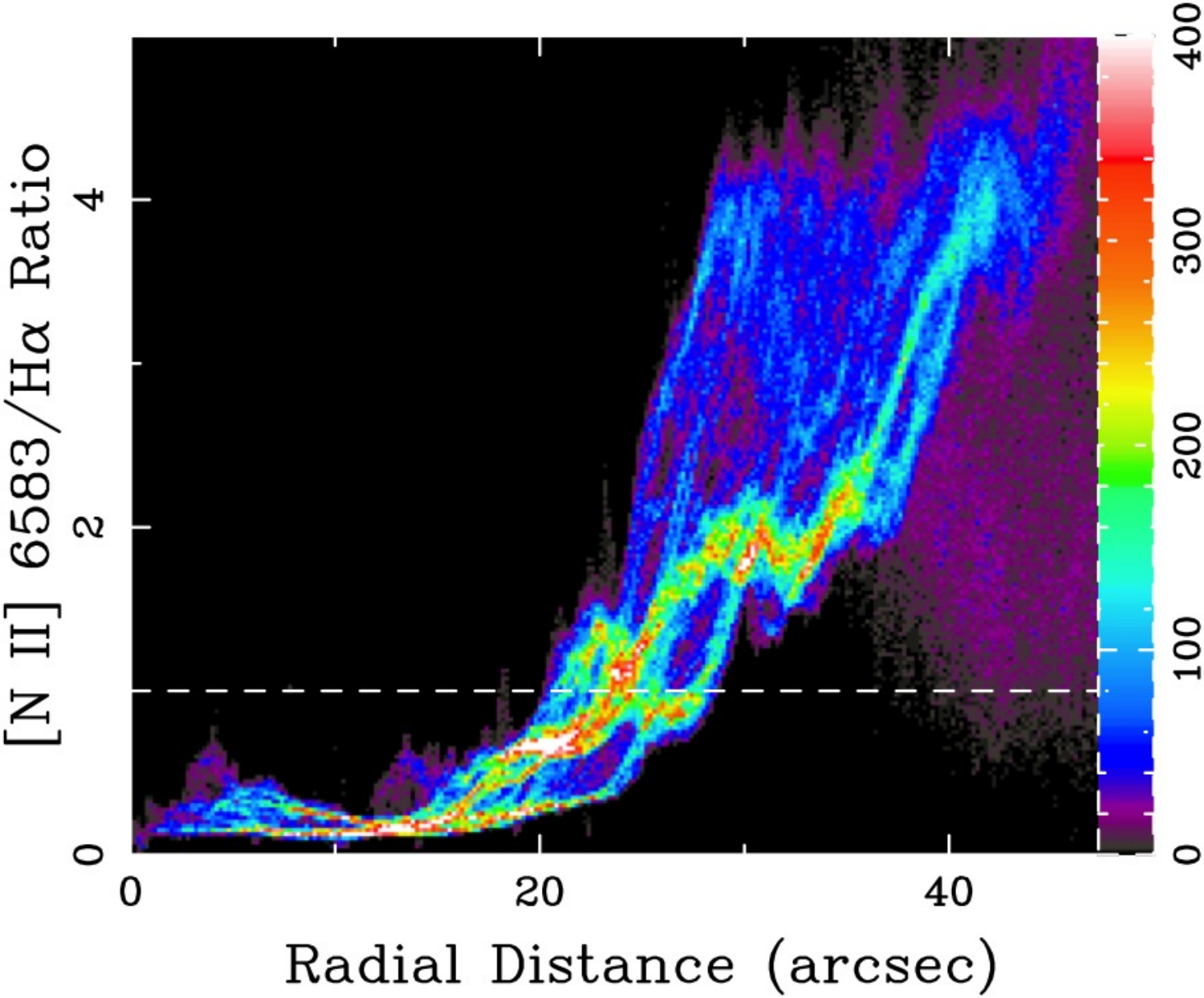}
	\caption{[Top] The {\nii} 6583-to-{\ha} line ratio map
	distinguishes the ionized region ($< 1$)
	from the PDR ($> 1$).
	Image conventions follow those of Fig.\,\ref{fig:cHb}.
	[Bottom] The radial density distribution map of the 
	{\nii} 6583-to-{\ha} line ratio shows the IF
	as the region of the steepest gradient between 
	$20^{\prime\prime}$ and $40^{\prime\prime}$.
	The dashed line is where the ratio is unity.}
    \label{fig:N2Ha}
\end{figure}

Comparisons between two line maps also provide a wealth of information.
{\ha} emission is representative of the ionized region, 
while {\nii} emission is of the PDR.
Hence, the {\nii}\,6583\,{\AA}-to-{\ha} line ratio map 
helps to spatially separate these regions (Fig.\,\ref{fig:N2Ha}, top),
i.e., to identify where IF is located. 
This is one of the reasons why blending of {\ha} and {\nii} lines 
in the {\ha} band can be a critical issue.
IF is most likely located where the {\nii}\,6583\,{\AA}-to-{\ha} 
gradient becomes the steepest.

The radial density distribution map (Fig.\,\ref{fig:N2Ha}, bottom)
indicates that the steepest gradient happens at various radii
depending on the azimuthal angle.
Around the short axis, the ratio remains low in the inner part
of the main ring, and precipitously increases at $\sim$25$^{\prime\prime}$.
\citep{lame1994} and \citep{odell2013a} noted that 
the apparent physical thinness of the {\nii}-and-{\sii}-bright region
at the periphery of the main ring (roughly a few arcsec width) 
indicated that at the line of sight in this region IF intersects 
nearly at parallel with our line of sight.
This means that the inclination of the hollow ellipsoidal shell 
of the main ring of NGC\,6720
(cf.\ Fig.\,14 by \citealt{Guerrero1997}; Fig.\,11 by \citealt{odell2013a})
is pivoted around the short axis in the plane of the sky
\citep{Guerrero1997,odell2007,Martin2016}.
This further means that along the long axis we are looking into 
the inclined inner wall of the hollow ellipsoidal shell (i.e.\ IF),
where the {\nii}\,6583\,{\AA}-to-{\ha} ratio starts to rise 
at small radii but increases rather slowly, meandering around 2.
This meandering suggests that the IF surface is not necessarily 
of uniform density and may be sprinkled with micro-structures 
(e.g.\ those associated RT/KH instabilities; \S\,\ref{sect:ext}).
Hence, the size distribution of these micro-structures,
may yield useful constraints for numerical simulations
to determine hydrodynamical IF properties.

\begin{figure}
	\includegraphics[width=\columnwidth]{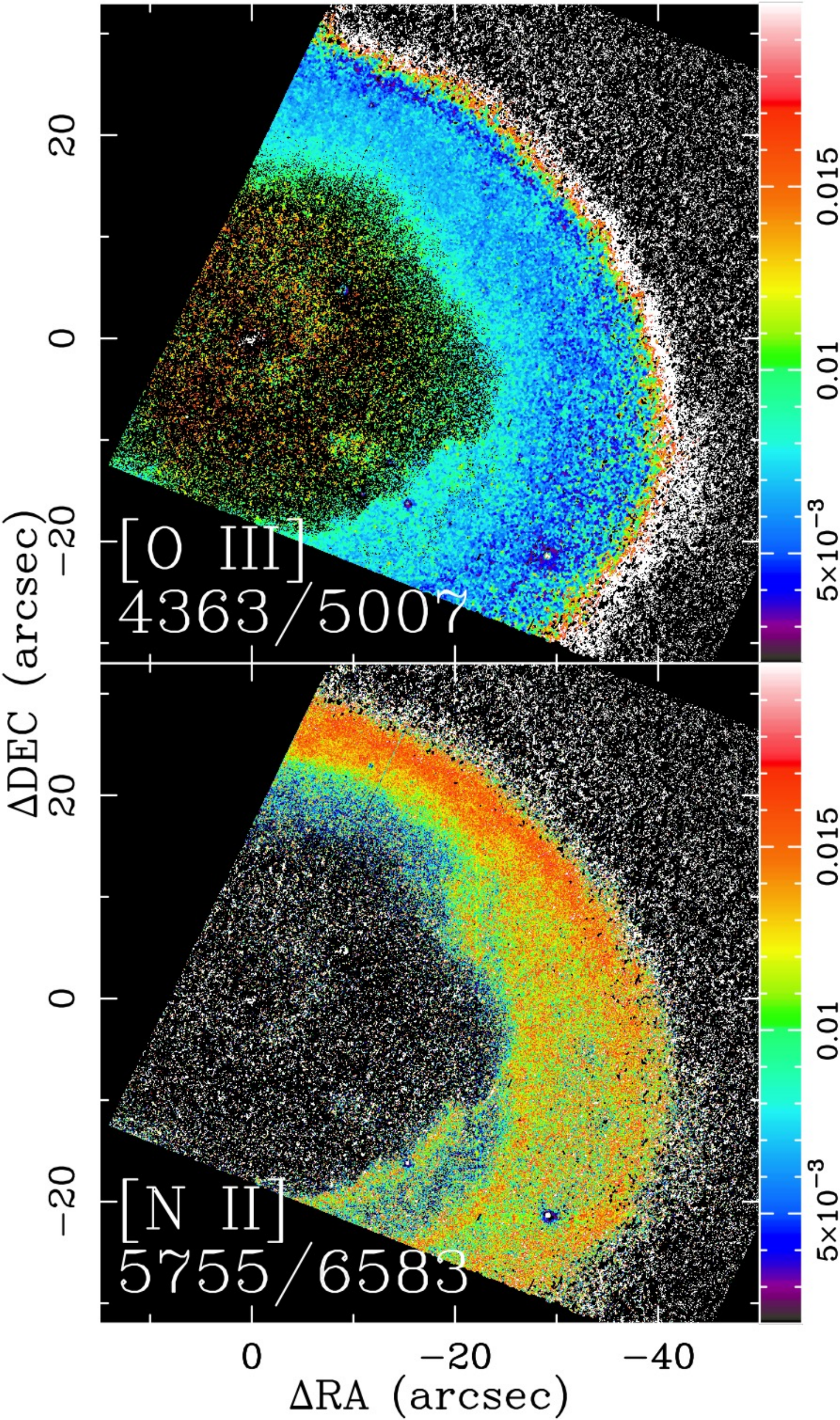}
	\caption{The {\oiii} 4363-to-5007\,{\AA} (top)
	and {\nii} 5755-to-6583\,{\AA} (bottom) line ratio maps
	both compare the spatial distribution of 
	the $^1$S$_0$ to $^1$D$_2$ transition
	with that of the $^1$D$_2$ to $^3$P$_2$ transition,
	but reveal opposite spatial variations
	(though the range of ratio is the same as indicated by the color wedge)
	because of the ``{\te} inversion'' by self-shielding
	of ionizing photons.
	Image conventions follow those of Fig.\,\ref{fig:cHb}.}
    \label{fig:O3N2}
\end{figure}

Another interesting comparison is between 
the {\oiii} 4363-to-5007\,{\AA} and {\nii} 5755-to-6583\,{\AA} line ratio maps
(Fig.\,\ref{fig:O3N2}).
These line ratios compare the $^1$S$_0$ to $^1$D$_2$ transition
with the $^1$D$_2$ to $^3$P$_2$ transition of the O$^{2+}$ and N$^{+}$ ions.
The former transition probes higher temperature regions than 
the latter transition.
However, interestingly, these line ratios exhibit opposite trending.
The {\oiii} 4363-to-5007\,{\AA} line ratio map 
(Fig.\,\ref{fig:O3N2}, top) shows larger ratios in the inner region 
of the main ring than in the outer region.
This is expected from the radially-decreasing general 
temperature structure of the nebula.

On the other hand, the {\nii} 5755-to-6583\,{\AA} ratio map
(Fig.\,\ref{fig:O3N2}, bottom)
shows relatively larger values in the outer region of the main ring 
than in the inner region.
This is exactly the same ``{\te} inversion'' in the main ring 
seen above in \S\,\ref{sect:denstemp} (Fig.\,\ref{fig:NeTe}, right)
as a consequence of self-shielding of low-energy ionizing radiation 
by metals in the inner part of the main ring (\citealt{Kewley2019}).

This comparison plainly demonstrates the importance of adopting 
diagnostic lines that arise from the very region of interest.
The {\sii} and {\nii} lines probe the same energy regimes, i.e., the 
resulting {\Ne} and {\te} would be likely co-spatial.
However, the {\oiii} lines probe higher temperature regimes than 
the {\nii} lines do (i.e., the {\oiii} lines are more suited to probe 
the inner cavity).
Hence, plasma diagnostics done with the {\sii} and {\oiii} lines 
would not make much sense as the {\Ne} and $T_{\rm e}$([O\,{\sc iii}])
are not co-spatial, and $T_{\rm e}$([O\,{\sc iii}]) most likely 
exhibits opposite trending with respect to {\te}.

\subsection{N$^{+}$ and S$^{+}$ Relative Abundance Maps}
\label{sect:abund}

\begin{figure}
    \centering
	\includegraphics[width=\columnwidth]{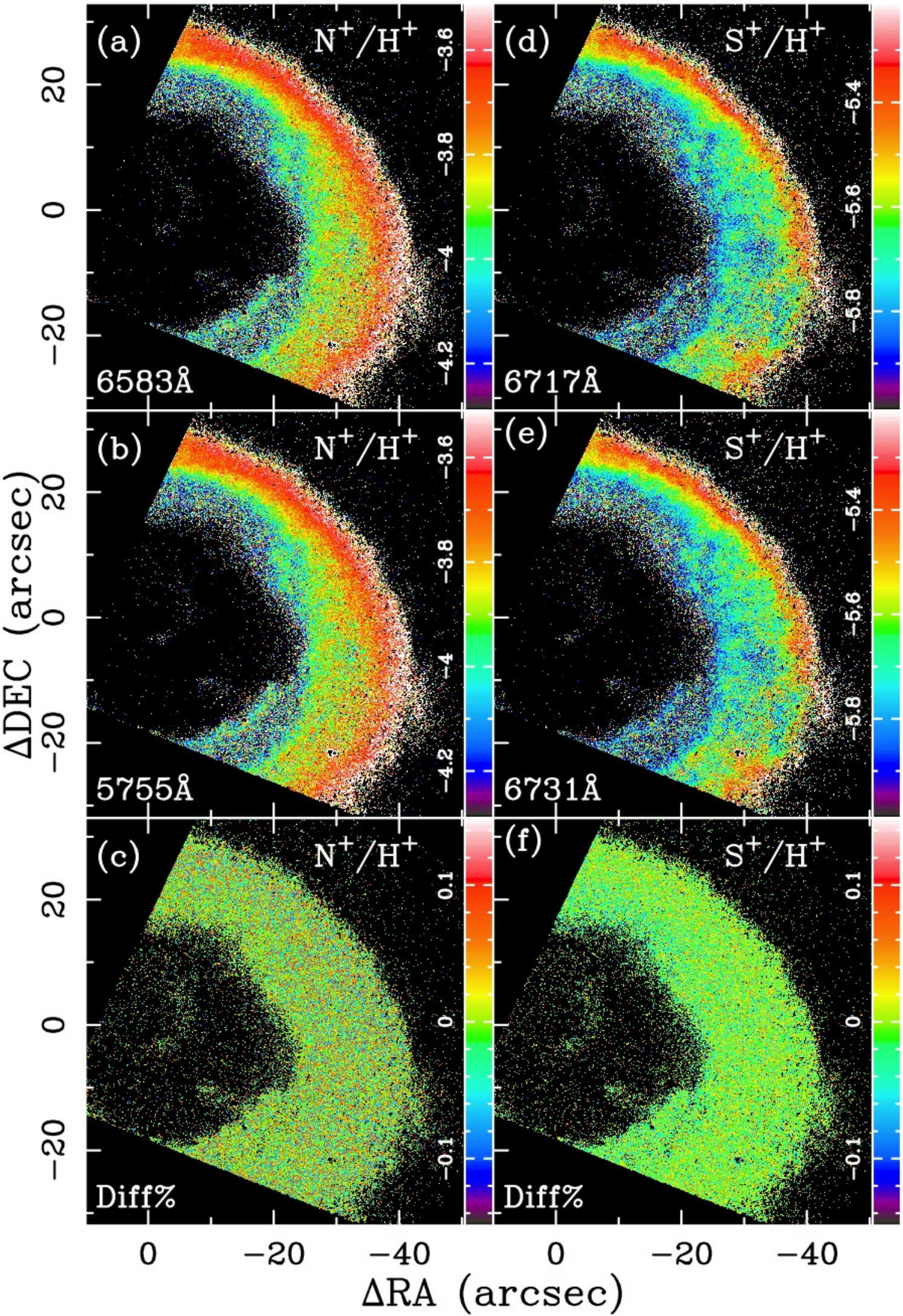}
	\caption{Relative ionic abundance distribution maps of 
	$n({\rm N}^{+})$/$n({\rm H}^{+})$
	derived from 
	the (a) {\nii}\,6583\,{\AA} and (b) {\nii}\,5755\,{\AA} lines
	and (c) their percentage difference map, and 
	and of $n({\rm S}^{+})$/$n({\rm H}^{+})$
	derived from 
	the (d) {\sii}\,6717\,{\AA} and (e) {\sii}\,6731\,{\AA} lines
    and (f) their percentage difference maps.
	These demonstrate that 
	the derived relative ionic abundance maps are identical.
	The wedge shown in each frame indicates 
	the adopted log color scale from 
	$5 \times 10^{-5}$ to $3.2 \times 10^{-4}$ for $n({\rm N}^{+})$ and
	from $10^{-6}$ to $6.3 \times 10^{-6}$ for $n({\rm S}^{+})$
    for the abundance map
	and 
	the adopted linear color scale from
	$-0.15$\,\% to $+0.15$\,\% for the percentage difference maps.
	Image conventions follow from those of Fig.\,\ref{fig:cHb}.}
    \label{fig:abund}
\end{figure}

The final step of plasma diagnostics is the determination of ionic 
and elemental abundance distributions.
Theoretically, for a given set of {\Nee} and {\tee},
ionic populations for a particular ionic species 
should be uniquely determined from any line emission/transition 
of the adopted $N$-level system.
For the present study, 
it is natural to compute the ionic abundance distribution maps of 
$n({\rm N}^{+})$ and $n({\rm S}^{+})$ relative to $n({\rm H}^{+})$
using any of the {\nii} and {\sii} line maps,
as our derivation of {\Nee} and {\tee} is based on the {\nii} and {\sii} lines
(\S\,\ref{sect:denstemp}; Figs.\,\ref{fig:NeTe}, \ref{fig:NeTeradial}).

Here, we use the PyNeb {\sc getIonAbundance} function
with the extinction-corrected line maps 
at {\nii}\,5755 and 6583\,{\AA} and {\sii}\,6717 and 6731\,{\AA}
(Figs.\,\ref{fig:linemaps}d,f--h).
If the resulting relative N$^{+}$ and S$^{+}$ abundance maps 
come out to be identical for each of the {\nii} and {\sii} pairs, 
we can safely say that the robustness of the proposed PPAP is guaranteed.
The results of this experiment are summarized in Fig.\,\ref{fig:abund}.
The left column shows
the $n({\rm N}^{+})/n({\rm H}^{+})$ maps 
(Fig.\,\ref{fig:abund}a,b) derived from 
the {\nii}\,6583 and 5755\,{\AA} line maps (Figs.\,\ref{fig:linemaps}d,f),
and their difference in percentage (Fig.\,\ref{fig:abund}c).
Similarly, the right column presents the same for ${\rm S}^{+}$
from the {\nii}\,6717 and 6731\,{\AA} lines 
(Figs.\,\ref{fig:abund}g,h).

The derived $n({\rm N}^{+})/n({\rm H}^{+})$ and $n({\rm S}^{+})/n({\rm H}^{+})$
vary from about $5 \times 10^{-5}$ and $10^{-6}$ at the inner edge 
to about $3.2 \times 10^{-4}$ and $6.3 \times 10^{-6}$ 
at the outer edge of the main ring, respectively.
These values are consistent with previously determined values
at various positions in the nebula
presented by 
\citet{Barker1987} 
($n({\rm N}^{+})/n({\rm H}^{+})$ of 
$9.5 \times 10^{-6}$ to $2.21 \times 10^{-4}$
from multi-position aperture spectroscopy)
and 
\citet{liu2004} ($n({\rm N}^{+})/n({\rm H}^{+})$
of $6.61 \times 10^{-5}$
and $n({\rm S}^{+})/n({\rm H}^{+})$ of $1.23 \times 10^{-6}$
based on line intensities for the entire nebula obtained 
by scanning a long slit across the nebula during exposure),
for example.
However, readers are reminded that direct comparisons of numerical values 
do not carry significant weight, as previous analyses 
were not fully spatially resolved and involved inconsistencies.

What is remarkable here is that
the spatial distribution of N$^{+}$ and S$^{+}$ ions derived from
different lines/transitions appears very much identical to each other
(Fig.\,\ref{fig:abund}a,b for $n({\rm N}^{+})/n({\rm H}^{+})$
and Fig.\,\ref{fig:abund}d,e for $n({\rm S}^{+})/n({\rm H}^{+})$).
The percentage difference turns out to be practically nil:
$0.0004 \pm 0.053$\,\% and $-0.005 \pm 0.036$\,\%
for $n({\rm N}^{+})/n({\rm H}^{+})$ and 
$n({\rm S}^{+})/n({\rm H}^{+})$, respectively
(Fig.\,\ref{fig:abund}c,f).

We see a greater concentration of N$^{+}$ and S$^{+}$ ions
in the outer part of the main ring.
This is expected from the fact that N$^{+}$ and S$^{+}$ are
species of lower excitation energies.
Comparing N$^{+}$ and S$^{+}$, 
the lower-excitation S$^{+}$ ions are distributed in a physically 
narrower extent than N$^{+}$ ions at the periphery of the main ring.
These ionic abundance distribution maps would certainly
allow empirical analyses of their spatial variations in more depth,
providing excellent constraints for photoionization and PDR models.

In the past, the abundance derivation for a particular ionic species
from different lines/transitions usually yielded different results,
and their average was adopted as the final abundance value.
However, it is not too difficult
to understand that one obtains different abundance values from different 
lines/transitions,
given how analyses were typically done, using 
{\Ne} and {\te} that are not necessarily consistent 
(discrepancies typically at $\sim$10\,\% or greater)
with {\chb} or {\ha}-to-{\hb}, and hence, not necessarily consistent 
with $n({\rm N}^{+})$ and $n({\rm S}^{+})$.

\begin{figure}
    \centering
	\includegraphics[width=\columnwidth]{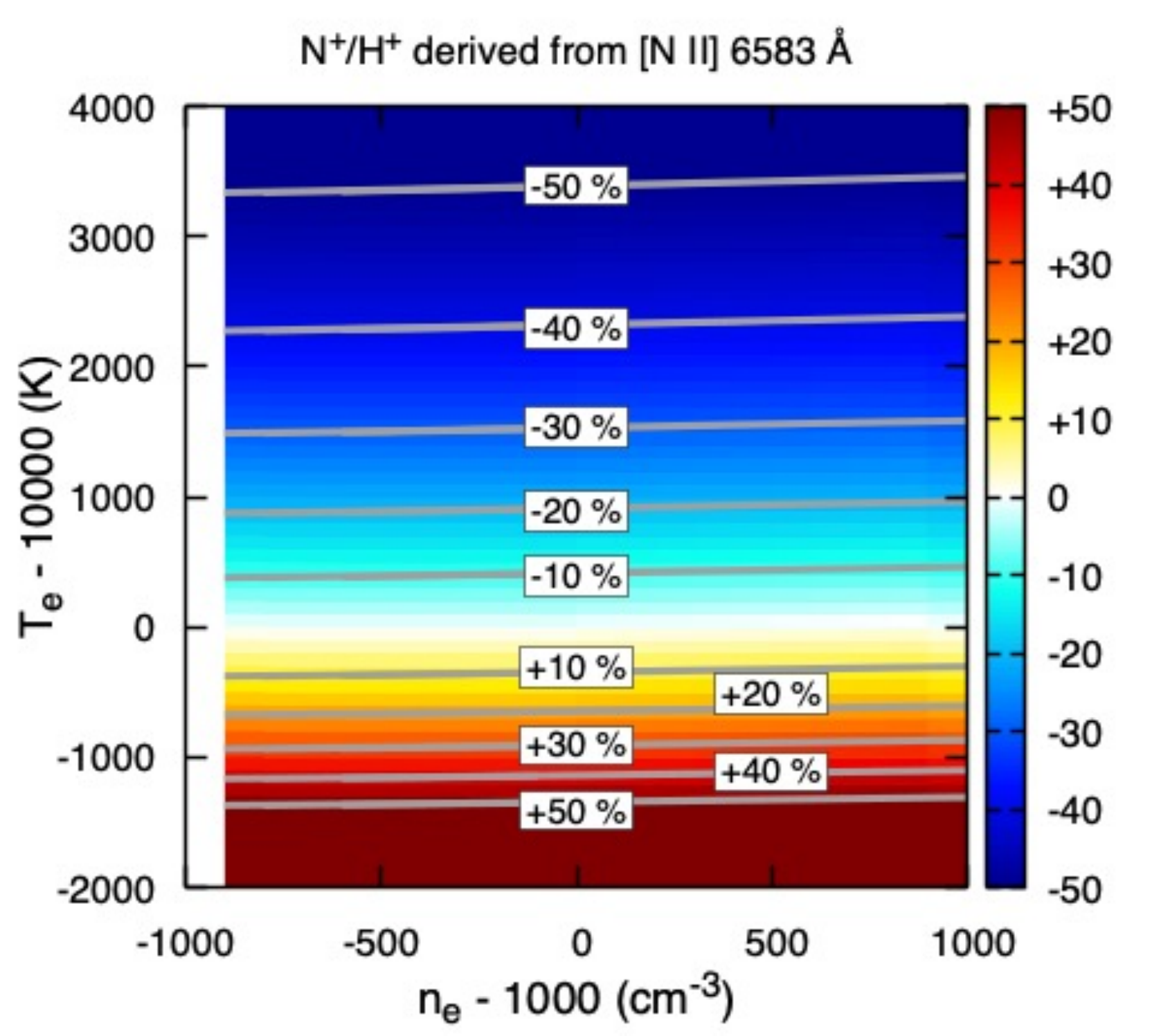}
	\includegraphics[width=\columnwidth]{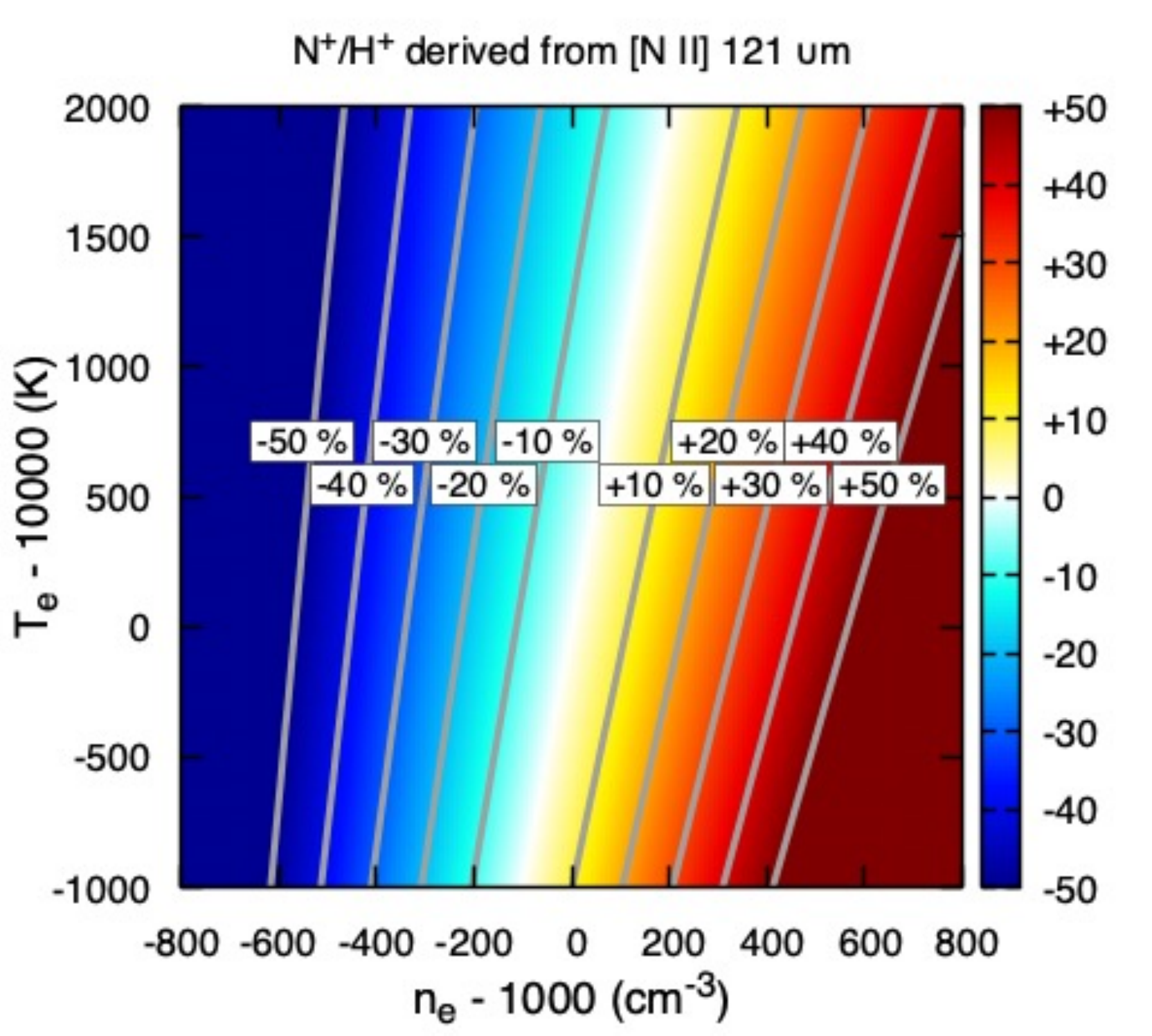}
	\caption{Relative \% differences in the $n$(N$^{+}$)/$n$(H$^{+}$) 
	abundance when {\Nee} and {\tee} are deviated from 
	the reference values (of $10^{3}$\,cm$^{-3}$ and $10^{4}$\,K)
	based on the {\nii}\,6583\,{\AA} (top) and 
	{\nii}\,121\,$\mu$m (bottom) diagnostics.}
    \label{fig:abundchanges}
\end{figure}

Such practice is intolerable in performing plasma diagnostics for
extended objects using spectral maps.
Here, PPAP is followed diligently, paying particular 
attention to self-consistency for both the interstellar extinction 
correction and plasma diagnostics, to yield identical abundances 
from multiple transitions of the same ionic species.
Thus, Fig\,\ref{fig:abund} demonstrates clearly that 
spatially-resolved plasma diagnostics can be performed
as rigorous numerical analyses if we adopt PPAP 
by seeking a converged self-consistent solution iteratively.

Before concluding this section, 
we emphasize the importance of self-consistency 
in abundance analyses from a different perspective.
Fig.\,\ref{fig:abundchanges} illustrates how much discrepancy 
in relative $n$(N$^{+}$)/$n$(H$^{+}$) abundance can arise 
when {\Nee} and {\tee} values are deviated from the exact solution 
(set to be $10^3$\,cm$^{-3}$ and $10^4$\,K) 
for diagnostics with the {\nii}\,6583\,{\AA} 
(which is known to be temperature diagnostic)
and {\nii}\,121\,$\mu$m
(which is known to be density diagnostic) lines.
The top panel of Fig.\,\ref{fig:abundchanges} shows that 
deviations of tens of \% from the ``true'' value 
are possible when {\te} is altered even with a few hundred K in {\tee}.
Similarly, 
the bottom panel of Fig.\,\ref{fig:abundchanges} proves
discrepancies in tens of \% can arise 
if {\Nee} is varied by only a couple of hundred cm$^{-3}$.

In the present analyses above, 
we see that the exact converged/optimized solution for {\Ne} 
ranges from $\sim$500 to $\sim$1,500\,cm$^{-3}$
(Fig.\,\ref{fig:NeTe}, left), while
{\te} ranges from $\sim$8,000 to $\sim$1,1000\,K 
(Fig.\,\ref{fig:NeTe}, right).
The radial density distribution for these values shows
at any given radial bin, there is a spread of 
$\sim$1,000\,cm$^{-3}$ in {\Ne} and 
$\sim$2,000\,K in {\te} because of the spatial variation
in the azimuthal direction (Fig.\,\ref{fig:NeTeradial}).
According to Fig.\,\ref{fig:abundchanges},
the {\te} spread of $\sim$2,000\,K corresponds 
to $-20$ to $+30$\,\% discrepancy in the resulting N$^+$
abundance, while 
the {\Ne} spread of $\sim$1,000\,cm$^{-3}$ corresponds 
to roughly $\pm50$\,\% discrepancy.
Hence, as soon as we allow simplifications and/or rounding of 
{\Nee} and {\tee} in the course of extinction correction
and plasma diagnostics,
we are destined to be compromised by discrepancies 
at tens of \% in the derivatives.

\subsection{Overall Quantitative Assessment}
\label{sect:assess}

We have thus established that the proposed PPAP
(1) allows us to perform both the extinction correction
and plasma diagnostics as a streamlined single process, and
(2) offers more self-consistent and exact solutions of 
{\chb} and ({\Nee}, {\tee}), plus other derivatives,
than any of the previous procedures in the literature. 
The beauty of PPAP may rest on its straightforwardness,
as it is based solely on a set of relevant line emission 
distribution maps obtained by some spectral mapping observations,
with just the initial choice of the extinction law and $R_V$ 
value to adopt.
There is no need to assume anything else.

To reiterate, PPAP is borne out because there is reciprocal, 
but subtle, dependence of critical parameters between the 
determination of extinction ({\chb}) and plasma diagnostics 
({\Nee} and {\tee} of the same energy regime).
That is, {\chb} is necessary to correct observed line emission maps 
for extinction in order to perform plasma diagnostics,
while {\chb} cannot be obtained unless we obtain {\Nee} and {\tee}
by performing plasma diagnostics with extinction-corrected line 
emission maps.
Thus, none of these quantities can be determined independently,
and hence, by carefully following through these mutual dependence
to the end of the analyses via an iterative search for convergence, 
we can find {\chb} and ({\Nee}, {\tee}) that are consistent with
each other.

\begin{figure*}
    \centering
	\includegraphics[width=\textwidth]{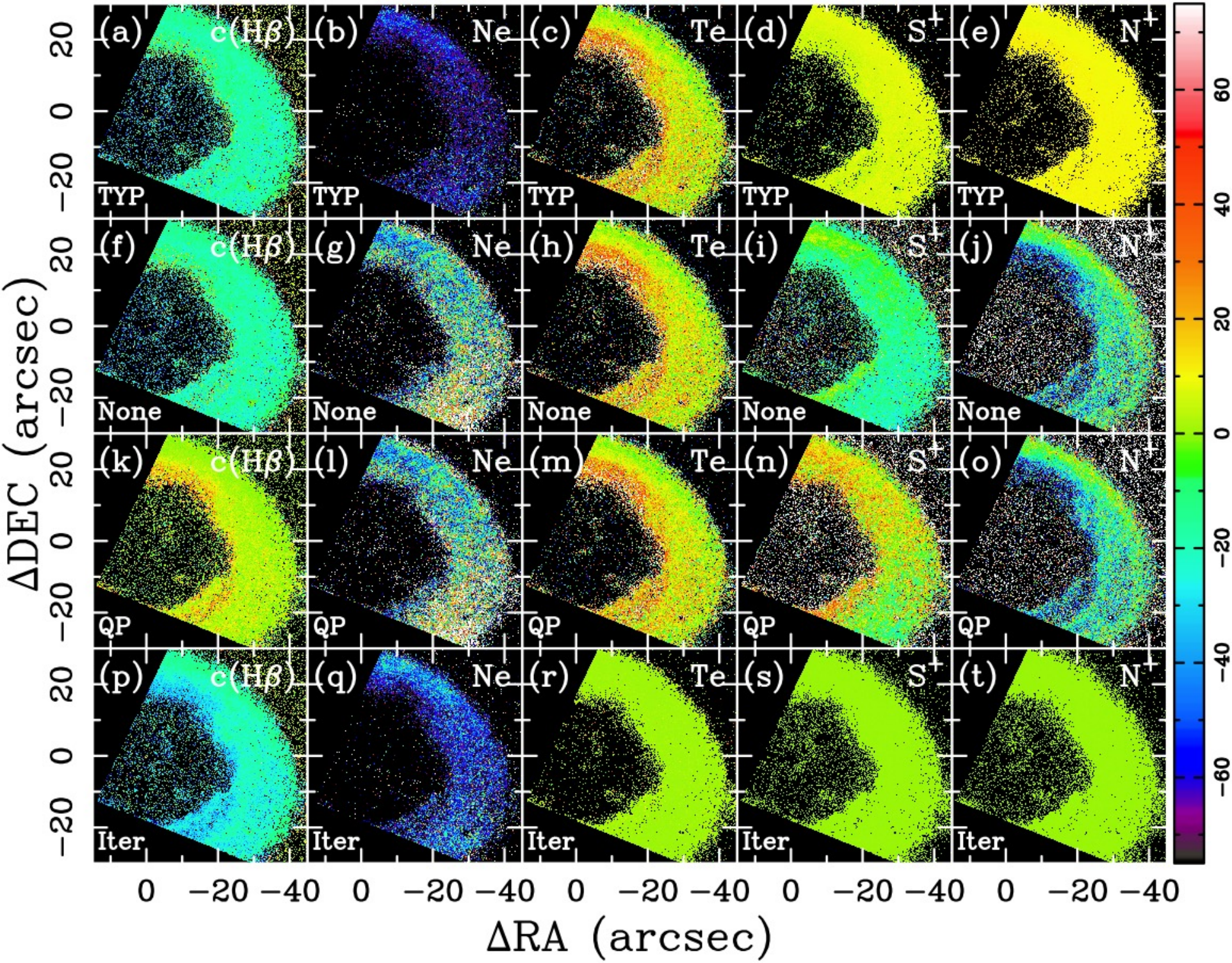}
	\caption{Graphical summary of discrepancies that result
	from various inconsistencies injected over the course of PPAP.
	The observed discrepancies are displayed
	in terms of the relative percentage difference distribution
	in {\chb} (left-most column), {\Nee} (second column from left), and 
	{\tee} (middle column) compared to the full PPAP results 
	(\S\,\ref{sect:ext}, Fig.\,\ref{fig:cHb}; 
	\S\,\ref{sect:denstemp}, Fig.\,\ref{fig:NeTe})
	and in $n({\rm S}^+)/n({\rm H}^+)$ (second column from right), 
	and $n({\rm N}^+)/n({\rm H}^+)$ (right-most column) between 
	the individual results from each of the two diagnostic lines
	(\S\,\ref{sect:abund}, Fig.\,\ref{fig:abund}).
	The four different permutations of the source of inconsistencies 
	are (1) an emulated ``typical'' simplified method  
	in the literature (top row; panels a--e, marked as TYP),
	(2) omitting both the QP line extraction and iterative search 
	for convergence (second row; panels f--j, marked as None),
	(3) executing only the QP line extraction 
	(third row; panels k--o marked as QP),
	and
	(4) executing only the itrative search for convergence  
	(fourth row; panels p--t marked as Iter).
	The color wedge on the right indicates the displayed range of 
	the relative percentage difference between $-75$\,\% to 75\,\%.   
	The median and standard deviation of the relative percentage difference 
	for each case are also summarized in Table\,\ref{tab:compall}.
	Other image conventions follow those of Fig.\,\ref{fig:cHb}.}
    \label{fig:compall}
\end{figure*}

\begin{table*}[]
    \begin{tabular}{lccccc}
    \hline
    PPAP  & $\Delta${\chb}  & $\Delta${\Nee} & $\Delta${\tee} & $\Delta$($n({\rm S}^+)/n({\rm H}^+)$) & $\Delta$($n({\rm N}^+)/n({\rm H}^+)$)  \\
    Permutation & ($\%$) & ($\%$) & ($\%$) & ($\%$) & ($\%$) \\
    \hline\hline
    TYP   & $-14\pm11$ & $-65\pm17$ & $\phantom{-1}7\pm17\phantom{\,(\times 10^{-1})}$ & $\phantom{-1}6\pm\phantom{11}1\phantom{\,(\times 10^{-1})}$ & $\phantom{-1}8\pm\phantom{11}1\phantom{\,(\times 10^{-1})}$  \\
    None  & $-14\pm11$ & $-22\pm42$ & $\phantom{-1}7\pm17\phantom{\,(\times 10^{-1})}$ & $-12\pm\phantom{1}13\phantom{\,(\times 10^{-1})}$ & $-19\pm\phantom{1}34\phantom{\,(\times 10^{-1})}$  \\
    QP    & $\phantom{-1}1\pm\phantom{1}4$ & $-33\pm42$ & $\phantom{-1}7\pm17\phantom{\,(\times 10^{-1})}$ & $\phantom{-1}6\pm\phantom{1}21\phantom{\,(\times 10^{-1})}$ & $-19\pm\phantom{1}33\phantom{\,(\times 10^{-1})}$  \\
    Iter  & $-18\pm12$ & $-53\pm22$ & $\phantom{-1}4\pm\phantom{1}5\,(\times 10^{-1})$ & $\phantom{-1}6\pm343\,(\times 10^{-4})$ & $\phantom{-1}6\pm522$ $(\times 10^{-4})$    \\
    \hline
    \end{tabular}
    \caption{Quantitative summary of discrepancies that result
	from various inconsistencies injected over the course of PPAP
	in terms of the median and standard deviation 
	of the relative percentage difference distribution
	for each case shown in Fig.\,\ref{fig:compall}.}
    \label{tab:compall}
\end{table*}

To further demonstrate the robustness of PPAP, 
here we assess how much improvement is offered by PPAP
by quantifying discrepancies that arise by not following PPAP properly.
As an example of such a ``wrong'' practice, 
first we emulate ``typical'' plasma diagnostics 
widely practiced in the literature by adopting the following
assumptions and simplifications:
\begin{enumerate}
    \item[(1)] No QP line extraction\\
    The raw F656N and F658N images are adopted 
    as the {\ha} and {\nii} 6583\,{\AA} line emission maps, respectively.
    Hence, the adopted {\ha} and {\nii} 6583\,{\AA} line maps are 
    compromised by mutual line contamination.
    \item[(2)] Uniform theoretical {\ha}-to-{\hb} ratio of 2.858\\
    Uniform distribution of {\Nee} at $10^3$\,cm$^{-3}$ and {\tee} at $10^4$\,K 
    is assumed. Then, it follows that the theoretical {\ha}-to-{\hb} ratio
    is 2.858 (e.g.\,\citealt{Hummer1987,storey1995}). Hence, the adopted
    ratio is applicable only when $n_{\rm e}=10^3$\,cm$^{-3}$ and 
    $T_{\rm e}=10^4$\,K.
    \item[(3)] Simplified {\Ne} diagnostic\\
    For an {\Nee} diagnostic using the {\sii} 6717-to-6731\,{\AA} line ratio,
    a simple analytic form of $\log n_{\rm e} = 4.71-2.00\times I(6717)/I(6731)$
    is adopted.
    This expression reproduces the logistic functional relation 
    between {\Nee} and the {\sii} line ratio fairly well 
    under the assumption of $T_{\rm e}=10^4$\,K 
    (e.g.\,\citealt{osterbrock2006,odell2013b}).
    Hence, if the true {\tee} deviates from $10^4$\,K, 
    the relation would not yield correct {\Ne}.
    \item[(4)] Simplified {\te} diagnostic\\
    For a {\tee} diagnostic using the {\nii} the 5755-to-6583\,{\AA} line ratio,
    an analytic expression of 
    $I(5755)/I(6583) 
    = 5.17 \exp(2.5\times10^4/T_{\rm e})/(1+2.5\times T_{\rm e}^{-1/2})$
    is adopted.
    This is an approximated function that relates {\tee} to the 
    {\nii} line ratio  
    under the assumption of $N_{\rm e}=10^3$\,cm$^{-3}$
    (e.g.\,\citealt{osterbrock2006,atomicastro}).
    Hence, if the true {\Nee} deviates from $10^3$\,cm$^{-3}$, 
    the relation would not yield correct {\te}.
\end{enumerate}
In short, in this example, the assumption of uniform {\Ne} of $10^3$\,cm$^{-3}$ 
and {\te} of $10^4$\,K is forced, and 
the issue of line contamination is not properly mitigated.
Hence, unless uniform {\Nee} and {\tee} come out from PPAP as assumed, 
the results of the analyses have to be regarded as suspect
because inconsistencies between the initial assumptions and the outcomes
are not resolved.

As another example of a ``wrong'' practice, 
we perform PPAP only partially, 
by omitting both or either one of the two components of PPAP: 
the QP line extraction and iterative search for convergence.
We can then quantify how much discrepancies can arise in the end
by failing to remove line contamination and/or
leaving inconsistencies among the reciprocally dependent 
critical parameters in the problem.

The resulting discrepancies from all of these trial cases are 
graphically presented in Fig.\,\ref{fig:compall}
as relative percentage difference distribution maps.
For {\chb} and ({\Nee}, {\tee}), 
the outcomes from the ``wrong'' analyses are compared with 
the results from full PPAP (\S\,\ref{sect:ext}, \S\,\ref{sect:denstemp}).
For the N$^+$ and S$^+$ abundances,
a comparison is made between the resulting two abundance maps obtained 
from the adopted two diagnostic lines individually (\S\,\ref{sect:abund}).
The mean and standard deviation of the distribution of 
the relative percentages are also summarized in
Table\,\ref{tab:compall}.

In general, {\Nee} and {\tee} anti-correlate with 
the theoretical {\ha}-to-{\hb} line ratio, and hence, {\chb}.
This is rather intuitive as the presence of attenuating dust grains
(higher {\chb}) suggests 
more contrasted {\ha}-to-{\hb} line ratios and 
lower degrees of ionization (lower {\Nee} and {\tee}).
However, when iterative adjustment of {\chb} and ({\Nee}, {\tee}) are
not performed, the initial assumption of ({\Nee}, {\tee}), i.e., 
how they differ from the true ({\Nee}, {\tee}) values, greatly 
affects the magnitude of discrepancy.

Nonetheless, inspection of Fig.\,\ref{fig:compall} reveals 
a great deal of information. 
The first row of Fig.\,\ref{fig:compall} and Table\,\ref{tab:compall} 
show discrepancies between the outcomes from 
the emulated ``typical'' procedure in the literature
and full PPAP.
By comparing the absolute value of the median
and the standard deviation of the spread, 
we can say that
{\Nee} is significantly underestimated, and 
{\chb} to a lesser extent.
Plus, the abundance maps derived from different lines differ 
by 10\,\%, roughly speaking.

The major issue of this ``typical'' procedure boils down to 
the forced uniform {\Nee} and {\tee} assumption.
The discrepancy between the imposed {\Nee} and {\tee} and
actual {\Nee} and {\tee} culminates as offsets in {\chb}.
For the present case, given the spread in the derived
({\Ne}, {\te}) values (Fig.\,\ref{fig:NeTe}), 
the forced uncertainties already amount roughly to 10--20\,\%.
As the adopted {\Nee} diagnostic curve is very much steeper
than the {\tee} counterpart, even slight offsets in {\chb}
(and hence, in diagnostic line ratios) would influence 
{\Nee} more than {\tee}.
The greatest discrepancies of all at 65\,\% in {\Nee} in this 
procedure reminds us of its shoddiness.
{\tee} is off by 7\,\% on average, 
but we can still consider that {\tee} meanders around the true value.
Similarly, this method is the only one among all that 
produces abundance maps that are not consistent with each other.
Thus, this exercise exemplifies the inappropriateness of 
such a ``typical'' method of plasma diagnostics.

The other rows of Fig.\,\ref{fig:compall} compare
discrepancies among different permutations of PPAP.
The difference between ``TYP'' and ``None'' is that in 
``None'' we rigorously consider the {\Nee} and {\tee} dual dependence 
on both of the {\Nee} and {\tee} diagnostic functions. 
With ``QP'', the mutual contamination between {\ha} and {\nii} is 
addressed, and hence, the resulting {\chb} is among the best
(Fig.\,\ref{fig:compall}f,k,p).
This indeed proves the significance of the QP process where appropriate.
{\Nee} turns out to be a difficult quantity to determine. 
But, it may be expected from the fact that the {\Nee} diagnostic
curve is a rather steep function of the adopted diagnostic line ratio.
With ``Iter'',
what is striking is the goodness of the match we see in
{\tee} and N$^+$ and S$^+$ abundances:
the iterative process alone recovers the correct solutions
for these quantities.
Nevertheless, it alone cannot resolve everything.
It appears that the correctness of {\chb} and {\Ne}
is sacrificed at the expense of the other quantities
during the iterative optimization of the solution.

All in all, lessons learned from this exercise are very straightforward.
The widely practiced simplifications 
of uniform {\Nee} and {\tee} assumption 
in ``typical'' plasma diagnostics 
without considering the extinction correction simultaneously
would only feed discrepancies.
Both the extinction correction and plasma diagnostics 
must be performed altogether as an integrated process.

As for PPAP, 
it is {\em necessary} to perform the QP process
(or anything takes care of spatially resolved line calibration)
iterative search for converged solutions.
If not, imposed inconsistencies would amplify 
as relative discrepancies at tens of \%
over the course of the whole analyses in one way or another, 
affecting various results seemingly at random. 
In other words, 
if self-consistency is maintained 
without imposing obviously unrealistic uniform {\Nee} and {\tee}
assumptions
in performing 
plasma diagnostics including the extinction correction,
it is possible to obtain rather exact solution 
for {\chb} and ({\Nee}, {\tee}) as well as 
other derivatives such as abundances
and carry out any subsequent quantitative analyses quite rigorously.

In extreme cases where the goal is determining just {\tee}, however,
it {\em may be} tolerable as long as extinction correction
and plasma diagnostics are performed iteratively for converged solutions.
This is possible when the {\tee} diagnostic curve happens to be 
only a weak function of the diagnostic line ratios.
In that case, it is critical to remember that {\chb} and {\Nee} are 
compromised (i.e., the remaining discrepancies are absorbed 
as offsets in {\chb} and {\Nee}).

\subsection{Other Caveats}
\label{sect:others}

\subsubsection{Plasma Diagnostics with {\oiii} Lines}
\label{sect:O3}

\begin{figure*}
\gridline{\fig{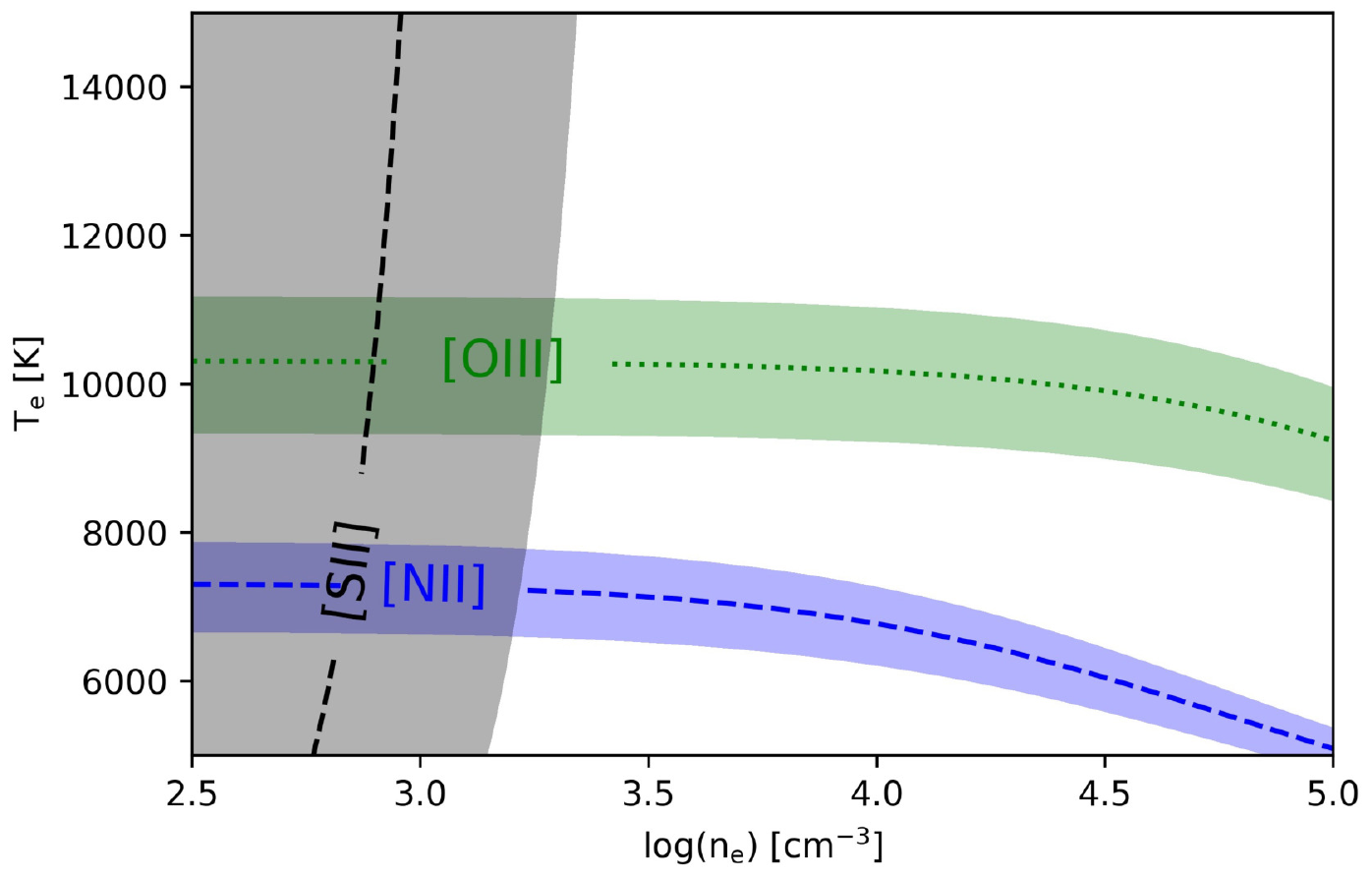}{0.5\textwidth}{Position (a)}
          \fig{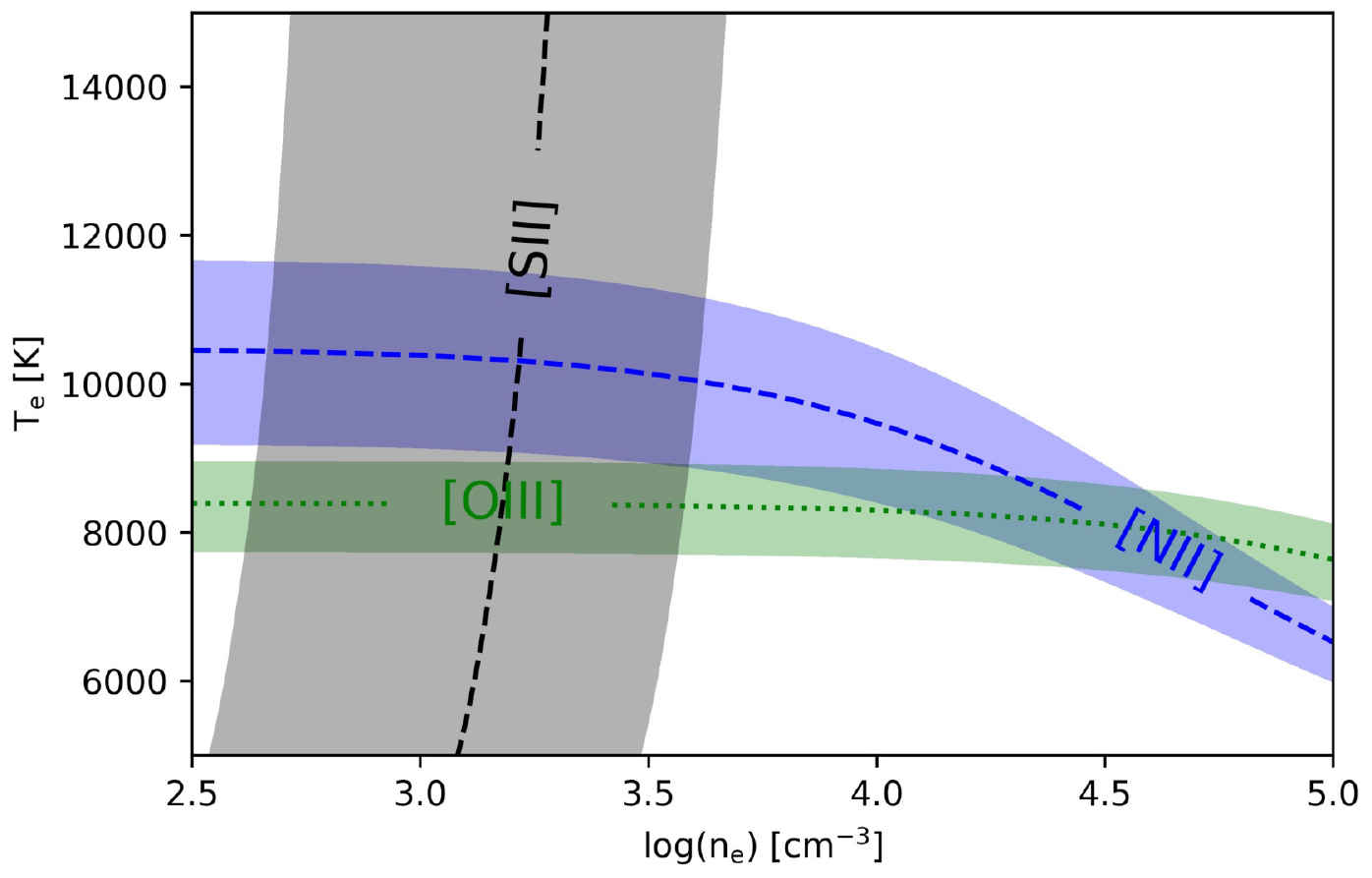}{0.5\textwidth}{Position (b)}}
\gridline{\fig{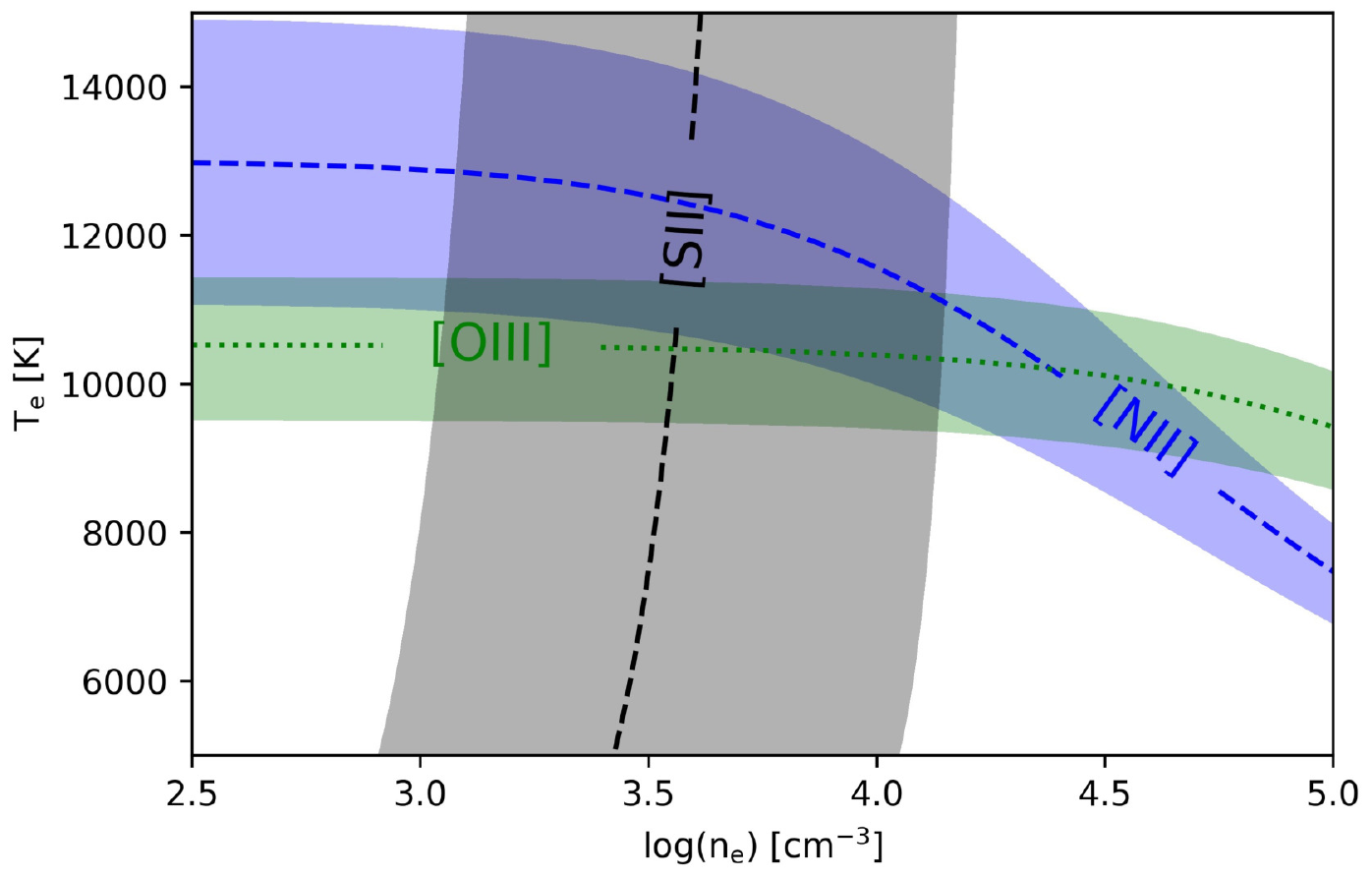}{0.5\textwidth}{Position (c)}
          \fig{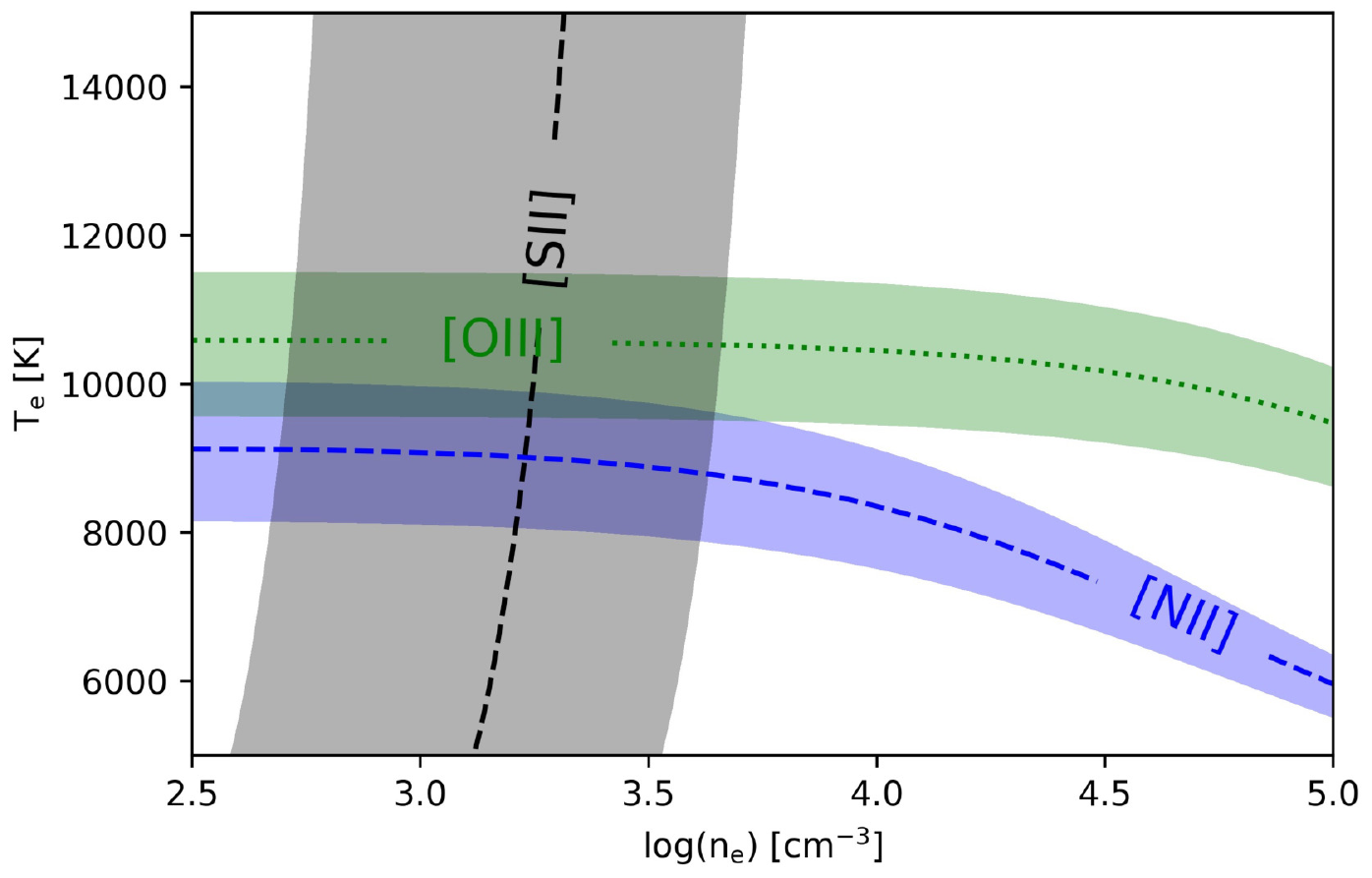}{0.5\textwidth}{Position (d)}}
    \caption{The {\Nee}-{\tee} diagrams 
    for the range of $10^{2.5} < n_{\rm e} < 10^{5}$\,cm$^{-3}$
    and $5,000 < T_{\rm e} < 15,000$\,K,
    showing the {\sii}, {\nii}, and {\oiii} diagnostic curves
    at four local emission peaks in NGC\,6720:
    (a) the inner edge of the main ring along the short axis,
    (b) the outer edge of the main ring along the short axis,
    (a) the inner edge of the main ring along the long axis, and
    (b) the outer edge of the main ring along the long axis.
    The thickness of each diagnostic curve represents uncertainties
    propagated from the assumed 20\,\% uncertainty in the input line ratio.}
    \label{fig:nTdiag}
\end{figure*}

For the present analyses, we adopt {\sii} and {\nii} line ratios
as our primary diagnostics.
Naturally, these diagnostics are suited to probe low-excitation regions
in the outer part of the main ring where these emission lines are strong
(Fig.\,\ref{fig:linemaps}d--h).
The present data set also includes the {\oiii} line maps at
4363 and 5007,{\AA} (Fig.\,\ref{fig:linemaps}a,c), 
which provide another diagnostic line ratio
for higher-excitation regions in the inner part of the main ring.
The difference in the spatial coverage among these diagnostics is
evident from the line ratio maps (Fig.\,\ref{fig:O3N2}).
This spatial anti-correlation is actually a critical point to consider 
in plasma diagnostics, especially when selecting two diagnostics as a pair.
However, this view was often neglected in the past because spectra
were rarely taken in a spatially resolved manner.
Therefore, it is instructive to closely examine how plasma diagnostics 
work out taking into account the spatial distribution of
diagnostic lines, especially when one deals with spatially resolved
spectral images.

Fig.\,\ref{fig:nTdiag} presents the {\Nee}-{\tee} diagram 
at four positions in the main ring.
Panels (a) and (b) are of a local peak in {\oiii} (Fig.\,\ref{fig:linemaps}e,g)
near the inner and outer edge of the main ring along the short axis 
(i.e.\ NNW direction).
Similarly, panels (c) and (d) are of a local peak in {\nii} (Fig.\,\ref{fig:linemaps}d--f) and {\sii} (Fig.\,\ref{fig:linemaps}g,h)
as well as {\oiii}
near the inner and outer edge of the main ring along the long axis
(i.e.\ NSW direction), respectively.
These {\Nee}-{\tee} diagrams evidently show how each of the 
{\sii}, {\nii}, and {\oiii} diagnostic line ratios behaves
differently at each position because of the local physical conditions
(Fig.\,\ref{fig:nTdiag}).

As reviewed in \S\,\ref{sect:plasma},
any given diagnostic line ratio can be expressed 
as a non-linear function of {\Nee} and {\tee}. 
Hence, each line ratio takes the form of a curve in the {\Nee}-{\tee} space
as shown in Fig.\,\ref{fig:nTdiag}.
The shaded thickness for each line represents uncertainties 
propagated from the assumed 20\,\% uncertainty in the extinction correction.
Then, the intersection of two diagnostic curves would specify 
{\Nee} and {\tee}, provided that the corresponding emission lines 
arise from the same region of the target nebula.
Because these curves are non-linear functions,
intersections can only be determined numerically.
This is indeed why plasma diagnostics must be done as an iterative process.

In addition, for such a numerical {\Nee} and {\tee} search to proceed optimally,
an {\Nee} diagnostic should cover a wide range
of {\tee} at nearly constant {\Nee}. 
Similarly, a {\tee} diagnostic should cover a wide range of {\Nee} 
at nearly constant {\tee}.
This is because we would like an {\Nee} diagnostic immune to 
the {\em wrong} initial {\tee} guestimate so that the inherent 
{\tee} dependence of the diagnostic can be ignored, and vice versa.
Still, for the optimal results, 
both {\Nee} and {\tee} should be simultaneously determined 
by seeking convergence through a nested iterative process.

As demonstrated by Fig.\,\ref{fig:nTdiag}, diagnostic curves 
intersect with each other at multiple positions in the {\Nee}-{\tee} space.
This exemplifies the fact that we cannot indiscriminately choose 
any pair of diagnostic lines, especially because 
differences of a few hundred\,cm$^{-3}$ and a few thousand\,K
can easily cause at least tens of \% difference in the 
resulting abundances as shown above at the end of \S\,\ref{sect:abund}.
Ideally, we should like to select {\Nee} and {\tee} diagnostics that 
are of similar transition energies arising from the same part of 
the target object.
If not, the resulting {\Nee} and {\tee} can easily be off by 
a few thousands K and/or a couple of hundreds of cm$^{-3}$,
and hence, several tens of \% discrepancies in the resulting 
abundances. 

\subsubsection{Contamination by {\heii} Lines}
\label{sect:he2}

As shown in Fig.\,\ref{fig:linemaps}b,
the {\heii}\,4686\,{\AA} line is reasonably strong: 
its relative strength to {\hb} is as strong as
the {\sii} 6717/31\,{\AA} doublet 
and roughly $\sim$10\,\% of {\ha}.
Thus, the effect of the {\heii} line contamination in the {\ha}
and {\hb} maps needs to be assessed. 
If it is significant,
the resulting {\chb}, and subsequently, {\Ne}, and {\te},
will be compromised, and will have to be corrected for accordingly.
To this end, we perform 
the iterative derivation of {\chb}, {\Ne}, and {\te}
one more time using the {\ha} and {\hb} maps from which 
their respective contaminating {\heii} lines at 6560 and 4859\,{\AA} removed.

This consideration actually adds another layer of complication in PPAP,
because now we have to know the extinction, $c(\lambda)$, 
in order to derive the very extinction itself at Step (2).
More specifically, 
the observed {\ha}-to-{\hb} ratio that is
needed to derive {\chb} (Eq.\,(\ref{eq:chb})) has to be replaced 
by the version of the {\ha}-to-{\hb} ratio from which {\heii} 
contamination is subtracted.
However, to calculate the {\heii} line flux distribution 
at 6560 and 4859\,{\AA}
(the contaminants to be subtracted from the {\ha} and {\hb} maps,
respectively),
the {\em extinction-corrected} {\heii}\,4686\,{\AA} map 
needs to be scaled by an appropriate
theoretical line ratio (e.g.\,\citealt{Hummer1987,hummer1998,storey1995}),
i.e.\
\begin{equation}
\begin{aligned}
 &\frac{I(\mathrm{H}\alpha)-I(\mathrm{He\,{\sc II}})_{6560}}{I(\mathrm{H}\beta)-I(\mathrm{He\,{\sc II}})_{4859}}
 \nonumber\\
 &\quad =
 \frac{I(\mathrm{H}\alpha)-I_0(\mathrm{He\,{\sc II}})_{6560} \cdot 10^{-c(6560)}}
 {I(\mathrm{H}\beta)-I_0(\mathrm{He\,{\sc II}})_{4859} \cdot 10^{-c(4859)}}
 \nonumber\\
 &\quad =
 \frac{I(\mathrm{H}\alpha)-I_0(\mathrm{He\,{\sc II}})_{4686} \cdot R\left(\frac{6560}{4686}\right) \cdot 10^{-c(6560)}}
 {I(\mathrm{H}\beta)-I_0(\mathrm{He\,{\sc II}})_{4686} \cdot R\left(\frac{4859}{4686}\right) \cdot 10^{-c(4859)}},
 \nonumber\\
 &\quad =
 \frac{I(\mathrm{H}\alpha)-I(\mathrm{He\,{\sc II}})_{4686}
 \cdot 10^{c(4686)} \cdot R\left(\frac{6560}{4686}\right) \cdot 10^{-c(6560)}}
 {I(\mathrm{H}\beta)-I(\mathrm{He\,{\sc II}})_{4686} 
 \cdot 10^{c(4686)} \cdot R\left(\frac{4859}{4686}\right) \cdot 10^{-c(4859)}},
 \nonumber\\
 &\quad =
 \frac{I(\mathrm{H}\alpha)-I(\mathrm{He\,{\sc II}})_{4686}
\cdot R\left(\frac{6560}{4686}\right) \cdot 10^{c(4686)-c(6560)}}
 {I(\mathrm{H}\beta)-I(\mathrm{He\,{\sc II}})_{4686} 
\cdot R\left(\frac{4859}{4686}\right) \cdot 10^{c(4686)-c(4859)}},
\end{aligned}
\end{equation}
where $R(\lambda_1/\lambda_2)$ is the theoretical {\heii} 
line ratio between wavelengths at $\lambda_1$ and $\lambda_2$ 
(which is a function of {\Ne} and {\te}; 
e.g.\,\citealt{Hummer1987,hummer1998,storey1995})
and the subscripted values refer to the line wavelengths.

\begin{figure}
    \centering
	\includegraphics[width=\columnwidth]{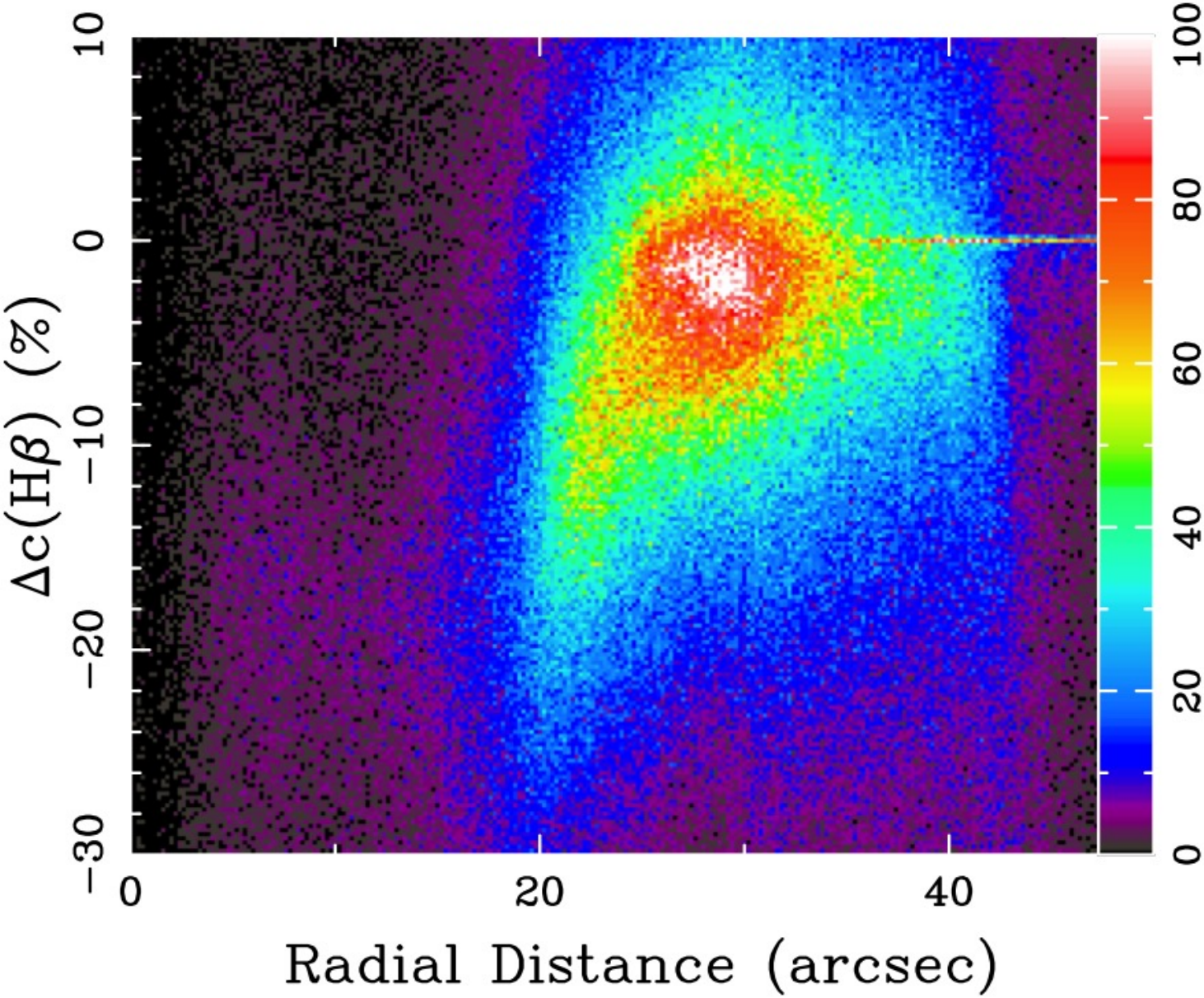}
	\caption{The radial density distribution of the percentage difference 
	in {\chb} compared to the results without the {\heii} contamination.
	The original {\chb} result (with the {\heii} contamination) 
	is underestimated especially in the inner half of the main ring 
	interfacing with the high-excitation central cavity ($<25^{\prime\prime}$). 
	The 0\,\% region seen beyond 35$^{\prime\prime}$ 
	is the no-data part beyond the outer edge of the main ring.}
    \label{fig:cHbHe2comp}
\end{figure}

Using this {\ha}-to-{\hb} ratio without the {\heii} contamination, 
we perform the same iterative processing to obtain 
the {\chb}, {\Ne}, and {\te} maps.
This modification tends to increase the {\ha}-to-{\hb} ratio slightly,
and hence, tends to increase the resulting {\chb} slightly.
The difference of the {\chb} values before and after applying 
this {\heii} contamination removal is quantified as the percentage
difference ratio, 
$\left(c(\mathrm{H}\beta)-c(\mathrm{H}\beta)_{\mathrm{no\,He\,{\sc II}}}\right)/c(\mathrm{H}\beta)_{\mathrm{no\,He\,{\sc II}}}$,
and plotted as a radial profile in Fig.\,\ref{fig:cHbHe2comp}.

The revised {\chb} with the {\heii} contamination
mitigated comes out slightly greater.
Thus, the {\heii} contamination turns out greater in {\hb} than in {\ha}
in the present case.
As a result, the percentage difference ratio comes out to be negative.
The difference is minor in the outer part of the main ring
($>25^{\prime\prime}$; at $-2.9\pm12.7$\,\%), and hence,
the revised {\Ne} and {\te} distributions do not show much difference.
On the other hand,
the difference increases progressively toward the inner
edge of the main ring ($<25^{\prime\prime}$):
the difference becomes as large as $\sim$20\,\% at the inner
edge ($\sim$20$^{\prime\prime}$).

It appears that the relative amount of the {\heii} contamination 
grows greater toward the center of the main ring into the inner cavity.
This agrees with the expectation that the effects of the {\heii} 
contamination is greater where the degree of excitation is higher.
For the present case of NGC\,6720, 
it appears that only the inner cavity 
is where the {\heii} contamination needs to be seriously mitigated,
as also suggested from its spatial distribution
(Fig.\,\ref{fig:linemaps}b).
Thus, we conclude that the {\heii} contamination must be 
properly assessed when determining quantities pertaining to 
high-excitation ionic species (such as {\heii}, of course).

\subsubsection{Contribution by Recombination}
\label{sect:recombination}

We should also bear in mind that strong recombination lines may 
influence the {\tee} determination (e.g.\ \citealt{2000MNRAS.312..585L,gl2020}).
For the present case, the {\nii}\,5755\,{\AA} line flux 
may be enhanced by recombination from the ${\rm N}^{2+}$ lines to 
the said line, and that the resulting {\te} may not solely be of 
collision.

This is, however, not likely for NGC\,6720.
\citet{zhang2004} reported that such contribution by recombination was 
extremely small in this nebula. 
Also, based on the far-IR line emission maps obtained with the
{\sl Herschel Space Observatory} as part of the Herschel
Planetary Nebula Survey \citep{ueta2014}, 
we can see that the spatial distributions of
the {\nii}\,122\,$\mu$m and {\niii}\,57\,$\mu$m lines
(not shown here; Ueta et al.\ {\sl in preparation})
show similar spatial differences as we see between 
the {\nii}\,5755/6548/6583\,{\AA} lines and 
the {\oiii}\,4363/5007\,{\AA} lines (Fig.\,\ref{fig:linemaps}).

When such contribution by recombination is expected to be 
significant, one needs to assess and remove the ${\rm N}^{2+}$ 
contribution by recombination.
Under PPAP, we can do so by adding another iterative loop,
as done to mitigate contamination by {\heii} lines (see \S\,\ref{sect:he2}). 
For example, using {\tee}(\oiii), 
{\Nee}([Cl\,{\sc iii}]), and {\niii} collisional excitation line maps
(e.g.\ at 1750\,{\AA} or 57\,$\mu$m), 
one can synthesise the map from N$^{2+}$ recombination 
contribution to {\nii}\,5755\,{\AA} 
by following Eq.(1) of \citet{2000MNRAS.312..585L}.
Subsequently, the updated {\nii}\,5755\,{\AA} line flux, 
which represents purely collisionally excited line flux,
is used to compute {\tee}(\nii) and {\Nee}(\sii) and 
derive {\chb}, which would then update the extinction corrected 
line maps for the next round of the iteration.

\section{Summary}
\label{sect:summary} 

We have established a proper plasma analysis practice (PPAP), 
a streamlined iterative procedure that integrates 
the extinction correction and plasma diagnostics 
for spatially extended targets in one go
(\S\,\ref{sect:iteration}; Fig.\,\ref{fig:schema}).
The major strength of PPAP is that it requires nothing other 
than just the input spectral images of critical diagnostic lines
plus the choice of the extinction law and $R_V$ value to adopt.
In other words, there is absolutely no need to assume anything. 

This work is motivated by the recognized but subtle dependence of 
the power-law extinction index, $c(\lambda)$, on both the electron 
density and temperature, ({\Nee}, {\tee}), which are the very 
quantities that must be known to determine $c(\lambda)$.
Such circular dependence must be resolved via an iterative process
looking for the optimum {\chb} (which is $c(\lambda)$ at {\hb}) and 
({\Nee}, {\tee}) values for convergence.
Unfortunately, however, these interdependent analyses have rarely 
been considered altogether as one in the past.

With this in mind, 
using a suite of narrowband filter images of NGC\,6720 taken
with WFC3 on {\sl HST\/}, we have derived by following PPAP
a self-consistent and spatially-resolved solution of
{\chb} (\S\,\ref{sect:ext}; Fig.\,\ref{fig:cHb}) and 
({\Nee}, {\tee}) (\S\,\ref{sect:denstemp}; Fig.\,\ref{fig:NeTe}) 
for the object simultaneously.
In the present exercise, the obtained solution pertains to 
the NW quadrant of the main ring structure in NGC\,6720,
for which critical diagnostic line maps are available
(e.g.\ the low-excitation {\sii} and {\nii} diagnostic line ratio maps).

The derived {\chb} and ({\Ne}, {\te}) maps clearly reveal 
spatial variations within the nebula,
which would not have been uncovered if uniform 
({\Nee}, {\tee}) were assumed, in unprecedented detail.
Also, the {\chb} map correctly accounts for both 
the ISM and circumsource components (Fig.\,\ref{fig:cHbradial})
to the whole extent of the nebula for which we have measurements.
The ionized gas-to-dust mass ratio in the main ring is found to be
fairly spatially varying with the median of $437\pm357$, much greater 
than the canonical value of 100 (\S\,\ref{sect:G2D}; Fig.\,\ref{fig:G2D}).
The total gas-to-dust mass ratio in the main ring of NGC\,6720 is
estimated to be about 1600.
Moreover, we have obtained 
properly extinction-corrected line emission maps
(\S\,\ref{sect:linemaps}; Fig.\,\ref{fig:linemaps})
as well as the relative ionic abundance distribution 
maps (N$^{+}$ and S$^{+}$ relative to H$^{+}$;
\S\,\ref{sect:abund}; Fig.\,\ref{fig:abund}) as the 
essential products of plasma diagnostics.

We have demonstrated that difference of only a few 
hundred\,cm$^{-3}$ in {\Nee} or a few thousand K in {\tee}
(commonly seen as the amount of spatial variations;
Fig.\,\ref{fig:NeTeradial})
can introduce differences in tens of \% in the resulting 
metal abundances (Fig.\,\ref{fig:abundchanges}).
Therefore, in the era of spatially-resolved spectroscopy,
we cannot afford to continue this old tradition of 
assuming constant {\Nee} and {\tee} (among other things)
without paying attention to the interdependence between
extinction correction and plasma diagnostics as well as
self-consistency among parameters. 
If we perform plasma diagnostics without PPAP, 
we will not be able to maintain the integrity of data 
at each spatial element.
This is because results of such analyses will always suffer
from uncertainties at tens of \% that can easily arise 
from inconsistencies introduced by the unnecessary assumptions.
This will simply defeats the purpose of conducting 
spatially-resolved spectroscopy in the first place.

We have also evaluated how much discrepancies can arise at each
step of the process if PPAP is not strictly followed
(\S\,\ref{sect:assess}; Fig.\,\ref{fig:compall}; Table\,\ref{tab:compall}).
Our analyses have demonstrated that plasma diagnostics
typically practiced in the literature 
and/or neglecting PPAP even partially 
would sustain uncertainties greater than 10\,\%.
We have also estimated that for PPAP to be effective
{\Nee} and {\tee} diagnostic line fluxes need to be 
determined at uncertainties better than 10\,\% in general.

The strength of securing self-consistent {\Nee} and {\tee} 
(and {\chb}) has been vindicated by the fact that 
multiple $n$(N$^{+}$)/$n$(H$^{+}$) and $n$(S$^{+}$)/$n$(H$^{+}$) 
distribution maps 
derived from the corresponding {\nii} and {\sii} line maps of 
distinct transitions come out to be identical.
Such a feat has never been accomplished in plasma diagnostics 
with spectral imaging data in the past.
Only by adopting PPAP fully and not incorporating any assumptions
that would degrade the observational data in any way (especially spatially), 
we can obtain robust outcomes.

Furthermore, 
there are always certain sets of diagnostic lines 
that are suited to probe particular parts of target sources,
like the {\nii} and {\sii} line ratio pair that probes 
low-excitation regions.
One should always be conscious about 
which {\Nee} and {\tee} diagnostic lines to use  
depending on different local physical conditions to be probed
(\S\,\ref{sect:O3}). 
To that end, when probing high-excitation regions for which
the {\oiii} lines or other diagnostic lines of the same
and higher transition energy regimes work, 
one should like to remove contamination in H lines 
by {\heii} lines
and consider contribution by strong recombination lines in
metal lines for optimum results (\S\,\ref{sect:he2}).

Various lessons learned from the present exercise are; 
\begin{enumerate}
    \item Plasma diagnostics should really be performed 
    with the extinction correction as an 
    integrated iterative procedure for the best results.

    \item We should remind ourselves that the theoretical 
    {\ha}-to-{\hb} ratio (or any H\,{\sc i} line ratio) is 
    a function of {\Nee} and {\tee} and never a constant.
    
    \item Imposing ({\Nee}, {\tee}) constancy in plasma diagnostics 
    will inject uncertainties at tens of \% in the outcomes of the
    analyses. 
    In the literature, such uncertainties are often
    attributed to the local fluctuations of physical conditions.
    In reality, the initial constancy assumption is what amplifies
    uncertainties: PPAP can resolve such local fluctuations.
    
    \item As long as self-attenuation by the circumsource material
    is expected, it is wrong to adopt the ISM extinction value
    for target sources, 
    as the non-negligible circumsource extinction will be surely 
    missed (and make ({\Nee}, {\tee}) incorrect).
    
    \item By the same token, it is incorrect to indiscriminately 
    adopt the empirical $A(V)$-to-$N_{\rm H}$ relation for ISM
    when some non-negligible amount of circumsource attenuation is 
    expected, as the ISM relation implicitly assumes the gas-to-dust
    mass ratio of 100, which is not necessarily true for individual
    target sources.

\end{enumerate}


\appendix

\section{QP Line Extraction}
\label{sect:fullqp}

Using the adopted WFC3 images with the QP method \citep{ueta2019},
we can separate 
(i) the {\ha} map at 6563\,{\AA} and 
the {\nii} maps at 6548 and 6583\,{\AA}
from the F656N and F658N image pair,
(ii) the {\sii} maps at 6717 and 6731\,{\AA} 
from any pair among the FQ672N, F673N, and FQ674N images, and
(iii) the {\hg} map at 4340\,{\AA} and the {\oiii} map at 4363\,{\AA}
from the FQ436N and FQ437N image pair.
The nebular continuum is determined 
by taking the mean of the three ``continuum'' band images 
(F547M, F645N, and FQ750N)
as long as pixel values register with ${\rm S/N} \ge 3$.

Here, the continuum emission is assumed to be distributed 
more or less similarly across the relevant spectral range.
Then, to subtract the continuum from each emission band map, 
an appropriate scaling factor of the averaged continuum map 
is determined using the integrated ADU counts\footnote{The native 
counts of the archived {\sl HST\/} data, usually referred to as
the Data Numbers (DNs), in e$^{-1}$ or e$^{-1}$\,s$^{-1}$,
depending on the instrument \citep{desjardin2019}.}
of about 10 field stars 
that appear in the FoV of both the continuum and emission band images.
For this work, we implement the QP algorithm anew in Python 
by adopting the Operator Splitting Quadratic Program (OSQP) solver 
(\citealt{osqp}).\footnote{Available from https://github.com/oxfordcontrol/osqp}

\subsection{{\ha} and {\nii}}

\begin{figure}[b]
    \centering
	\includegraphics[width=\columnwidth]{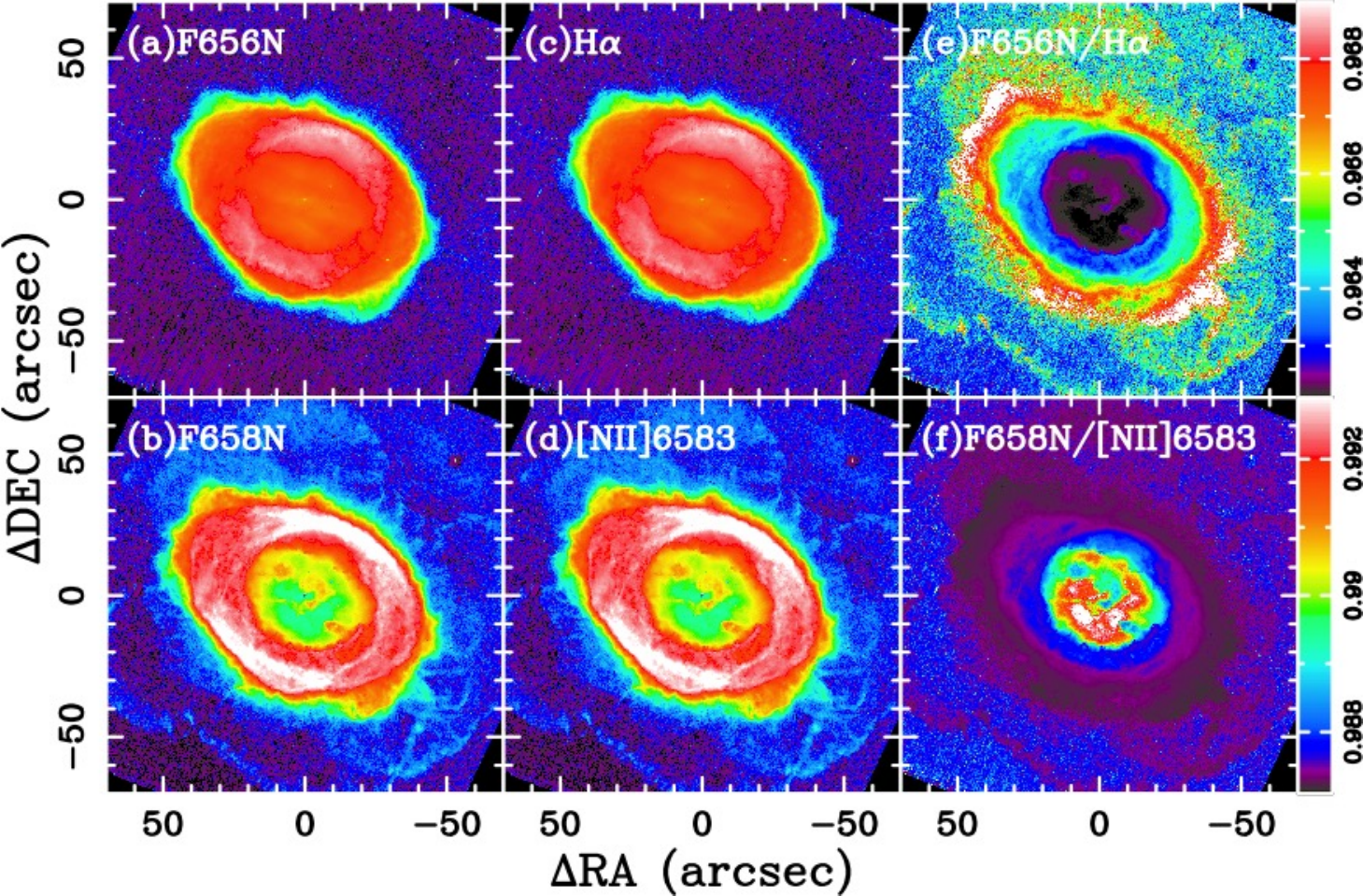}
    \caption{Summary of the QP extraction of the {\ha} and {\nii} 6583\,{\AA}
    line maps from the F656N and F658N images.
    From top-left to bottom-right,
    (a) F656N, (b) F658N, 
    (c) QP-extracted {\ha}, and  
    (d) QP-extracted {\nii}\,6583\,{\AA} line flux maps
    (in erg\,s$^{-1}$\,cm$^{-2}$\,pix$^{-1}$;
    log-scaled from $10^{-15.5}$ to $10^{-18.5}$)
    and
    (e) F656N-to-{\ha} and
    (f) F658N-to-{\nii}\,6583\,{\AA} line flux ratio maps
    (the wedge on the right indicating the range of the displayed ratio).
    Image conventions follow those for the {\chb} map in Fig.\,\ref{fig:cHb}.}
    \label{fig:ha}
\end{figure}

For the {\ha} and {\nii} line separation, 
we employ the following two conditions,
(i) the positivity condition (i.e.\ derived line fluxes cannot be negative) 
and
(ii) the theoretical line strength ratio of the {\nii} 6583\,{\AA}
line to the {\nii} 6548\,{\AA} line being 2.96 \citep{ueta2019}.
The second condition is relevant only in separating {\ha} and {\nii} lines
from the F656N and F658N pair:
we use only the positivity condition in separating other lines.

As required by the QP formulation, 
all the input images are taken to have been modulated 
by the system throughput.
Thus, we multiply the input WFC3 images 
from the archive by the bandpass unit response
(which is 
the value stored under the PHOTFLAM keyword
in the FITS image header and 
converts the original pixel count into the flux density at
erg\,s$^{-1}$\,cm$^{-2}$\,{\AA}$^{-1}$) and 
the bandpass equivalent width 
(which modulates the surface brightness of the bandpass 
by the corresponding system throughput)
computed via {\sc pysynphot} \citep{pysynphot}.
The QP process then extracts line flux distribution maps of
{\ha}\,6563\,{\AA},
{\nii}\,6548\,{\AA},
{\nii}\,6583\,{\AA},
{\sii}\,6717\,{\AA},
{\sii}\,6731\,{\AA},
{\hg} at 4340\,{\AA},
and
{\oiii}\,4363\,{\AA}
in erg\,s$^{-1}$\,cm$^{-2}$\,pix$^{-1}$
as observed (i.e.\
modulation by the system throughput is corrected for,
but the extinction is not yet corrected for).
Other narrowband images that can isolate each of the corresponding target lines
are multiplied by the bandpass unit response
and the bandpass rectangular width
(which is the equivalent width divided by the maximum system throughput 
in the bandpass) to convert the pixel units to flux
(erg\,s$^{-1}$\,cm$^{-2}$\,pix$^{-1}$;
modulation by the system throughput is corrected for,
but the extinction is not yet corrected for) 
for subsequent processes with the QP-extracted line maps.

Fig.\,\ref{fig:ha} shows 
the original WFC3 filter images of F656N (panel\,a) and F658N (panel\,b) 
and the QP-extracted {\ha} (panel\,c) and {\nii}\,6583\,{\AA} (panel\,d)
line flux maps,
along with the original-to-QP line flux ratio maps
(panels\,e and f) of NGC\,6720.
These results are consistent with the previous results \citep{ueta2019},
straightforwardly demonstrating the power of the QP line extraction.
For the present work, 
we display the surface brightness distribution in the ``petal''
structures beyond the main ring, in which the S/N ratio 
to show that the faint petal emission is largely of low-excitation {\nii}
\citep{lame1994,Martin2016}.

The F656N-to-{\ha} flux ratio tends to be low
in the central cavity of the main ring 
(up to $\sim$5\,\%; Fig.\,\ref{fig:ha}e).
The F658N-to-{\nii}\,6583\,{\AA} flux ratio map 
shows lower ratios near the periphery of the main ring
(a few\,\%; Fig.\,\ref{fig:ha}f).
This means that if the F656N map is blindly adopted to represent 
the {\ha} emission distribution, the {\ha} emission 
would be underestimated by at least 3\,\% in the periphery
of the main ring and up to $\sim$5\,\% in the central cavity.
Similarly, if the F658N map is hastily taken to represent 
the {\nii} 6583\,{\AA} emission distribution,
the {\nii} 6583\,{\AA} emission would be underestimated by
about 1\,\% in the central cavity and up to $\sim$2\,\%
in the main ring.
These differences may be small, but will be compounded
in the subsequent analyses to cause greater inconsistencies.

More importantly, we must remind ourselves that 
the spatial distribution of {\ha} and {\nii} is different 
to begin with.
{\ha} generally represents the ionized region, 
whereas {\nii} usually corresponds to the lower-temperature PDR.
Therefore, if both the {\ha} and {\nii} line maps suffer from
mutual line blending, the line ratio map between them 
tends to be marginalized, i.e., 
any structures we observe as {\ha}-to-{\nii} ratio variations
tend to be ``washed out'' as we see in the petal structures.
For example, this {\ha}-to-{\nii} marginalization would
blur the location of the ionization front (IF) that presumably
exists where the {\ha}-to-{\nii} gradient tends to be large.
Thus, keeping the spatial consistency is important in 
investigating spatially resolved line emission.

\subsection{{\sii} 6717 and 6731\,{\AA}}

\begin{figure}
    \centering
	\includegraphics[width=\columnwidth]{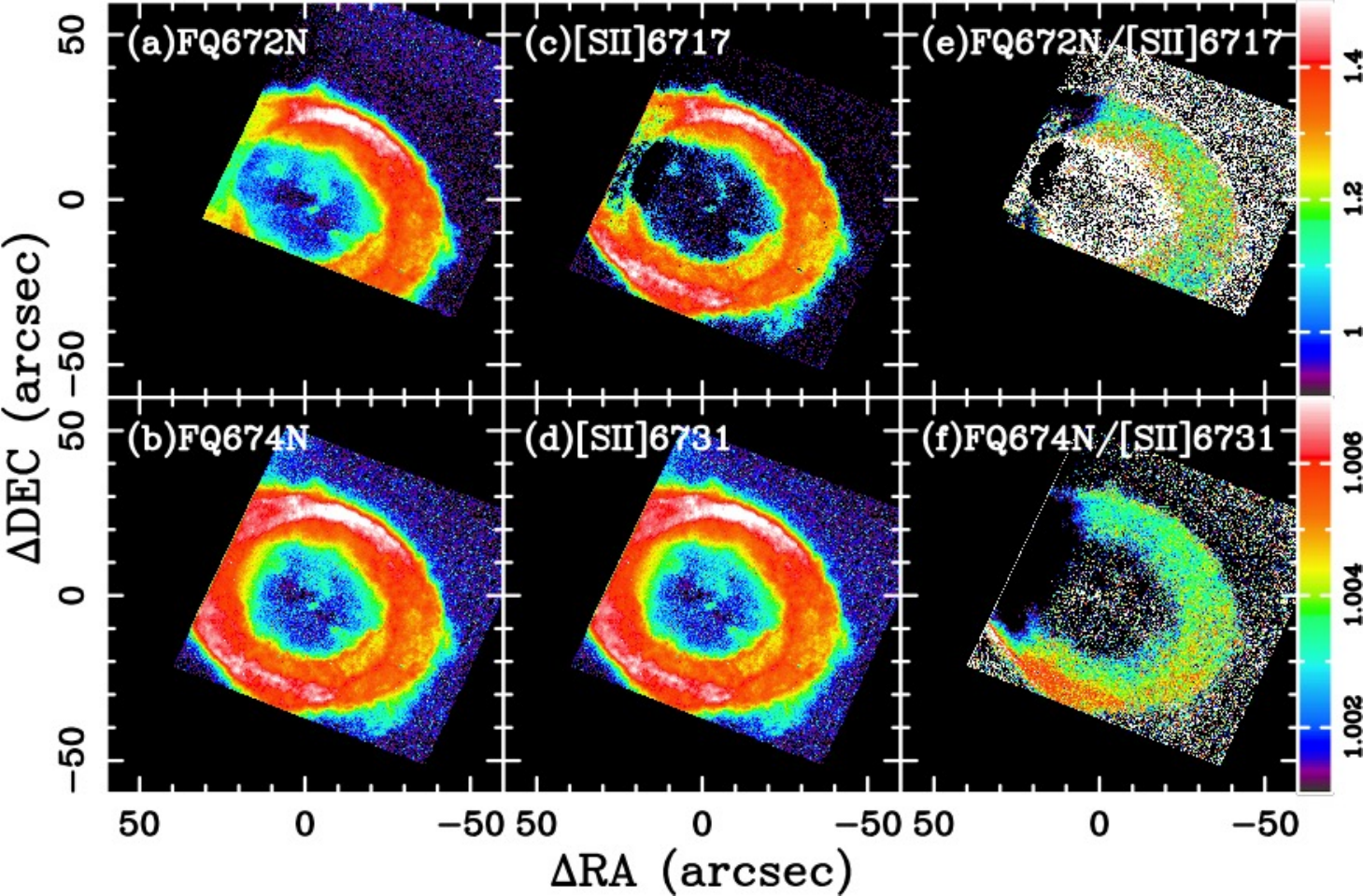}
	\caption{Same as Fig.\,\ref{fig:ha}, but for 
	the {\sii}\,6717 and 6731\,{\AA} line maps
	from the FQ672N and FQ674N images
	and
	log-scaled flux distribution 
	from $10^{-16.5}$ to $10^{-18.5}$ 
	erg\,s$^{-1}$\,cm$^{-2}$\,pix$^{-1}$.}
    \label{fig:s2}
\end{figure}

Fig.\,\ref{fig:s2} displays the QP results for
the {\sii}\,6717/31\,{\AA} line pair.
For this group, QP processing can be used with 
any pair of the three maps or all three maps
among the FQ672N (panel\,a), F673N (not shown), and FQ674N (panel\,b) filters.
For the present analysis, we opt to use the
F673N and FQ674N pair because this pair covers
the largest extent of the nebula.
The FQ674N filter isolates {\sii}\,6731\,{\AA} (panel\,d)
emission fairly well (less than 1\,\% difference; panel\,f),
while the FQ672N filter suffers from line blending 
as high as few 10s of \% (panel\,e).
We note that the NE and SE edges of the map are affected 
by the quad-filter edge effect 
(\S\,6.5 of \citealt{dressel2019}; panel\,f).

As we will see below, the {\sii} maps at 6717\,{\AA} (panel\,c) and 
6731\,{\AA} (panel\,d) play a critical role in determining {\Ne}.
Hence, the extent of {\sii} emission 
sets the maximum spatial extent 
where the present full 2-D plasma diagnostics would be valid. 
Practically, this means that results of the subsequent 
2-D plasma diagnostics are valid only in the NW quadrant of the main ring
(see \S\,\ref{sect:others}).

\subsection{{\hg} and {\oiii} 4363\,{\AA}}

\begin{figure}
    \centering
	\includegraphics[width=\columnwidth]{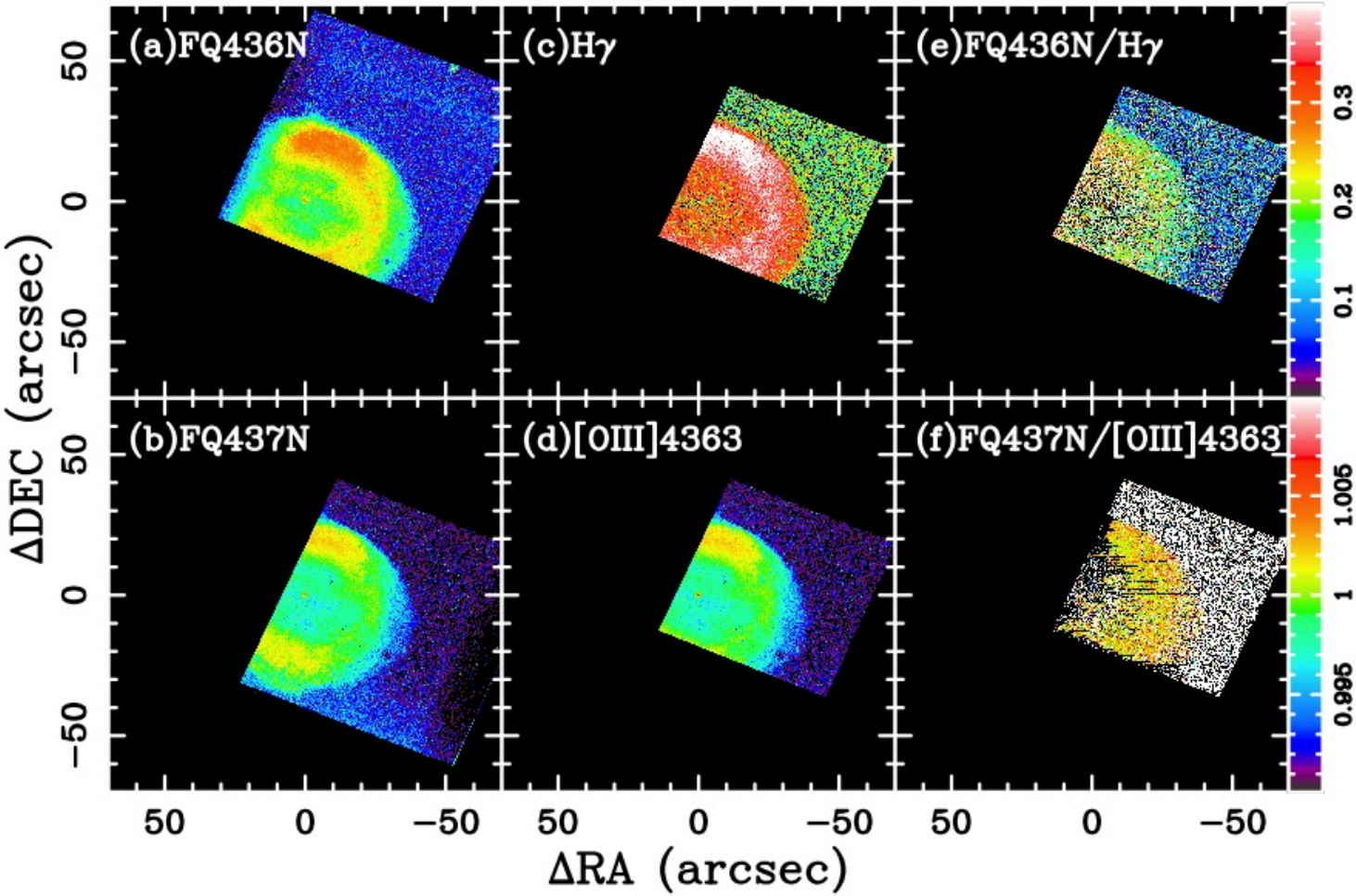}
	\caption{Same as Fig.\,\ref{fig:ha}, but for 
	the {\hg} and {\oiii}\,4363\,{\AA} line maps
	from the FQ436N and FQ437N images
	and
	log-scaled flux distribution
	from $10^{-16.5}$ to $10^{-18.5}$
	erg\,s$^{-1}$\,cm$^{-2}$\,pix$^{-1}$.}
    \label{fig:hg}
\end{figure}

Fig.\,\ref{fig:hg} presents the original WFC3 quad-filter images of 
FQ436N (panel\,a) and FQ437N (panel\,b)
and 
the QP-extracted line flux maps of 
{\hg} (panel\,c) and {\oiii}\,4363\,{\AA} (panel\,d),
as well as
the corresponding original-to-QP line flux ratio maps
(panels\,e--f).
The {\hg} emission captured by the FQ436N filter is 
very little as the system throughput at {\hg} is only 1.7\,\%.
Hence, the recovered {\hg} surface brightness 
is fairly uncertain (${\rm S/N} < 2$ even at the brightest region).
Even if a reasonably high S/N is achieved in the {\hg} map, 
because the transition energy of {\hg} is higher than {\ha} and {\hb},
the spatial extent covered by {\hg} is more restricted,
covering the higher-temperature regions than {\ha} and {\hb} do.
For the present work, therefore, we adopt the QP-recovered
{\ha}\,6563\,{\AA} (Fig.\,\ref{fig:ha}e), 
{\nii}\,6583\,{\AA} (Fig.\,\ref{fig:ha}f), 
{\sii} 6717\,{\AA} (Fig.\,\ref{fig:s2}e), and 
{\sii}\,6731\,{\AA} (Fig.\,\ref{fig:s2}f) maps
as well as
the original
{\hb} ($= {\rm F486N}$)
and 
{\nii}\,5755\,{\AA} ($= {\rm FQ575N}$; Fig.\,\ref{fig:linemaps}h)
maps because of their reasonable S/N and decent spatial coverage.

\begin{acknowledgments}
Based on observations made with the NASA/ESA Hubble Space Telescope, and obtained from the Hubble Legacy Archive, which is a collaboration between the Space Telescope Science Institute (STScI/NASA), the Space Telescope European Coordinating Facility (ST-ECF/ESA) and the Canadian Astronomy Data Centre (CADC/NRC/CSA).

This research made use of 
Astropy, a community-developed core Python package for Astronomy \citep{2013A&A...558A..33A,2018AJ....156..123A}
and PyNeb, a toolset dedicated to the analysis of emission lines \citep{pyneb},
as well as
OSQP, a convex quadratic programs solver \citep{osqp}.

TU was supported partially by 
the National Aeronautics and Space Administration under 
Grant No.\ NNX15AF24G issued through the Mission Directorate
and by
the Japan Society for the Promotion of Science (JSPS) 
through its invitation fellowship program (FY2020, long-term).
MO was supported by JSPS Grants-in-Aid for Scientific Research(C) (JP19K03914).

Authors thank Dr.\ Bob O'Dell for having email discussions about 
the use of the present {\sl HST\/}/WFC data set for NGC\,6720 and 
providing valuable comments after reading an early version of the 
manuscript.
\end{acknowledgments}

\vspace{5mm}
\facilities{HST(WFC3)}

\software{Astropy \citep{2013A&A...558A..33A,2018AJ....156..123A},  
          PyNeb \citep{pyneb},
          PySynphot \citep{pysynphot},
          OSQP \citep{osqp}
          }

\bibliography{ppap_pasp_pdf}{}
\bibliographystyle{aasjournal}

\end{document}